\documentclass{aa}
\newcommand{\rosi}{\textit{eROSITA}\xspace}
\newcommand{\xmm}{\textit{XMM-Newton}\xspace}
\newcommand{\chandra}{\textit{Chandra}\xspace}
\newcommand{\rosat}{\textit{ROSAT}\xspace}
\newcommand{\suzaku}{\textit{Suzaku}\xspace}
\usepackage{natbib,twoopt}
\usepackage{xcolor}
\usepackage[breaklinks=true, colorlinks=true, linkcolor=blue, citecolor=blue]{hyperref} 
\bibpunct{(}{)}{;}{a}{}{,}             
\makeatletter
\newcommandtwoopt{\citeads}[3][][]{\href{http://adsabs.harvard.edu/abs/#3}%
    {\def\hyper@linkstart##1##2{}%
     \let\hyper@linkend\@empty\citealp[#1][#2]{#3}}}
  \newcommandtwoopt{\citepads}[3][][]{\href{http://adsabs.harvard.edu/abs/#3}%
    {\def\hyper@linkstart##1##2{}%
     \let\hyper@linkend\@empty\citep[#1][#2]{#3}}}
  \newcommandtwoopt{\citetads}[3][][]{\href{http://adsabs.harvard.edu/abs/#3}%
    {\def\hyper@linkstart##1##2{}%
     \let\hyper@linkend\@empty\citet[#1][#2]{#3}}}
  \newcommandtwoopt{\citeyearads}[3][][]%
    {\href{http://adsabs.harvard.edu/abs/#3}
    {\def\hyper@linkstart##1##2{}%
     \let\hyper@linkend\@empty\citeyear[#1][#2]{#3}}}
\makeatother
\usepackage{graphicx}
\usepackage{multirow}
\usepackage{comment}
\usepackage{siunitx}  
\usepackage{txfonts}
\usepackage{subfig}
\usepackage{sidecap}
\sidecaptionvpos{figure}{c}
\begin{document}

   \title{The \rosi view of the Abell 3391/95 field: The Northern Clump}

   \subtitle{The largest infalling structure in the longest known gas filament
   observed with \rosi, \xmm, and \chandra}

   \author{Angie Veronica
          \inst{1}
          \and
          Yuanyuan Su\inst{2}
          \and
          Veronica Biffi\inst{5, 6, 7}
          \and
          Thomas H. Reiprich\inst{1}
          \and
          Florian Pacaud\inst{1}
          \and
          Paul E. J. Nulsen\inst{3, 13}
          \and
          Ralph P. Kraft\inst{3}
          \and
          Jeremy S. Sanders\inst{4}
          \and
          Akos Bogdan\inst{3}
          \and
          Melih Kara\inst{12}
          \and
          Klaus Dolag\inst{5}
          \and
          J\"urgen Kerp\inst{1}
          \and
          B\"arbel S. Koribalski\inst{9, 10}
          \and
          Thomas Erben\inst{1}
          \and
          Esra Bulbul\inst{4}
          \and
          Efrain Gatuzz\inst{4}
          \and
          Vittorio Ghirardini\inst{4}
          \and
          Andrew M. Hopkins\inst{8}
          \and
          Ang Liu\inst{4}
          \and
          Konstantinos Migkas\inst{1}
          \and
          Tessa Vernstrom\inst{11}
          }

   \institute{Argelander-Institut f\"ur Astronomie (AIfA), Universit\"at Bonn, Auf dem H\"ugel 71, 53121 Bonn, Germany\\
              \email{averonica@astro.uni-bonn.de}
         \and
              Physics and Astronomy, University of Kentucky, 505 Rose street, Lexington, KY 40506, USA
        \and      
         Center for Astrophysics | Harvard and Smithsonian, 60 Garden Street, Cambridge, MA 02138, USA
        \and
        Max-Planck-Institut f\"ur extraterrestrische Physik, Gießenbachstraße 1, 85748 Garching, Germany
        \and
        Universitaets-Sternwarte Muenchen, Ludwig-Maximilians-Universit\"at M\"unchen, Scheinerstraße, M\"unchen, Germany
        \and
        INAF - Osservatorio Astronomico di Trieste, via Tiepolo 11, I-34143 Trieste, Italy
        \and
        IFPU - Institute for Fundamental Physics of the Universe, Via Beirut 2, I-34014 Trieste, Italy
        \and
        Australian Astronomical Optics, Macquarie University, 105 Delhi Rd, North Ryde, NSW 2113, Australia
        \and
        CSIRO Astronomy and Space Science, P.O. Box 76, Epping, NSW 1710, Australia
        \and
        Western Sydney University, Locked Bag 1797, Penrith, NSW 2751, Australia
        \and
        CSIRO Astronomy and Space Science, PO Box 1130, Bentley, WA 6102, Australia
        \and
        Institute for Astroparticle Physics, Karlsruhe Institute of Technology, 76021 Karlsruhe, Germany
        \and
        ICRAR, University of Western Australia, 35 Stirling Hwy, Crawley, WA 6009, Australia}
   \date{Received ; accepted}

 
  \abstract
   {Galaxy clusters grow through mergers and the accretion of substructures along large-scale filaments. Many of the missing baryons in the local Universe may reside in such filaments as the warm-hot intergalactic medium (WHIM).}
   {SRG/\rosi performance verification (PV) observations revealed that the binary cluster Abell~3391/3395 and the Northern Clump (the MCXC J0621.7-5242 galaxy cluster) are aligning along a cosmic filament in soft X-rays, similarly to what has been seen in simulations before. We aim to understand the dynamical state of the Northern Clump as it enters the atmosphere ($3\times R_{200}$) of Abell~3391.}
   {We analyzed joint \rosi, \xmm, and \chandra observations to probe the morphological, thermal, and chemical properties of the Northern Clump from its center out to a radius of 988\:kpc ($R_{200}$).
   We utilized the ASKAP/EMU radio data, the DECam optical image, and the {\it Planck y}-map to study the influence of the wide-angle tail (WAT) radio source on the Northern Clump's central intracluster medium (ICM). Using \rosi data, we also analyzed the gas properties of the Northern Filament, the region between the virial radii of the Northern Clump and the A3391 cluster. From the Magneticum simulation, we identified an analog of the A3391/95 system along with an infalling group resembling the Northern Clump.} 
   {The Northern Clump is a weak cool-core cluster (WCC) centered on a WAT radio galaxy. The gas temperature over $0.2-0.5R_{500}$ is $k_BT_{500}=1.99\pm0.04$ keV. We employed the mass-temperature ($M-T$) scaling relation and obtained a mass estimate of $M_{500}=(7.68\pm0.43)\times10^{13}M_{\odot}$ and $R_{500}=(636\pm12)$ kpc.
   Its X-ray atmosphere has a boxy shape and deviates from spherical symmetry. We identify a southern surface brightness edge, likely caused by subsonic motion relative to the filament gas in the southern direction. At $\sim\! R_{500}$, the southern atmosphere (infalling head) appears to be 42\% hotter than its northern atmosphere. We detect a downstream tail pointing toward the north with a projected length of $\sim\!318\:\mathrm{kpc}$, plausibly the result of ram pressure stripping. Through a two-temperature fit, we identify a cooler component in the Northern Filament with $k_BT=0.68_{-0.64}^{+0.38}~\mathrm{keV}$ and $n_e=1.99_{-1.24}^{+0.88}\times10^{-5}~\mathrm{cm}^{-3}$, which are consistent within the expected ranges of WHIM properties. The analog group in the Magneticum simulation is experiencing changes in its gas properties and a shift between the position of the halo center and that of the bound gas, while approaching the main cluster pair.}
   {The Northern Clump is a dynamically active system and far from being relaxed. Its atmosphere is affected by an interaction with the WAT and by gas sloshing or its infall toward Abell~3391 along the filament, consistent with the analog group-size halo in the Magneticum simulation.}


   \maketitle
%

\section{Introduction}
\begin{SCfigure*}[][!h]
\includegraphics[width=0.65\textwidth]{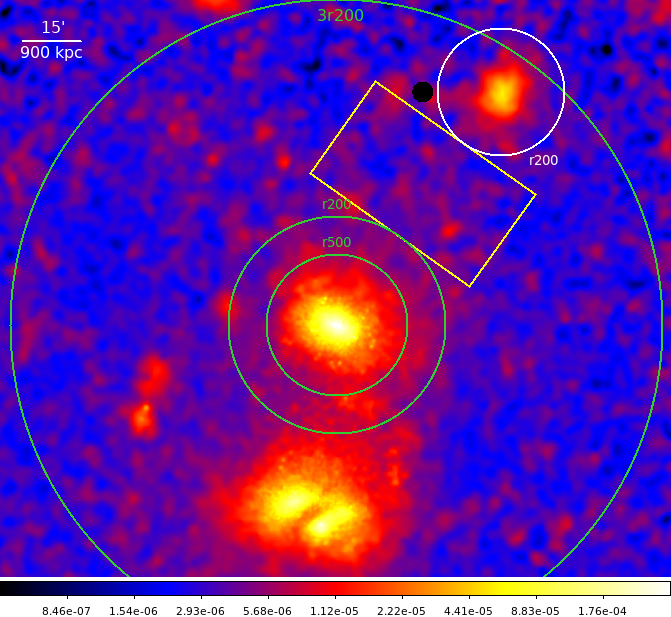}
\hspace{5pt}
\caption{\rosi particle-induced background subtracted, exposure corrected, absorption corrected image in the energy band $0.5-2.0~\mathrm{keV}$, zoomed-in to A3391 and the Northern Clump region. The image has been adaptively-smoothed with S/N set to 15. The point sources have been removed and refilled with their surrounding background values. The green circles depict some characteristic radii of A3391, while the white circle depicts the $R_{200}$ of the Northern Clump. The yellow box marks the region used for the spectral analysis of the Northern Filament.}
\label{fig:northclump_ero}
\end{SCfigure*}

Clusters of galaxies are formed by the gravitational infall and mergers of smaller structures \citep{Sarazin_2002,  Springel_2006, Markevitch_2007}. They are the high density nodes of the so-called Cosmic Web, connected by cosmic filaments \citep{Bond_1996}. These cosmic filaments are of importance, such that they are the passages for matter being accreted onto galaxy clusters \citep{West_1995, Bond_2010-filament}, as well as where the warm-hot intergalactic medium (WHIM) can be found \citep{Rost_2021}. The large-scale cosmological simulations of a $\Lambda$-dominated cold dark matter ($\Lambda$CDM) model \citep{Cen_1999, Dave_2001} predict that about $30-40\%$ of the total baryons in the Universe reside in the WHIM, and thus it is believed that the WHIM might be a solution to the missing baryon problem. However, with electron densities falling between $n_e\approx10^{-6}-10^{-4}~\mathrm{ cm}^{-3}$ and temperatures in the range of $T = 10^5-10^7$ K \citep{Nicastro_2017}, detecting the WHIM is rather difficult \citep{Werner_2008}. \cite{Bond_1996} predicted that the presence of the filaments are expected to be the strongest between highly clustered and aligned clusters of galaxies. Based on this prediction, it is sensible to conduct the search for the WHIM between a pair of galaxy clusters (e.g., \citealt{Kull_1999, Werner_2008, Yutaka_2008}), and also in the outskirts of galaxy clusters \citep[e.g.,][]{Reiprich_2013, Bulbul_2016, Nicastro_2018}. The binary galaxy cluster system Abell 3391 and Abell 3395 (A3391/95) is amongst the best candidates to conduct this search.

The A3391/95 system has been observed by numerous instruments of different wavelengths. Earlier studies of the system using \rosat and \textit{ASCA} show that the gas from the region between the northern cluster (A3391) and the double-peaked southern cluster (A3395S/N) is consistent with a filament \citep{Tittley_2001}. Meanwhile, more recent studies using \xmm, \chandra, and \suzaku show that the properties of the gas are more in agreement with tidally stripped intracluster medium (ICM) gas that originated from A3391 and A3395, indicating an early phase of the cluster merger \citep{Sugawara_2017, Alvarez_2018}.

The A3391/95 system is also one of the performance verification (PV) targets of the newly launched German X-ray telescope on board the Spectrum-Roentgen-Gamma (SRG) mission, the extended ROentgen Survey with an Imaging Telescope Array (\rosi). With large field-of-view (FoV) of $\sim\!1$ degree in diameter, the scan-mode observation, and superior soft X-ray effective area (the effective area of the combined seven \rosi telescope modules (TMs) is slightly higher than the one of \xmm (MOS1+MOS2+pn) in the $0.5-2.0~\mathrm{keV}$ energy band) \citep{Merloni_2012, Predehl_2021}, \rosi allows us to capture the A3391/95 system beyond $R_{100}$ with a good spectral resolution.

In total, four \rosi PV observations were performed on the A3391/95 system, including three raster scans and one pointed observation. By combining all the available observations, an area of $\sim\!15$ deg$^2$ is covered. This reveals a breathtaking view of the A3391/95 system that has only been seen in simulations before, such as the discovery of a long continuous filament ($\sim\!15$ Mpc) from the projected north toward the south of the A3391/95 system. This indicates that the system is part of a large structure. \rosi also found that the bridge gas consists of not only the hot gas, which further confirms the tidally stripped originated gas scenario, but also of warm gas from a filament \citep{Reiprich_2021}. A detailed \rosi study focusing on the bridge will be discussed in Ota et al. (in prep.) Additionally, some clumps that seem to fall into the system have also been discovered, two of which reside in the Northern and the Southern Filaments, namely the Northern Clump\footnote{As the extended objects detected in the A3391/95 field are called "clumps", we refer this specific galaxy cluster found in the north of the main system as the Northern Clump} (MCXC J0621.7-5242 cluster, \citealt{Piffaretti_2011} and also referred to as MS 0620.6-5239 cluster, \citealt{De_Grandi_1999, Tittley_2001}) and the Little Southern Clump, respectively \citep{Reiprich_2021}.

After \rosi's discovery, follow-up observations were performed by \xmm and \chandra on the Northern Clump. This biggest infalling clump in the A3391/95 field is a galaxy cluster, located at redshift $z=0.0511$ \citep{Tritton_1972, Piffaretti_2011}. It hosts a Fanaroff-Riley type I (FRI) radio source, PKS 0620-52 \citep{Trussoni_1999, Venturi_2000}. This source is associated with the cluster's brightest cluster galaxy (BCG), 2MASX J06214330–5241333 \citep{Brueggen_2020}. This source is featured with a pair of asymmetric wide-angle tail (WAT) radio lobes \citep{Morganti_1993, Trussoni_1999, Venturi_2000}, with the northeast lobe extending up to $4'$ (240 kpc), while the northwestern lobe up to $2.5'$ (150 kpc) \citep{Brueggen_2020}.

In this work, we investigate the morphology, thermal, and chemical properties of the Northern Clump utilizing all available \rosi, \xmm, and \chandra observations of the cluster. The cluster core is probed further with the help of the ASKAP/EMU radio data \citep{ASKAPEMU_2011}, the {\it Planck-y} map, and the DECam optical data. Furthermore, we identify an infalling galaxy group, resembling the observed Northern Clump, in the analog A3391/95 system from the Magneticum Simulation \citep{Biffi_2021} and discuss its properties in comparison to observations to support their interpretation.

This paper is organized as follows: in Sect. 2, we describe the observations, the data reduction steps, and the analysis strategy. In Sect. 3, we present and discuss the analysis results. In Sect. 4, we provide some insights about the analog A3391/95 system and the infalling group from the Magneticum Simulation. In Sect. 5, we summarize our findings and conclude the results.

Unless stated otherwise, all uncertainties are at the 68.3\% confidence interval. The assumed cosmology in this work is a flat $\Lambda$CDM cosmology, where $\Omega_m = 0.3$, $\Omega_\Lambda = 0.7$, and $H_0 = 70$ \si{km/s/Mpc}. At the Northern Clump redshift $z=0.0511$, $1''$ corresponds to 0.998 kpc.

\section{Data reduction and analysis}
We list all \rosi, \xmm, and \chandra observations used for imaging and spectral analysis in Table \ref{tab:obs}.

\begin{table*}[!h]
    \centering
    \caption{All observations of the Northern Clump used in this work.}
    \begin{tabular}{c c c c c}
    \hline
    \hline
Observing Date & ObsID & Exposure$^*$ [ks] & R.A. (J2000) & Dec. (J2000) \\
\hline
\multicolumn{5}{c}{\rosi}\\
\hline
October 2019 & 300005 (scan) & 55 & 06h 26m 49.44s  & -54d $04' 19.20''$  \\
October 2019 & 300006 (scan) & 54 & 06h 26m 49.44s  & -54d $04' 19.20''$  \\
October 2019 & 300016 (scan) & 58 & 06h 26m 49.44s  & -54d $04' 19.20''$  \\
October 2019 & 300014 (pointed) & 35 & 06h 26m 49.44s  & -54d $04' 19.20''$  \\
\hline
\multicolumn{5}{c}{\xmm}\\
\hline
March 2020 & 0852980601 & 73, 77, 65 & 06h 21m 42.55s & -52d $41' 47.4''$ \\
\hline
\multicolumn{5}{c}{\chandra}\\
\hline
October 2010 & 11499 & 20 & 06h 21m 43.30s & -52d $41' 33.3''$ \\
April 2020 & 22723 & 30 & 06h 21m 25.05s & -52d $43' 09.9''$ \\
May 2020 & 22724 & 30 & 06h 21m 42.53s & -52d $45' 24.7''$ \\
April 2020 & 22725 & 30 & 06h 21m 22.30s & -52d $42' 10.7''$ \\
\hline
\multicolumn{5}{l}{\footnotesize $^*$The exposure times listed for \rosi are the average filtered exposure time across the available TMs of each observation.}\\
\multicolumn{5}{l}{\footnotesize ~~The listed \xmm exposure times are the values after the SPF filtering for MOS1, MOS2, and pn cameras, respectively.}\\
\multicolumn{5}{l}{\footnotesize ~~The exposure times listed for \chandra are also the filtered exposure time.}\\
\hline
\hline
    \end{tabular}
    \label{tab:obs}
\end{table*}

\subsection{\rosi}
The A3391/95 PV observations are listed in Table \ref{tab:obs}. In this work we used \rosi data processing of configuration c001. The data reduction of all 16 \rosi data sets were realized using the extended Science Analysis Software (eSASS, \citealt{Brunner_2021}) version \texttt{eSASSusers\_201009}. The \rosi data reduction and image correction steps are described in great detail in Sect. 2.1 of \cite{Reiprich_2021}. The image correction includes particle-induced background (PIB) subtraction, exposure correction, and Galactic absorption correction across the FoV.

\subsubsection{Imaging analysis}
The energy band for the imaging analysis was restricted to $0.5-2~\mathrm{keV}$, with the lower energy limit used for the telescope modules (TMs) with on-chip filter (TM1, 2, 3, 4, 6; the combination of these TMs is referred to as TM8) set to 0.5 keV, while for the TMs without on-chip filter (TM5 and 7; the combination of these TMs is referred to as TM9) it was set to 0.8 keV due to the optical light leak contamination \citep{Predehl_2021}. The count rates of the final combined image correspond to an effective area given by one TM with on-chip filter in the energy band $0.5-2~\mathrm{keV}$. The final PIB subtracted, exposure corrected, and Galactic absorption corrected image is shown in Fig.~\ref{fig:northclump_ero}. The image has been adaptively-smoothed to enhance low surface brightness emission and large scale features, with signal-to-noise ratio (S/N) was set to 15. We note that the Northern Clump appears boxy, which could be the result of ram-pressure stripping as it interacts with the atmosphere of the A3391 cluster or the filament, or both. In this work, we focus on studying the properties of the Northern Clump out to its $R_{200}$, while a more detailed study about the filaments observed by \rosi around the A3391/95 system will be carried out by Veronica et al. (in prep).
\par
Thanks to the large \rosi FoV and the SRG scan mode, in which most of the observations were taken, the cosmic X-ray background (CXB) could be modeled from an annulus with an inner and outer radius of $25'$ and $35'$, respectively. The annulus is centered at the X-ray emission peak of the cluster determined by \chandra at $(\alpha,\delta)=(6\! :\! 21\! :\! 43.344,\: -52\! :\! 41\! :\! 33.0)$. Since the CXB region is at $1.5-2.1R_{200}$\footnote{All the cluster radii in this work are calculated using the relations stated in \cite{Reiprich_2013}, e.g., $r_{500}\approx0.65r_{200}$, $r_{100}\approx1.36r_{200}$, and $r_{2500}\approx0.28r_{200}$.}, it ensures minimum cluster emission in the CXB region. This region also stays beyond $R_{100}$ of A3391. Within the $13.5'$ radius, we utilized the \xmm point source catalog and beyond this radius we used \rosi detected point source catalog, which was generated following the same procedure as \xmm (see Sect. \ref{sec:xmm_datareduction}). The point sources are excluded both in imaging and spectral analyses.

\subsubsection{Spectral analysis}
All \rosi spectra, the Ancillary Response Files (ARFs), and the Response Matrix Files (RMFs) for the source and background region were extracted using the eSASS task \texttt{srctool}. Unless stated otherwise, the spectral fitting for all \rosi spectra was performed with \texttt{XSPEC} \citep{XSPEC} version: 12.10.1.
\par
We use the third \rosi scan observation (ObsID: 300016), where all seven \rosi TMs are available, for the spectral analysis of the Northern Clump. Due to the contamination by the nearby bright star, Canopus, some portions of the data were rejected from the analysis \citep[see][]{Reiprich_2021}, which further reduces the number of photons for the Northern Clump. To ensure enough photon counts, the only \rosi spectral analysis done for the Northern Clump in this work is to constrain the $k_BT_{500}$ from an annulus with inner and outer radii of $2.4'$ and $6'$ (corresponding to $0.2-0.5R_{500}$, see Subsect. \ref{sec:xmm_spectralanalysis}). The same definition of CXB region as in the imaging analysis was used.
\par
Spectral analysis was also performed for the filament region between the $R_{200}$ of the Northern Clump and the A3391 cluster, the Northern Filament. The filament region was chosen as a box centered at $(\alpha,\delta)=(6\! :\! 23\! :\! 54.654,\: -53\! :\! 05\! :\! 26.637)$ with height and width of $25'$ and $50'$, respectively. The configuration of the Northern Filament box is shown in Fig.~\ref{fig:northclump_ero} (yellow box). The CXB region for this analysis is a circular region with a radius of $30'$ centered at $(\alpha,\delta)=(6\! :\! 18\! :\! 18.275,\: -54\! :\! 33\! :\! 37.880)$. The average hydrogen column density at the box and its corresponding CXB region is $N_H = 0.048\times10^{22}~\mathrm{atoms~cm}^{-2}$.
\par
\begin{figure}[h!]
\centering
\includegraphics[width=\columnwidth]{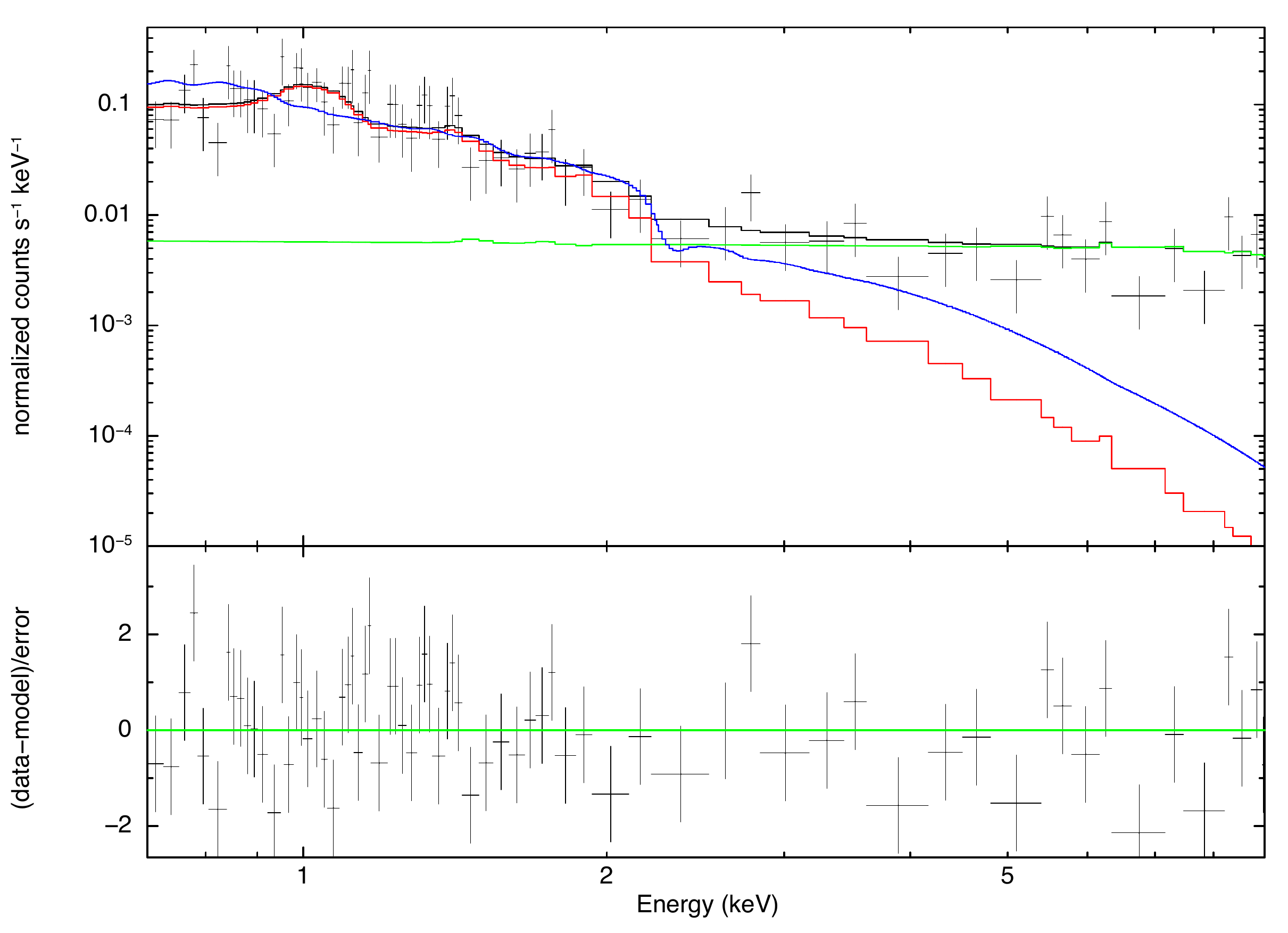}
\caption{\rosi spectrum of the Northern Clump from an annulus of $2.4-6'$, fitted in the energy range $0.7-9.0$ keV. The spectra and the corresponding response files of TM3 and TM4 are merged for a better visualization. The black points are the Northern Clump spectral data, while the black, red, green, and blue lines represent the total model, the cluster model, the instrumental background model, and the CXB model, respectively.}
\label{fig:eROSITA_spectrum}
\end{figure}

We followed the X-ray spectral fitting procedure described in \cite{Ghirardini_2021} and \cite{Liu_2021} with some modifications for consistency with the \xmm and \chandra spectral fitting.
The total \rosi model includes the absorbed thermal emission for the cluster, which is modeled using \texttt{phabs}$\times$\texttt{apec}, two absorbed thermal models for the Milky Way Halo (MWH) and the Local Group (LG), an unabsorbed thermal model for the Local Hot Bubble (LHB), and an absorbed power-law for the unresolved AGN. The parameters of the \rosi CXB components are listed in Table \ref{tab:ero_sky_BG}. The instrumental background was modeled based on the results of the \rosi Filter Wheel Closed (FWC) data analysis. The normalizations of the instrumental background components were thawed during the fit. The fitting was restricted to the $0.7-9.0~\mathrm{keV}$ band, with the lower energy limit set to 0.8 keV for TM5 and TM7 and to 0.7 keV for the other TMs. Source and CXB spectra for the individual TMs were fitted simultaneously. An example of \rosi spectrum and its fitted model is shown in Fig.~\ref{fig:eROSITA_spectrum}.
\par
The solar abundance table from \cite{Asplund_2009} was adopted and the C-statistic \citep{Cash_1979} was implemented.

\begin{table}[!h]
    \centering
    \caption{Information of the parameters and their starting fitting values of the \rosi CXB components for the Northern Clump $R_{500}$ analysis.}
    \begin{tabular}{c c c c }
    \hline
    \hline
Component & Parameter & Value & Comment\\
    \hline
$phabs$ &  $N_H^\dagger$ & $0.04$  & fix\\
\hline
$apec$ (LHB) & $k_BT$ [keV] & 0.099 & fix\\
& $Z~[Z_\odot]$ & 1& fix\\
& $z$ & 0 & fix\\
& $norm^*$ & $1.0\times10^{-8}$ & vary\\
\hline
$apec$ (MWH) &  $k_BT$ [keV] & 0.225 & fix\\
& $Z~[Z_\odot]$ & 1& fix\\
& $z$ & 0 & fix\\
& $norm^*$ & $1.0\times10^{-8}$ & vary\\
\hline
$apec$ (LG) &  $k_BT$ [keV] & 0.5 & vary\\
& $Z~[Z_\odot]$ & 1& fix\\
& $z$ & 0 & fix\\
& $norm^*$ & $1.0\times10^{-8}$ & vary\\
\hline
$pow$ & $\Gamma$ & 1.4 & fix\\
(unresolved AGN) & $norm^{**}$ & $3.4\times10^{-7}$ & vary\\
\hline
\multicolumn{4}{l}{\footnotesize $^\dagger$[10$^{22}$ atoms cm$^{-2}$]}\\
\multicolumn{4}{l}{\footnotesize $^*[\mathrm{cm}^{-5}/\mathrm{arcmin}^2]$}\\
\multicolumn{4}{l}{\footnotesize $^{**}$[photons/keV/cm$^2$/s/arcmin$^2$ at 1 keV]}\\
    \hline
    \hline
    \end{tabular}
    \label{tab:ero_sky_BG}
\end{table}

\subsection{\xmm}\label{sec:xmm_datareduction}
The data reduction steps of the \xmm data sets of the Northern Clump observation were realized using HEASoft version 6.25 and the Science Analysis Software (SAS) version 18.0.0 (\texttt{xmmsas\_20190531\_1155}). For all \xmm instruments, we filtered out the time intervals contaminated by soft proton flares (SPF) (e.g., \citealt{deLuca_2004, Kuntz_2008, Snowden_2008}). The procedure started by constructing the light curve in the energy band $0.3-10.0~\mathrm{keV}$ from the entire FoV. The mean value of the counts was estimated by fitting a Poisson law to the counts histogram. Then a $\pm3\sigma$ clipping was applied, such that any time intervals above and below the $3\sigma$ threshold were considered contaminated by the SPF and rejected from further steps.

Light curves from each of \xmm's detectors are shown in Appendix \ref{app:xmm}. To check whether there was still any persisting soft proton background in our data, we calculated the ratio of count rates within the FoV (IN) to those of the unexposed region of the MOS detectors (OUT) \citep{deLuca_2004}. The obtained IN/OUT ratios for each detector are around unity, which implies that the retained data are not contaminated by soft protons. After this SPF filtering, we are left with $\sim\!73$ ks, $\sim\!77$ ks, and $\sim\!65$ ks for MOS1, MOS2, and pn, respectively.

We checked for ``anomalous state'' in the detectors' chips. This is usually indicated by a low hardness ratio and high total background rate \citep{Kuntz_2008}. Since, in such cases, there is not a robust method yet to model the spectral distribution of the low energy background, any CCD affected by this anomalous state was rejected in further analysis. This resulted in the conservative exclusion of MOS1-CCD4 and MOS2-CCD5. In addition, MOS1-CCD6 and MOS1-CCD3 were not operational during this observation, due to the previous micro-meteorite impacts \citep{XMM_handbook}.

Utilizing the FWC data, the quiescent particle background (QPB) \citep{Kuntz_2008} was modeled and then subtracted from the observation. We followed the procedure described in \cite{Ramos-Ceja_2019}. First, the cleaned observation data were matched with the suitable FWC data and then this FWC template was re-projected to the sky position. Afterwards, rescaling factors were calculated to renormalize the QPB template to a level comparable with that of the observation. This renormalization was calculated separately for each MOS CCD and pn quadrant based on their individual count rates in unexposed regions. The renormalization of the central MOS CCD chips that do not have any unexposed regions relies on the unexposed regions of the best correlated surrounding CCD chips as determined by \cite{Kuntz_2008}. For spectroscopic analysis, global renormalization factors were calculated in the energy bands $2.5-5.0~\mathrm{keV}$ and $8.0-9.0~\mathrm{keV}$ for MOS cameras, and $2.5-5.0~\mathrm{keV}$ for pn camera. This energy selection aims to include the bands where the instrumental background would have the most impact on the spectral analysis, while avoiding the fluorescence X-ray (FX) emission lines present for the EPIC cameras. For imaging analysis, on the other hand, the template renormalization process relies on the exact same band that was used for the science analysis. 

Another contamination when studying the ICM gas might come from the point sources that are distributed across the FoV. Therefore, it is required to detect and excise the point sources, for example, in the calculation of surface brightness profiles and spectral analysis. To detect the point sources in the observation, we implemented the procedure described in \cite{Pacaud_2006} and \cite{Ramos-Ceja_2019}. Firstly, wavelet filtering was applied on the combined detector image in the soft energy band ($0.5-2.0~\mathrm{keV}$). Afterwards, an automatic detection of point sources and catalog generation were performed on the wavelet-filtered image by the Source Extractor software \citep[\texttt{SExtractor},][]{SExtractor}. Lastly, we carried out a manual inspection on the soft and hard energy band images ($2.0-10.0~\mathrm{keV}$) to list any point sources that were missed by the automatic detection.
To generate a surface brightness image, the areas where the point sources were removed were refilled with random background photons, that is the ghosting procedure. The ghosted surface brightness images were used only for visual inspection.

\subsubsection{Imaging analysis}
\begin{figure}[h!]
\centering
\includegraphics[width=\columnwidth]{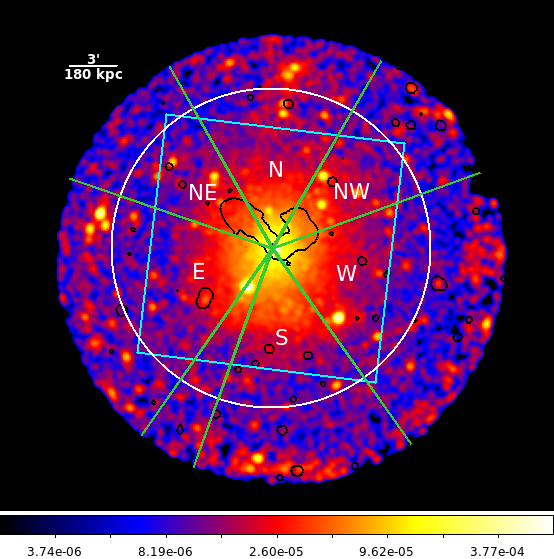}
\caption{\xmm count rate image of the Northern Clump in the energy band $0.5-2.0$ keV, overlaid with the ASKAP/EMU radio contour (black). The anomalous CCD chips are removed and the QPB is subtracted from the image. The sectors are south (S), west (W), northwest (NW), north (N), northeast (NE), and east (E). The white circle and cyan square depict the outermost annuli of corresponding setups. The radius of the circle is $10'$ and the half-width of the square is $7.5'$. Gaussian smoothing with kernel radius of 7 pixels is applied to the image.}
\label{fig:xmm_SB_sectors}
\end{figure}

We constructed \xmm's X-ray surface brightness ($S_X$) profiles using the combined photon image and exposure map from the \xmm detectors (MOS1, MOS2, and pn) in the energy band $0.5-2.0$ keV. Since the ghosting procedure used for generating surface brightness images does not represent the exact true counts from the ICM gas, we preferred to mask these point source regions in the calculation of the $S_X$ profile.
The first $S_X$ profile was calculated from 60 concentric annuli placed around the \chandra X-ray emission peak at $(\alpha,\delta)=(6\! :\! 21\! :\! 43.344,\: -52\! :\! 41\! :\! 33.0)$.
With $10''$ increments between each annulus, the whole arrangement covers radial distance of $10'$ from the center. While the QPB was subtracted from the data, the CXB was estimated from an annulus with an inner radius of $12'$ and an outer radius of $13.25'$, respectively. Then, the estimated $S_X$ from this background region was subtracted from the source regions.

A second $S_X$ profile was also constructed. The setup consists of 45 concentric box-shaped ``annuli'' centered at the same peak coordinates. Between each annulus there is a $20''$ width increment. This implies that the width of each box-shaped annulus corresponds to the diameter of each circular annulus. The half-width of the outermost box annulus is $7.5'$. The setup was angled at $173^\circ$, which was chosen visually to follow the shape of the Northern Clump. The same X-ray background region as in the circular annuli setup was used. To investigate, whether the boxiness could be an artifact originating from the MOS detectors' central chips, a $S_X$ profile from pn-only data and using the box annuli setup was calculated as well.
\par
We fit each of the setups with a single $\beta$-model \citep{Cavaliere_1976}, which is written as 

\begin{equation}
    S_X(R) = S_X(0)\left(1 + \frac{R^2}{r_c^2} \right)^{-3\beta + \frac{1}{2}},
\end{equation}
where $R$ is the projected distance from the center, $S_X(0)$ is the normalization factor, $r_c$ is the estimated core radius, and $\beta$ is the surface brightness slope.
\par
To better address the features in any particular direction, the $S_X$ profiles in the projected south, west, northwest, north, northeast, and east directions were constructed, as well. For these sectorized $S_X$ profiles, all \xmm detectors were employed in both circular and box annuli setups. The configuration of the sectors is displayed in Fig.~\ref{fig:xmm_SB_sectors}.
\begin{figure*}[!ht]
\centering
\includegraphics[width=\textwidth]{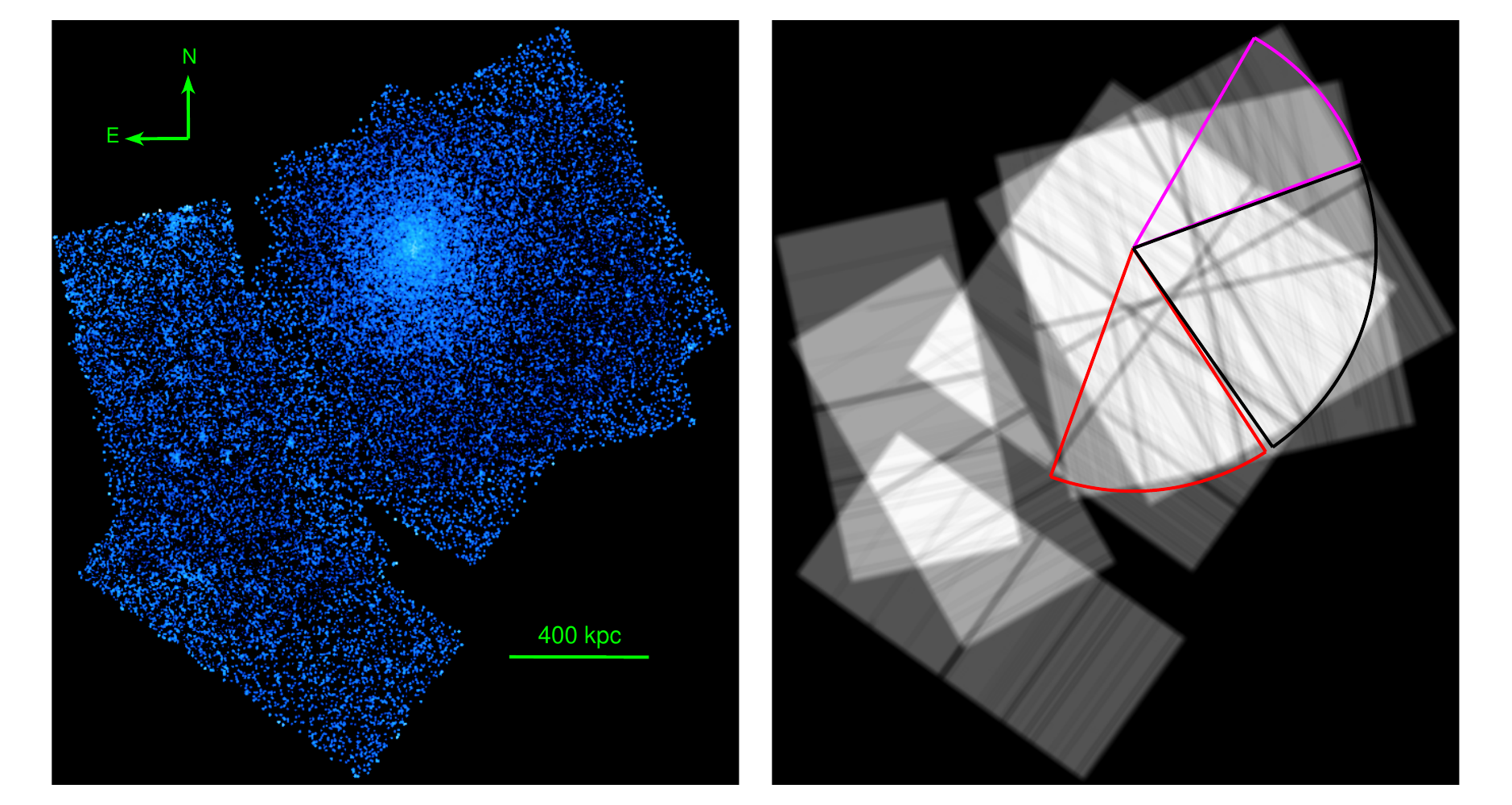}
\caption{{\it Left:} Mosaic \chandra ACIS-I image of the northern clump. The image is blanksky background subtracted, exposure corrected, and point-source refilled.
{\it Right:} The mosaic exposure map corresponding to the left image. Red, black, and magenta pie regions represent the regions used to derive the surface brightness profiles of the south, west, and northwest directions as shown in Fig~\ref{fig:chandra_profile}.}
\label{fig:chandra_image}
\end{figure*}

\subsubsection{Spectral analysis}\label{sec:xmm_spectralanalysis}
Unless stated otherwise, the spectral fitting for all \xmm spectra was performed with \texttt{XSPEC} \citep{XSPEC} version: 12.10.1 using the following model,
\begin{equation}
\begin{split}
\mathtt{Model_{XMM} =} &\quad\mathtt{constant\times(apec_1 + phabs\times(apec_2 +}\\
&\quad\mathtt{powerlaw) + gauss_1 + gauss_2 + gauss_3 + }\\
&\quad\mathtt{ gauss_4 + gauss_5)\, + phabs\times apec_3},\\
\end{split}
\label{eq:xmm_model}
\end{equation}
where the first term consists of the CXB and \xmm's known instrumental lines, scaled to the areas of the source regions (\texttt{constant}, specified in units of \si{arcmin^2}). The absorption parametrized by the hydrogen column density, $N_H$, along the line of sight is represented by \texttt{phabs}. The $N_H$ values used in this work are taken from the A3391/95 total $N_H$ map described in Sect. 2.5 of \cite{Reiprich_2021}. The X-ray emission components to account for the LHB and the MWH are represented by \texttt{apec$\mathtt{_1}$} and \texttt{apec$\mathtt{_2}$}, respectively. The power-law component, \texttt{powerlaw}, is to account for the unresolved AGN. The information regarding the CXB components and their starting fitting parameter values are listed in Table \ref{tab:sky_BG}. Since the FX emission lines of the \xmm cameras are not convoluted in the ARF, we modeled the lines with the other background components as the \texttt{gauss$\mathtt{_i}$}'s. The modeled FX lines include the Al-K$_\alpha$ at 1.48 keV (for MOS and pn cameras), Si-K$_\alpha$ at 1.74 keV (MOS cameras), Ni-K$_\alpha$ at 7.47 keV (pn camera), Cu-K$_\alpha$ at 8.04 keV (pn camera), and Zn-K$_\alpha$ at 8.63 keV (pn camera), respectively. The second term, \texttt{phabs}$\times$\texttt{apec$\mathtt{_3}$}, represents the absorbed cluster emission from the source region. The temperature $k_BT$, metallicity $Z$, and normalization $norm$ of this cluster emission component were left to vary and linked across the detectors.
\par
The \texttt{apec}'s parameter $norm$ is defined as

\begin{equation}
norm = \frac{10^{-14}}{4\pi[D_A(1+z)]^2}\int n_e n_H \mathrm{d}V,
\label{eq:norm}
\end{equation}
where $D_A$ is the angular diameter distance to the source in the unit of cm, while $n_e$ and $n_H$ are the electron and hydrogen densities in the unit of cm$^{-3}$, respectively.

\begin{table}[!h]
    \centering
    \caption{Information of the parameters and their starting fitting values of the \xmm CXB components.}
    \begin{tabular}{c c c c }
    \hline
    \hline
Component & Parameter & Value & Comment\\
    \hline
$phabs$ &  $N_H^\dagger$ & $0.046$  & fix\\
\hline
$apec_1$ (LHB) & $k_BT$ [keV] & 0.099 & fix\\
& $Z~[Z_\odot]$ & 1& fix\\
& $z$ & 0 & fix\\
& $norm^*$ & $1.7\times10^{-6}$ & vary\\
\hline
$apec_2$ (MWH) &  $k_BT$ [keV] & 0.225 & fix\\
& $Z~[Z_\odot]$ & 1& fix\\
& $z$ & 0 & fix\\
& $norm^*$ & $7.3\times10^{-7}$ & vary\\
\hline
$pow$ & $\Gamma$ & 1.4 & fix\\
(unresolved AGN) & $norm^{**}$ & $5\times10^{-7}$ & vary\\
\hline
\multicolumn{4}{l}{\footnotesize $^\dagger$[10$^{22}$ atoms cm$^{-2}$]}\\
\multicolumn{4}{l}{\footnotesize $^*[\mathrm{cm}^{-5}/\mathrm{arcmin}^2]$}\\
\multicolumn{4}{l}{\footnotesize $^{**}$[photons/keV/cm$^2$/s/arcmin$^2$ at 1 keV]}\\
    \hline
    \hline
    \end{tabular}
    \label{tab:sky_BG}
\end{table}

To ensure a cluster-emission-free background region, the \rosat All-Sky Survey (RASS) data were utilized to constrain all of the sky background components, in particular the MWH as well as CXB normalization. The RASS data were extracted using the X-ray background tool\footnote{\href{https://heasarc.gsfc.nasa.gov/Tools/xraybg_help.html}{\texttt{HEASARC: X-Ray Background Tool}}}. The specified background region is an annulus centered at the \chandra center, with an inner radius of $0.58^\circ$ and an outer radius of $0.9^\circ$. To avoid having cluster emission from A3391, we manually removed some of the RASS background region that is within $R_{100}$ of the A3391 cluster. This reduces the number of detected pixels from 38 to 34.
\par
The variety of overlap area between each annulus and each individual camera and CCD combination can result in some very low S/N spectra being extracted, for which a background model can not always be robustly constrained. For this reason, we opted to combine the spectra of all CCDs for a given annulus, and statistically subtract the rescaled FWC spectrum rather than modeling their spectra. The co-addition of rescaled spectra for the different CCDs implies that our background spectra no longer follow Poisson statistics and therefore we used the $\chi^2$-statistic for parameter estimation, after ensuring that each bin contains enough photons (>25). Additionally, since we use the RASS background, which is Gaussian data, to estimate the CXB normalizations, the choice is believed to be sensible. However, we also performed several tests using pgstat statistic from \texttt{XSPEC}, where the source data were treated as Poisson data (C-statistic) and the background as Gaussian. The resulting cluster parameters of these tests are always consistent within the $1\sigma$ error bars of the $\chi^2$-statistic fitting results. Hence, the choice of the used statistics for \xmm spectral fitting would not change any main conclusions. We present the results of the pgstat tests in Appendix \ref{App:NC_profles}, Fig.~\ref{fig:pgstat}. To minimize the effect of the detector noise in the lower energy, photons with $E<0.7$ keV were ignored. The solar abundance table from \cite{Asplund_2009} was adopted.
\par
We determined the $k_BT_{500}$ in order to calculate the $M_{500}$, consequently the $R_{500}$. We note that $k_BT_{500}$ was calculated through an iterative procedure, such that we extracted and fit spectra from an annulus centered at the \chandra center, varying the inner and outer radii until they correspond roughly to the $0.2-0.5R_{500}$ cluster region. Based on this, we obtained $k_BT_{500}=1.99\pm0.04$ keV from an annulus with inner and outer radii of $2.4'$ and $6.0'$, respectively. We input this value into the mass-temperature ($M-T$) scaling relation by \cite{Lovisari_2015},

\begin{equation}
\log(M/C1) = a \cdot \log(T/C2) + b,
\label{eq:scaling}
\end{equation}
where $a = 1.65\pm0.07$, $b=0.19\pm0.02$, $C_1=5\times10^{13}h_{70}^{-1} M_{\odot}$, and $C_2=2.0$ keV. Assuming spherical symmetry and taking 500 times the critical density of the Universe at the Northern Clump redshift as $\rho_{500}(z=0.0511)=4.82\times10^{-27}$ g cm$^{-3}$, we obtain $M_{500}=(7.68\pm0.43)\times10^{13}M_{\odot}$ and $R_{500}=(10.62\pm0.20)'\approx(636.03\pm12.03)$ kpc. These values are 19\% and 7\% lower than the reported values in \cite{Piffaretti_2011} ($M_{500}=9.47 \times 10^{13} M_{\odot} $ and $R_{500}=682$ kpc), which are determined using the $L-M$ relation, instead.
\par
We also analyzed the spectra in three different directions of the cluster, such as the south, west, and north with position angle (P.A.) from $250^\circ$ to $305^\circ$, $20^\circ$ to $305^\circ$, and  $70^\circ$ to $100^\circ$, respectively. At larger radii, i.e., $4-12.5'$, due to the removed MOS1-CCD chips we used pn-only spectra in the north, and MOS2+pn spectra in the south and west.
\par
We always keep the first bin as a full annulus, instead of a sector in any specific directions. We conducted a more detailed analysis to take into account the emission originated from the AGN of the BCG, such that we implemented three different fitting methods for this particular bin.
The first method is to mask the AGN with a circle of $15''$ radius. This is the method that is implemented to the central bin during the simultaneous fitting described above. The second method involves constraining the AGN photon index and normalization from \chandra observation. Here, the masking was lifted out. Since \xmm and \chandra observations of the Northern Clump were performed almost simultaneously, little to no temporal variation from the AGN can be expected \citep[e.g.,][]{Maughan_2019}. The method and the resulting AGN parameter values from the spectral fitting are discussed in Subsect. \ref{chandra_centralAGN}. We froze \chandra's parameter values during the \xmm spectral fitting of the first bin. Lastly, to account for the expected multi-temperature structure in this region, we performed a two-temperature structure fit in addition to the AGN component. Due to the statistics, the normalization and the metallicity of the multi-temperature components were linked, while the second temperature were set to be half of the first.

\subsection{\chandra}
We analyzed a total of 110 ks \chandra observations on the Northern Clump as listed in Table \ref{tab:obs}, including three ACIS-I observations taken in 2020 and one ACIS-S observation in 2010. CIAO 4.12 and CALDB 4.9.3 were used for the Chandra data reduction. All the observations were reprocessed from level 1 events using the CIAO tool \texttt{chandra\_repro} to guarantee the latest and consistent calibrations. 
Readout artifacts from the central bright AGN were removed from the reprocessed events using \texttt{acisreadcorr}.
We filtered background flares beyond $3\sigma$ using the light
curve filtering script \texttt{lc\_clean}; clean exposure times are listed
in Table \ref{tab:obs}. 
We combined all the observations to resolve faint point sources. The \chandra PSF varies significantly across the FoV. 
For each observation, we produced a PSF map for an energy of 2.3 keV and an enclosed count fraction of 0.393. 
We obtain an exposure-time weighted average PSF map from these four observations.
Point sources were detected in a $0.3-7.0$ keV mosaic image with \texttt{wavdetect}, supplied with this exposure-time weighted PSF map. The detection threshold was set to $10^{-6}$. The scales ranged from 1 to 8 pixels, increasing in steps of a factor of 2.
We adopted the blank sky background for the imaging and spectral analyses. A compatible blank sky background file was customized to each of the event file using the \texttt{blanksky} script. As shown in Subsect. \ref{sec:surb}, 
the surface brightness of our blank sky background is consistent with the sum of the stowed background and the astrophysical background.

\subsubsection{Imaging analysis}
We only include the three ACIS-I observations taken in 2020 for the imaging analysis due to the different backgrounds, responses, FoVs, and calibrations between these observations and the ACIS-S observation taken in 2010.  
All the observations were reprojected to a common tangent point using reproject\_obs.
Images in the $0.5-2.0$ keV energy band and their exposure maps were produced using \texttt{flux\_obs}. Point sources identified by \texttt{wavdetect} were replaced with the surface brightness from the immediate surrounding regions using \texttt{dmfilth}.  
Scaled blanksky images were produced using \texttt{blanksky\_image}. A final background subtracted and exposure corrected mosaic image of the Northern Clump is shown in Figure~\ref{fig:chandra_image}. 

\subsubsection{Spectral analysis}
We include all four existing \chandra observations for the spectral analysis. Spectra were extracted using \texttt{specextract} from point-source excised event files for regions of consecutive sectional annuli in the south and west directions, ranging from $\delta r=0.5'$ at the cluster center to $\delta r=1.67'$ at $r=6.68'$.
The corresponding response files and matrices were produced for each spectrum using \texttt{specextract}.
Blanksky background normalized to the count rate in the $9.5-12.0$\:keV energy band of each observation was applied. 
All the spectra were grouped to have at least one count per energy bin and restricted to the $0.7-7.0$\:keV range. The spectral fitting was performed with \texttt{PyXspec 2.0.3}. Our spectral model takes the form of \texttt{phabs$\times$apec}, representing a foreground-absorbed thermal component. The total $N_H$ value of $4\times10^{20}$\:cm$^{-2}$ is taken from the A3391/95 total $N_H$ map described in \cite{Reiprich_2021}.
C-statistics \citep{Cash_1979} was used for parameter estimation. The solar abundance table of
\cite{Asplund_2009} was adopted. Due to low photon counts, we found it necessary to fix the metallicity at $0.3Z_{\odot}$.

\subsubsection{Central AGN}\label{chandra_centralAGN}
We characterize the property of the central AGN detected at equatorial coordinates of $(\alpha,\delta)=(6\!:\!21\!:\!43.304,\: -52\! :\! 41\! :\! 33.153)$.
Spectra of the central $r=2^{\prime\prime}$ were extracted from clean events files with point sources included. Spectra from the surrounding $r=2-5^{\prime\prime}$ were used as the local background.
We fit the spectrum to a simple absorbed power-law model and fixed the column density $N_{H}=4\times 10^{20} cm^{-2}$. We obtain a steep photon index of $2.5\pm0.07$ and a best-fit luminosity of $5.7\pm0.3\times10^{41}$\:erg/s in the $2.0-10.0$\:keV when using the observation taken in 2020, which is used for the AGN modelling in the \xmm analysis.

\subsection{ASKAP/EMU, DECam, and Planck Data}
\label{emu}
We use the ASKAP/EMU radio image and DECam images, as well as the {\it Planck y}-map for overlays generated as described in detail in \cite{Reiprich_2021}.

\section{Results}
\subsection{Boxiness, tail, and X-ray cavity in images}\label{sec:image_features}
\begin{figure}[!ht]
   \centering
   \includegraphics[width=\columnwidth]{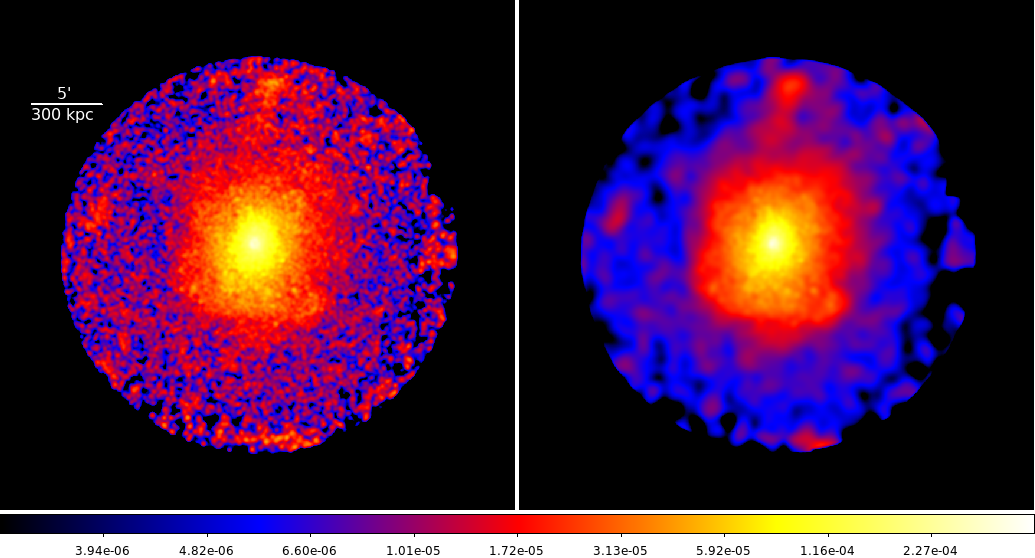}
   \includegraphics[width=\columnwidth]{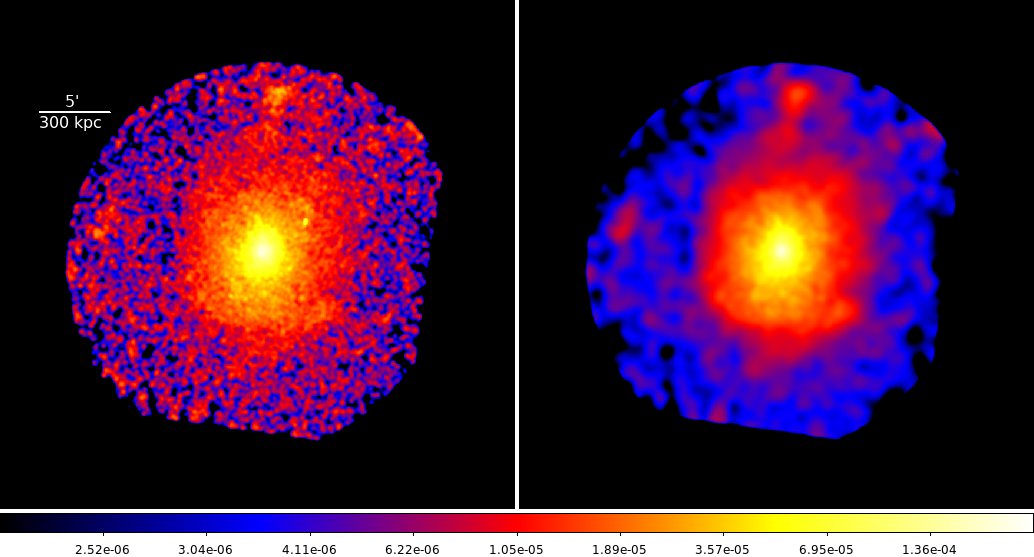}
   \caption{\xmm surface brightness images (\textit{left}) and adaptively-smoothed images (\textit{right}) in the energy band $0.5-2$ keV. The S/N of the adaptively-smoothed images is set to 15. \textit{Top:} MOS1+2+pn combined. \textit{Bottom:} pn-only.}
\label{fig:XMM_asmooth}%
\end{figure}

\begin{figure*}[!ht]
\centering
\includegraphics[width=\textwidth]{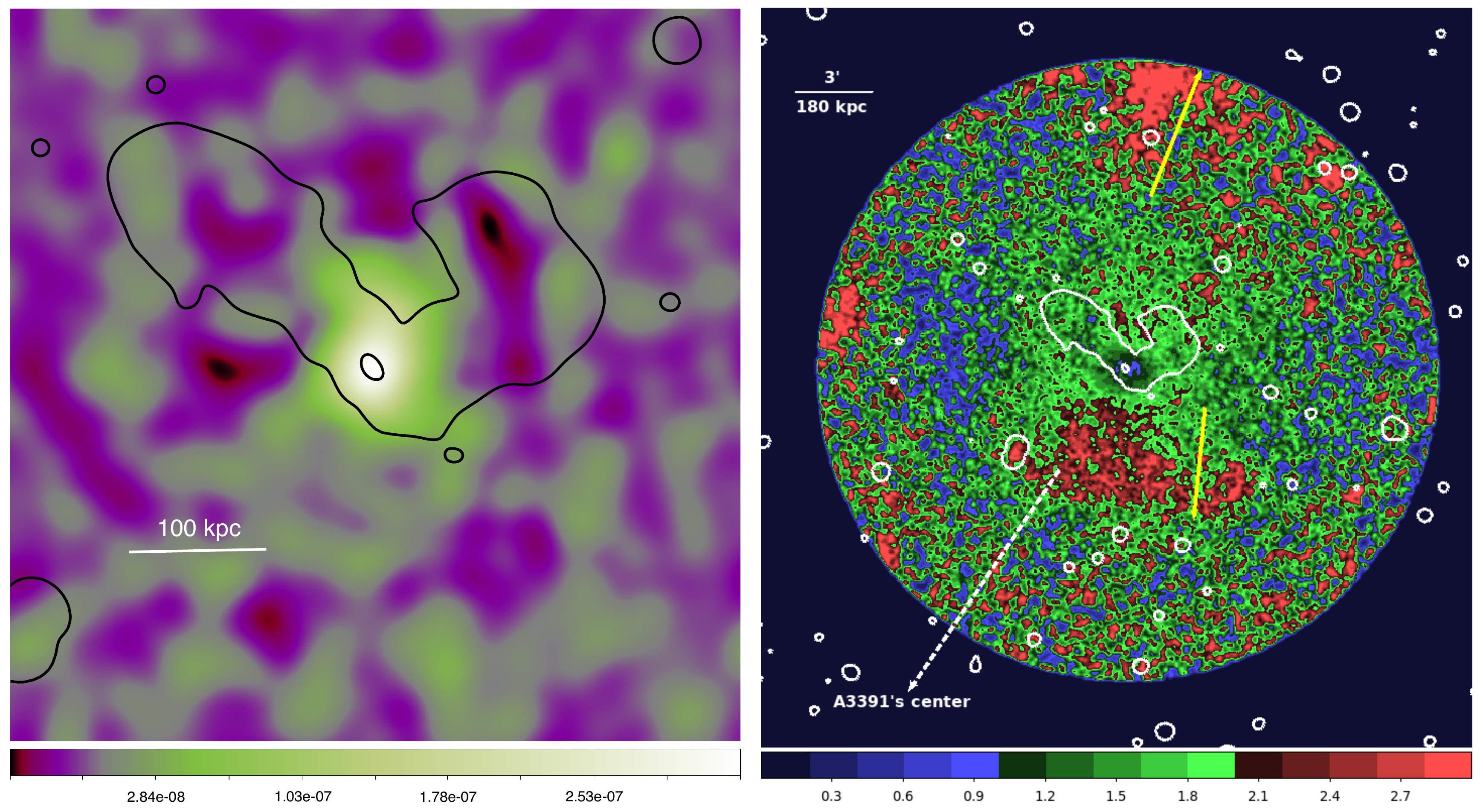}
\caption{\textit{Left:} \chandra unsharp masked image of the central region of the Northern Clump. \textit{Right:} \xmm residual image. The northern yellow arrow shows the $5.3'$ projected length of the tail emission. The enhancement in the south is indicated by the southern yellow arrow and the center of A3391 cluster is pointed by the dashed-line white arrow. Gaussian smoothing with kernel radius of 3 pixels is applied to the image. Black and white contours in both images represent the ASKAP/EMU radio continuum emission.}
\label{fig:chandra_xmm_residual}
\end{figure*}

As observed in the clean images shown in Fig.~\ref{fig:XMM_photonimages}, the main body of the Northern Clump is well-embedded in the central chips of MOS1 (top left) and MOS2 (top right) cameras. A question arises, whether the boxyness of the system is intrinsic to the cluster itself, or rather an artifact caused by the central MOS chips. The boxy shape of the cluster can still be recognized in the pn image (bottom left), although not as prominent.
To further investigate this question, we generated exposure corrected images, as well as adaptively-smoothed images using MOS1+2+pn data and pn-only data, where the latter should not be affected by the CCD chips placement. The adaptive smoothing procedure was realized using the SAS task, \texttt{asmooth}, \texttt{smoothstyle}=`adaptive'. We set the S/N of the adaptive smoothing to 15. Through adaptive smoothing we also aim to enhance any soft emission, especially in the cluster outskirts. The resulting images are displayed in Fig.~\ref{fig:XMM_asmooth}. We observe similarity in the morphology of the cluster in the two types of images, such that the boxy shape and indication of tail-like emission projected northward.
\par
We present a composite X-ray+radio+optical image in Fig.~\ref{fig:xmm_SZ_askap} (left), focusing on the central region of the Northern Clump. We utilized the \xmm count rate image in soft band (blue color and white contour), the ASKAP/EMU radio image (red color and yellow contour), and DECam Sloan g-band image (green). The denser X-ray emission of the Northern Clump shows an asymmetrical morphology and has a sharp edge pointing at the projected southwest direction -- towards the A3391 center. We notice that the opening angle of the WAT radio source indicates movement towards the south. While looking at the bent direction of both radio lobes towards the northeast, the central radio source seems to move towards the southwest. We speculate that the radio lobes might have ''escaped'' from hot gas confinement and the change in direction is due to motion through a lower density medium, as indicated by the contours. The central radio source may be moving locally within the Northern Clump towards the southwest as a result of, for example, sloshing, while globally, the Northern Clump moves towards the southeast.
\par
We used unsharp masks in the \chandra image to highlight substructures in the ICM. The original image was smoothed using two different scales and one smoothed image was subtracted from the other to make the edges more visually appealing. This technique can be mathematically expressed as
\begin{equation}
    \hat{U}=I*g-I*h
\end{equation}
where $I$ stands for the original image and $g$ and $h$ are the two smoothing scales, for which we chose $g=30$ and $h=100$ pixels. As shown in Fig.~\ref{fig:chandra_xmm_residual} (left), the unsharp masked image reveals regions with a possible deficient surface brightness $r\approx 2.09'\: (125\:\mathrm{kpc})$ to the east and west of the cluster center. We further overlaid the ASKAP/EMU radio continuum emission of the cluster center. A seeming picture of X-ray cavities filled with radio lobes starts to emerge, as observed for other clusters \citep[e.g.,][]{Birzan_2004, McNamara_2007}.
\par
From the presented \xmm and \chandra images, and with the help from ASKAP/EMU radio image, as well as the DECam optical image, we identify clear indications of Northern Clump's boxyness, tail-like emission in the north, and hint of X-ray cavities that coincide with the radio lobes.

\subsection{X-ray surface brightness ($S_X$) profiles}\label{sec:surb}
\subsubsection{\xmm}

\begin{figure*}[!ht]
\centering
\includegraphics[width=0.49\textwidth]{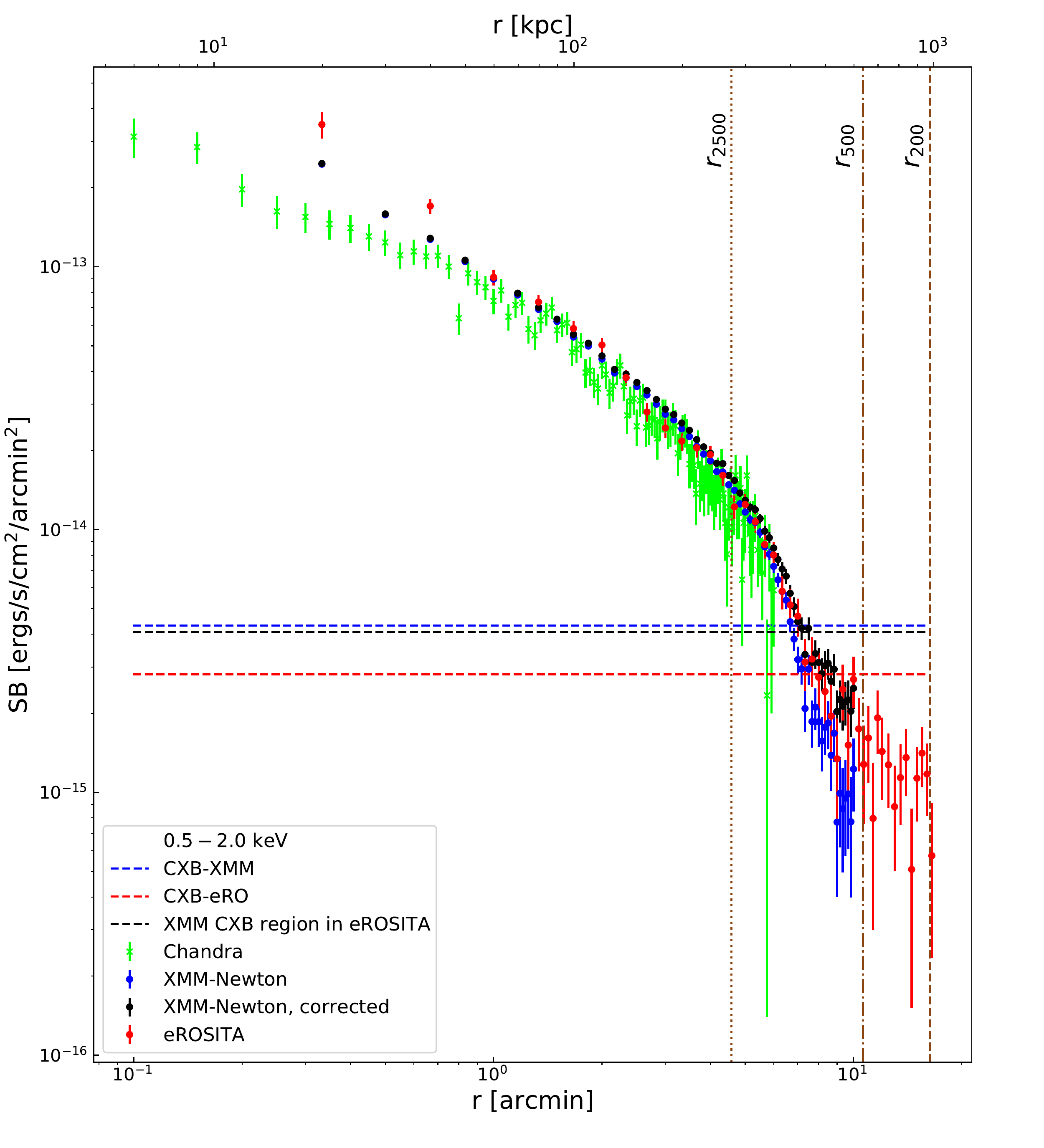}
 \includegraphics[width=0.49\textwidth]{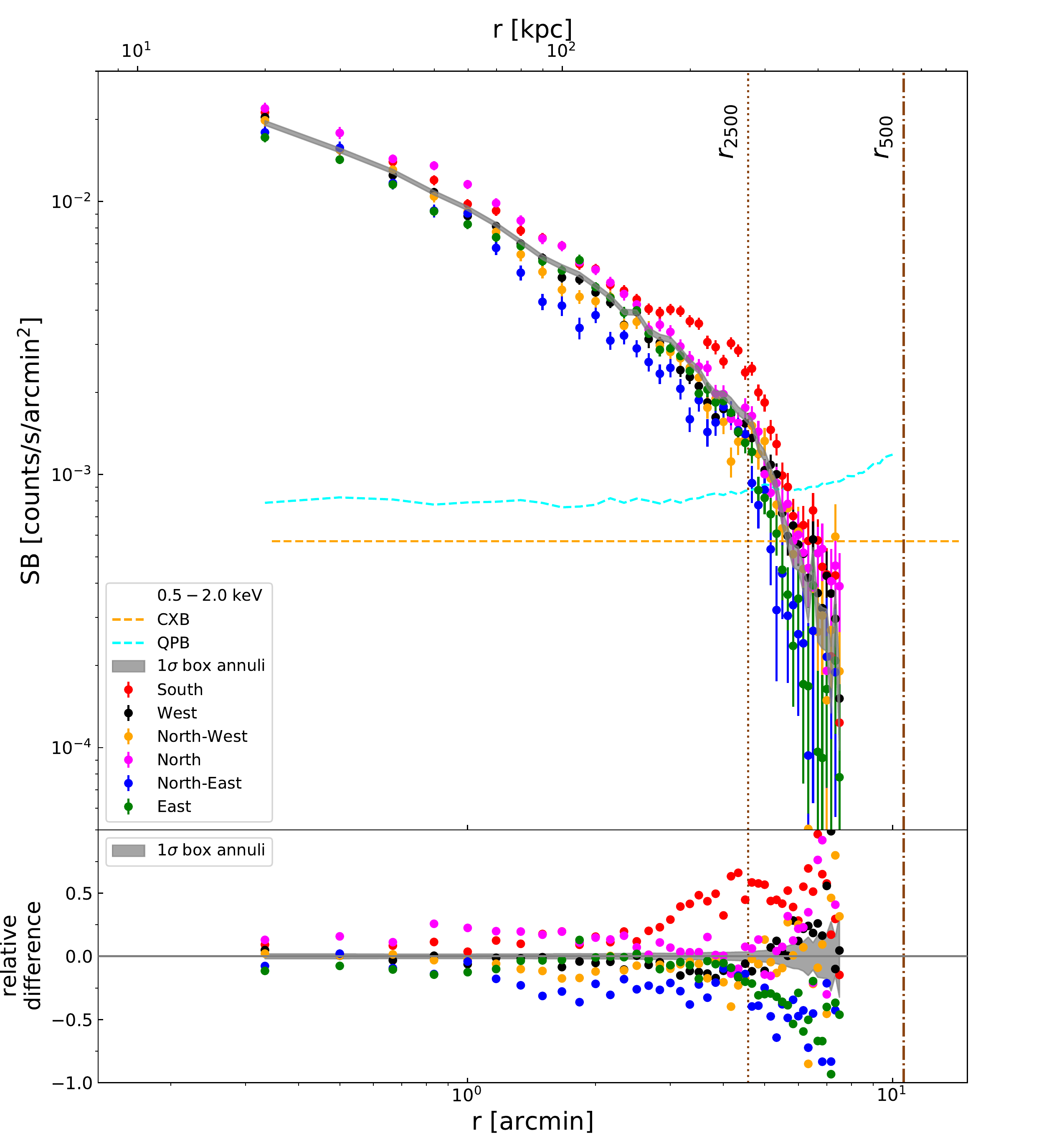}
\caption{\textit{Left:} Instrument independent, absorption corrected, and background subtracted $S_X$ profiles of \chandra (green), \xmm (blue), and \rosi (red) in the energy band $0.5-2.0$ keV. The blue, red, and black horizontal dashed lines are the CXB levels of \xmm, \rosi, and \rosi with \xmm defined CXB region, respectively. Black points are the CXB corrected \xmm $S_X$. The various radii are plotted in brown. \textit{Right:} \xmm $S_X$ profiles in different directions of the Northern Clump using the box annuli setup (top) and the relative differences with respect to the average surface brightness (bottom). The grey shaded areas in both plots are the $1\sigma$ confidence region of the average $S_X$. The various radii are plotted in brown. The CXB and QPB level are plotted in orange and cyan dashed lines.}
\label{fig:xmm_SB}
\end{figure*}

The full azimuthal $S_X$ profiles of the three different setups can be found in Appendix \ref{App:SB_sec_circ}. In all three profiles we observe a steep drop at around $5'<r<6'$ ($299.4\mathrm{\:kpc}<r<359.28\mathrm{\:kpc}$). Moreover, the overall shapes and residuals seen in the $S_X$ profiles of the box annuli setups using all \xmm detectors and pn-only data support the case that the boxy appearance of the cluster is not an artifact caused by the placement of the central MOS chip.
\par
We present the sectorized $S_X$ profiles (using the box annuli setup) in Fig.~\ref{fig:xmm_SB} (right, top plot). The bottom plot of the Figure shows the relative differences with respect to the corresponding $S_X$ of the full azimuthal setups, which is to examine any excess or deficit of the surface brightness. The $1\sigma$ confidence region of the full azimuthal $S_X$ is plotted in gray. Hereafter, the full azimuthal $S_X$ profile will be referred to as the average $S_X$ profile. The sectorized $S_X$ profiles using the circular annuli setup can be found in Appendix \ref{App:SB_sec_circ}.
\par
Some noticeable features among the different directions are present, such as the $S_X$ depression in the northeast, excess of the $S_X$ in the south, and the enhancement towards the north. These features are seen in both setups, but they are more pronounced in the box annuli setup.
\par
The depression in the northeast is at $1'<r<3'$ ($59.88\mathrm{\:kpc}<r<179.64\mathrm{\:kpc}$). The $S_X$ gets closer to the average at around $4'$ ($239.52\mathrm{\:kpc}$) and drops rapidly beyond this point. The largest fractional decrement in this direction, of 36\%, occurs at $1.67'$ ($100\mathrm{\:kpc}$).
\par
In the southern sector, $S_X$ enhancement is observed at around $2'<r<6'$ ($119.76\mathrm{\:kpc}<r<359.28\mathrm{\:kpc}$). The peak is at $4.33'$ ($259.20\mathrm{\:kpc}$) with a 66\% enhancement. We suggest that the origin of the surface brightness excess in this specific direction is either from compressed gas caused by ram pressure or an accreted clump of cold gas from the Northern Filament, where the cluster resides. In both scenarios, the Northern Clump should be falling towards the A3391 cluster along with the filament and within the region where its mean density dominates, collapses from the filament will continue onto the system.
\par
We observe significant enhancement at larger radii in the north direction consistent with the observed tail feature in the image. A relative difference of 170\% occurs at $7.5'$ ($449.1\mathrm{\:kpc}$).
\par
The features seen in the sectorized $S_X$ profiles are present in the residual image (Fig.~\ref{fig:chandra_xmm_residual}, right). The residual image is the result of dividing the surface brightness image by the spherically symmetric model. We overlaid the residual image with the ASKAP/EMU radio contour (white contour). Surface brightness depression at around $r\approx2.05'$ (122.75 kpc) to the northeast and northwest of the center are observed. They coincide with the lobes of the WAT radio source and also in agreement with what is found in Chandra unsharp masked image (Fig.~\ref{fig:chandra_xmm_residual}, left). In the south, a density enhancement relative to other directions is revealed at predominantly around $2'<r<5'$ ($119.76\mathrm{\:kpc}<r<229.40\mathrm{\:kpc}$). Furthermore, we detect a clear tail of emission north of the Northern Clump, $\sim\!7'$ ($419.16\mathrm{\:kpc}$) away from the center. This tail has a projected length of $\sim\!5.3'$ ($317.4\mathrm{\:kpc}$), as indicated by the northern yellow arrow in Fig.~\ref{fig:chandra_xmm_residual} (right).

\subsubsection{\chandra}\label{sec:chandra_SBanalysis}
\begin{figure*}[!ht]
\centering
\includegraphics[width=0.98\textwidth]{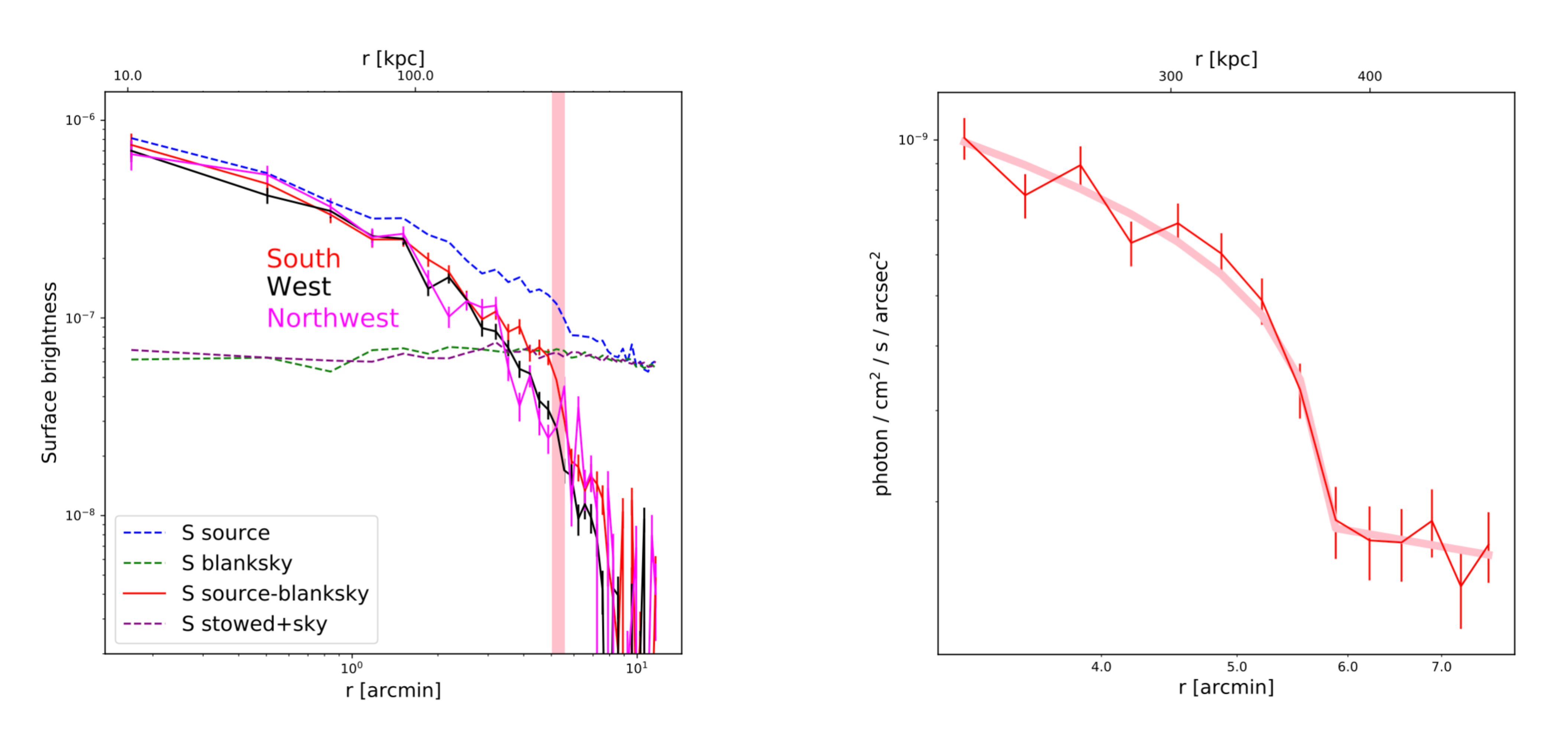}
\caption{{\it Left:} Blanksky background subtracted surface brightness profiles of the south (red), west (west), and northwest (magenta) directions in units of counts/s/arcsec$^2$ as measured with \chandra. The level of the tailored blanksky background (green dashed) and that of the stowed background plus the astrophysical background (purple dashed) for the south direction are consistent. The blue dashed line represents the total emission from both the source and background of the south direction. The vertical pink line marks the position of a possible density jump in the south direction. {\it Right:} surface brightness profile zoomed in around the surface brightness edge in the south direction in units of photon/s/cm$^2$/arcsec$^2$. We fit the data with a broken powerlaw density model (pink) and obtain a best-fit density jump of 1.45.
}
\label{fig:chandra_profile}
\end{figure*}

We derived the X-ray surface brightness profiles in south, west, and northwest directions, ranging from the cluster center to radii of $11.70'$ (700\:kpc) as shown in Fig.~\ref{fig:chandra_profile}. We note a region of enhanced brightness at around $r=5.84'$ (350\:kpc) in the south direction relative to other directions. We fit the  profile to a broken power law profile with a density jump projected along the line of sight 
\begin{equation}
    S(r)=norm\int F(\omega)^2dl, ~~{\rm where} ~\omega^2=r^2+l^2.
\end{equation}
$F(\omega)$ is the deprojected density profile:

\begin{equation}
F(\omega)=
\begin{cases}
\omega^{-\alpha_1} ~~{\rm if} ~\omega<r_{\rm break}\\
\frac{1}{jump}\omega^{-\alpha_2}~ {\rm otherwise,}
\end{cases}
\end{equation}
where $\alpha_1$, $\alpha_2$, $jump$, and $r_{\rm break}$ are free parameters.
We obtain a best-fit density jump of $1.45\pm0.32$ at $r=5.74\pm0.17'$ ($344\pm10$\:kpc).
As the ICM temperature appears to be uniform across the edge (Fig.~\ref{fig:profiles_05-7}), we assume that the pressure discontinuity equals the density jump and calculate the associated Mach number (Landau \& Lifshitz 1959):

\begin{equation}
    \frac{p_0}{p_1}=\left(1+\frac{\gamma-1}{2}\mathcal{M}\right)^{\gamma/(\gamma-1)}
\end{equation}
for which we obtain $\mathcal{M}=0.48^{+0.30}_{-0.34}$. Given a sound speed of 802\:km$/$s for a 2.5\:keV ICM, the central region of the Northern Clump is moving towards south at a speed of $386^{+239}_{-274}$ km$/$s relative to the motion of the gas in the southern direction.

\subsubsection{\rosi, \xmm, and \chandra}
We present $S_X$ profiles of all the used X-ray instruments in the energy band $0.5-2.0$ keV in Fig.~\ref{fig:xmm_SB} (left). To correct for the effective area of different instruments and  absorption, each of the three profiles had been multiplied by a conversion factor. For \rosi and \xmm, this conversion factor was calculated by taking the ratio of the unabsorbed energy flux to the absorbed count rates in the $0.5-2$ keV band. We supplied the best-fit cluster parameters in the $2.4-6'$, the $(k_BT_{500})$ region and the on-axis response files of an on-chip TM and MOS camera. To account for possible temperature changes across the radial ranges, we computed the conversion factor for 3 keV and 1.5 keV cluster emission, as well. The variation is found to be less than $3\%$. For \chandra, whose exposure maps already include a correction for the telescope effective area in addition to vignetting, the conversion factor is simply the ratio of the unabsorbed energy flux to the absorbed photon flux. The factor is $9.87\times10^{-12}~\mathrm{ergs/cm^2/counts}$ for \rosi, $7.55\times10^{-12}~\mathrm{ergs/cm^2/counts}$ for \xmm, and $1.72\times10^{-9}~\mathrm{ergs/photons}$ for \chandra.
\par
The annulus used to estimate the \xmm CXB level ($12.0-13.25'~(1.13-1.24R_{500})$, shown as the horizontal blue line) may still contain some cluster emission. In order to correct for this, we calculated the surface brightness of this same background region using \rosi data (horizontal black dashed lines) and estimated the excess cluster surface brightness as the difference between the two background estimate (horizontal red dashed line). We added back this excess surface brightness to obtain the corrected \xmm data points (plotted in black). The CXB correction results in better agreement between \xmm and \rosi, especially in the $7.0-10.0'$ radial range. The CXB subtracted surface brightness profile is nearly flat beyond the $R_{500}$ and out to the $R_{200}$. This may imply that at this regime, there is still some emission either from the cluster or the filament in which the Northern Clump resides, or both.
\par
Through this exercise, we show the complementary nature of the three instruments. For instance, \chandra (green points) with the best on-axis spatial resolution can cover as far as $3''$ from the center, \xmm (blue and black points) covers the intermediate radial range with good statistics, and \rosi (red points) with its wide FoV and scan mode delivers the cluster surface brightness out to $R_{200}$.
The discrepancies in the center can be well-explained by the PSF effect. Moreover, although masked, the bright point source in the cluster center is likely to contribute some of its emission to the central bins of \rosi and \xmm profiles. Other causes, such as point source detection and different background treatment could also play a role in the observed differences among the profiles.

\subsection{Spectral analysis}

\begin{figure*}[!ht]
   \centering
   \includegraphics[width=0.49\textwidth]{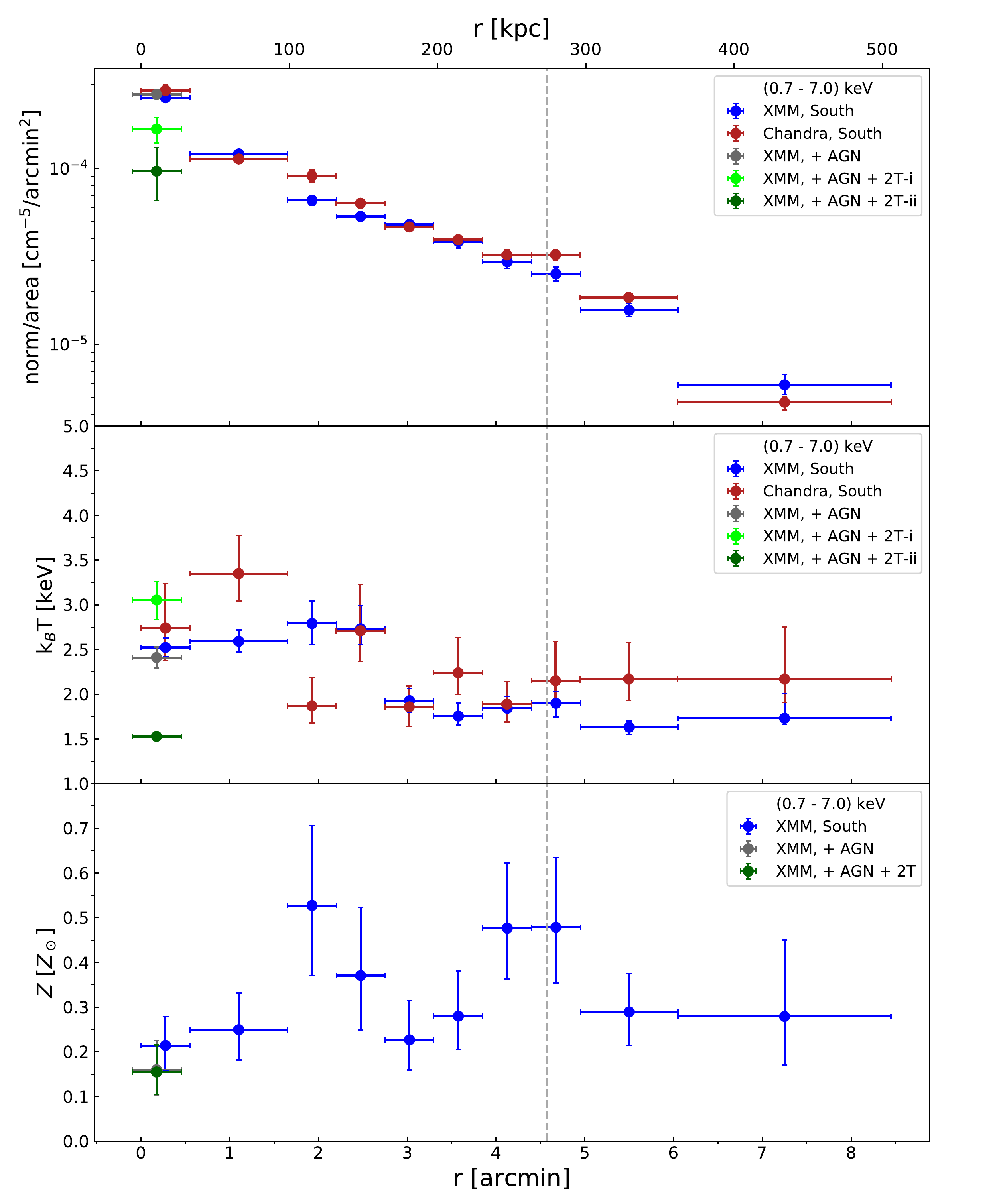}
   \includegraphics[width=0.49\textwidth]{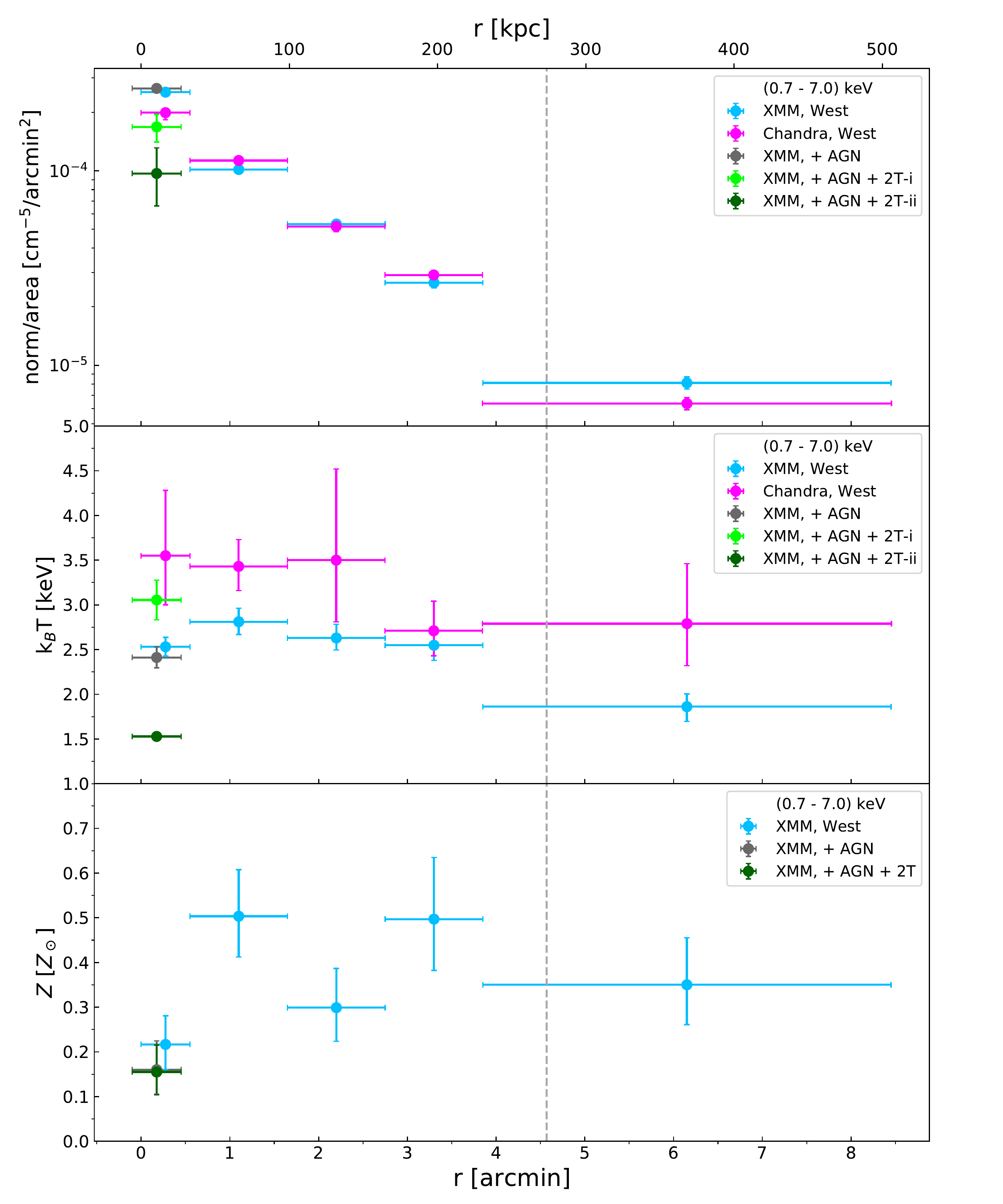}
   \caption{Normalization (top), temperature (middle), and metallicity (bottom) profiles of the Northern Clump derived in the energy band $0.7-7.0$ keV. \textit{Left:} Southern region profiles. The blue and red data points represent \xmm and \chandra, respectively. \textit{Right:} Western region profiles. The cyan and magenta data points represent \xmm and \chandra, respectively. Due to modest photon counts, \chandra metallicity could not be constrained in any sectors. The central data points fitted with AGN component are shown in grey, light green and dark green. These data points are shifted for visualization purposes. The grey dashed lines indicate the $R_{2500}$ of the cluster. The x-axis error bars are not the $1\sigma$ error range, but the full width of the bins.
}
\label{fig:profiles_05-7}
\end{figure*}

\subsubsection{\rosi cluster parameters}
We list the \rosi cluster parameters for the $k_BT_{500}$ region (an annulus of $2.4-6.0'$) in Table \ref{tab:eRO_T500}. The metallicity returned by \rosi is consistent with the one returned by \xmm, both in the wide and the small soft energy band, while the normalizations and temperatures are lower. Fitting in the soft band does improve the consistency between \rosi and \xmm temperatures while not fully solving the discrepancies. While this could be an effect of instrumental calibration uncertainties, it could also be caused by multi-temperature structure. As predicted by simulations, in the case of multi-temperature plasma \rosi that has superior soft response was shown to return lower temperatures in comparison to other instruments \citep{Reiprich_2013}. Fitting in the soft band only should minimize such effects as the relative effective areas are similar in this range for \rosi and \xmm. This issue will be addressed further in the future, employing also \rosi mock observations based on hydrodynamical simulations, and divising methodologies similar to the algorithm suggested by \cite{Vikhlinin_2006} to predict the temperature in the case of multi-temperature plasma.

\begin{table}[!h]
    \centering
    \caption{\rosi and \xmm cluster parameters from the $k_BT_{500}$ region ($0.2-0.5 R_{500}$).}
    \begin{tabular}{c c c c}
    \hline
    \hline
\multirow{2}{*}{Instrument} & $norm^\dagger$ & $k_BT$ & $Z$\\
 & & $[$keV$]$ & $[Z_{\odot}]$\\[5pt]
\hline
\multicolumn{4}{c}{full energy band$^*$}\\
\hline
\rosi & $1.59_{-0.19}^{+0.17}$ & $1.67_{-0.11}^{+0.20}$ & $0.36_{-0.12}^{+0.11}$\\[5pt]
\xmm & $2.35_{-0.045}^{+0.046}$ & $1.99\pm0.04$ & $0.31_{-0.026}^{+0.025}$ \\[5pt]
\hline
\multicolumn{4}{c}{$0.8-2.0$ keV}\\
\hline
\rosi &  $1.62_{-0.23}^{+0.26}$ & $1.70_{-0.11}^{+0.22}$ & $0.36_{-0.14}^{+0.19}$\\[5pt]
\xmm & $2.45_{-0.052}^{+0.053}$ & $1.99_{-0.054}^{+0.050}$ & $0.27_{-0.025}^{+0.027}$ \\[5pt]
\hline
\hline
\multicolumn{4}{l}{\footnotesize $^\dagger$ $[10^{-5}$ cm$^{-5}/$arcmin$^{2}]$} \\
\multicolumn{4}{l}{\footnotesize $^*$ $0.7/0.8-9.0$ keV for \rosi TM8/TM9.} \\
\multicolumn{4}{l}{\footnotesize ~~$0.7-7.0$ keV for \xmm} \\
    \hline
    \hline
    \end{tabular}
    \label{tab:eRO_T500}
\end{table}

\subsubsection{Inner of south and west}
We plot the results of the spectral analysis of \xmm and \chandra in the south and west directions of the Northern Clump in Fig.~\ref{fig:profiles_05-7} left and right panel, respectively. The top, middle, and bottom panels are the normalization per area, temperature, and metallicity profiles. Due to modest photon counts, the metallicities in \chandra could only be constrained using bins of full azimuthal direction. The profile can be found in Appendix \ref{App:NC_profles}, Fig.~\ref{fig:profiles_allZ}. The information regarding the source regions and their best-fit parameter values are listed in Table \ref{tab:xmm_spectralparameters}.
\par
In both directions, we observe a flattening towards the central region of the cluster. The temperature profile towards the south indicates an elevated temperature around $2'<r<3'$ ($119.76\mathrm{\:kpc}<r<179.64\mathrm{\:kpc}$). The metalicity also appears enhanced in this region. Beyond the fourth bin, the temperature profile flattens at an average temperature of 1.94 keV.

\subsubsection{Outskirts of north, south, and west}
\begin{figure}[!h]
\centering
\includegraphics[width=\columnwidth]{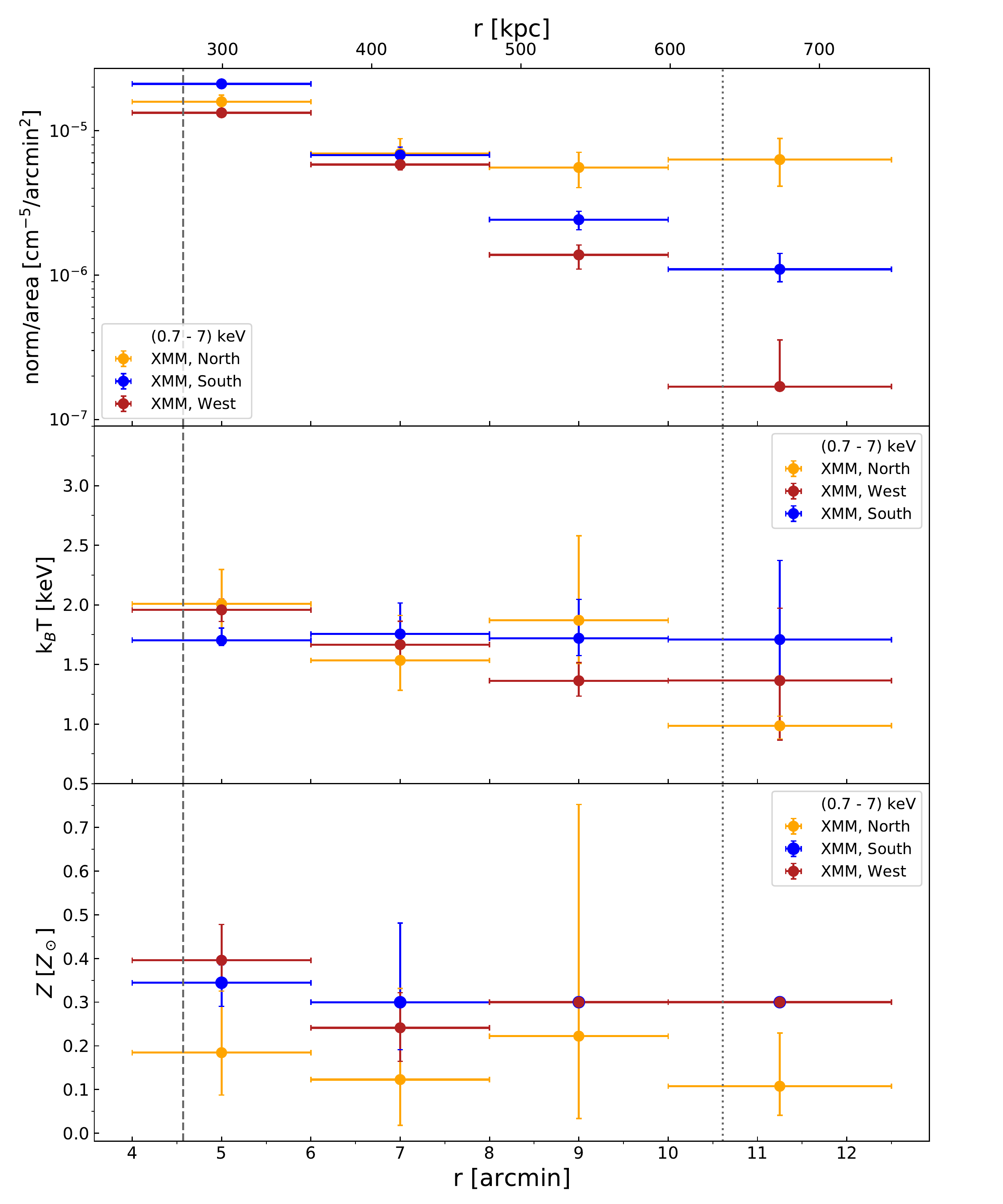}
\caption{\xmm normalization (top), temperature (middle), and metallicity (bottom) profiles of the outer regions of the Northern Clump in the different directions in the energy band $0.7-7.0$ keV. The northern, southern, and western regions are represented by orange, blue, and red data points, respectively. The grey dashed (dotted) lines indicate the $R_{2500}$ ($R_{500}$) of the cluster. The x-axis error bars are not the $1\sigma$ error range, but the full width of the bins. The metallicity in the last two bins of the south and west has been frozen to $0.3Z_{\odot}$.}
\label{fig:xmm_north}
\end{figure}

We compare spectral properties of the north, south, and west directions over $R_{2500}$ to $R_{500}$ ($r=4-12.5' = 239.52-733.53\mathrm{\:kpc}$) as shown in Fig.~\ref{fig:xmm_north}. We compare only \xmm values since, unfortunately, the outer north of the Northern Clump is not covered by the \chandra observations. We found it necessary to fix the metallicity in the last two bins of the south and west to $0.3Z_\odot$. The best-fit temperatures and normalizations vary within $1\sigma$ when we fixed the metallicity of south and west to the best-fit value of the north or fixed the metallicity of the north to $0.3Z_\odot$. In general, our treatment of metallicity does not change the following results.
\par
We note a higher normalization per area in the last two bins of the northern region. In comparison to the south (west) third and fourth data points, the normalization per area in the north are 56 (75)\% and 83 (97)\% higher, respectively. From \xmm residual image (Fig.~\ref{fig:chandra_xmm_residual}, right), we estimated that the excess emission in the north associated with the tail feature is located at $\sim\!7'$ ($419.16\mathrm{\:kpc}$) from the center and stretched out to $12.5'$ ($733.53\mathrm{\:kpc}$). Therefore, we see an agreement between the imaging and spectral analyses results. This tail feature is often observed in other merging systems as a result of ram pressure stripping. For example, the tail from M86 and M49 as they fall into the Virgo Cluster (\citealp{Randall_2008, Su2019}. \citealp[Also see][]{Su_2014, Su_2017a, Su_2017b}), the northeast tail feature observed in A2142 cluster (\citealp{Eckert_2014}. \citealp[Further confirmed through dynamical study of the member galaxies in][]{Liu_2018}), and a more extreme case of complete stripping is observed in the galaxy cluster Zwicky 8338 \citep{2015A&A...583L...2S}.
\par
In the south, higher temperature in comparison to the other two directions is apparent. At $11.25'$ ($673.65\mathrm{\:kpc}$), the temperature in the south is higher by 42 (20)\% than in the north (west). This elevated temperature may be associated with shock or compression heated gas as the Northern Clump falls towards the south, into the A3391 cluster. At $5.0'$ ($299.4\mathrm{\:kpc}$), the normalization per area is higher than the other two directions, which is in agreement with the residual image (Fig.~\ref{fig:chandra_xmm_residual}, right) and the southern $S_X$ profile (Fig.~\ref{fig:xmm_SB}, right).
\par
The properties of the outskirts of the Northern Clump are similar to those systems listed above, particularly the case of the M49 group residing beyond the virial radius of the Virgo Cluster \citep{Su2019}. In addition to the stripped tail, a temperature enhancement is noted in front of the cold front in M49, resembling the temperature elevation and surface brightness edge (see Subsect. \ref{sec:chandra_SBanalysis}) observed at $\sim\!11.25'$ ($673.65\mathrm{\:kpc}$) and $5'$ ($299.4\mathrm{\:kpc}$), respectively in the south of the Northern Clump. Our findings indicate that the Northern Clump may be undergoing the same infalling process with M49.

\begin{figure*}[!ht]
\centering
\includegraphics[width=0.98\textwidth]{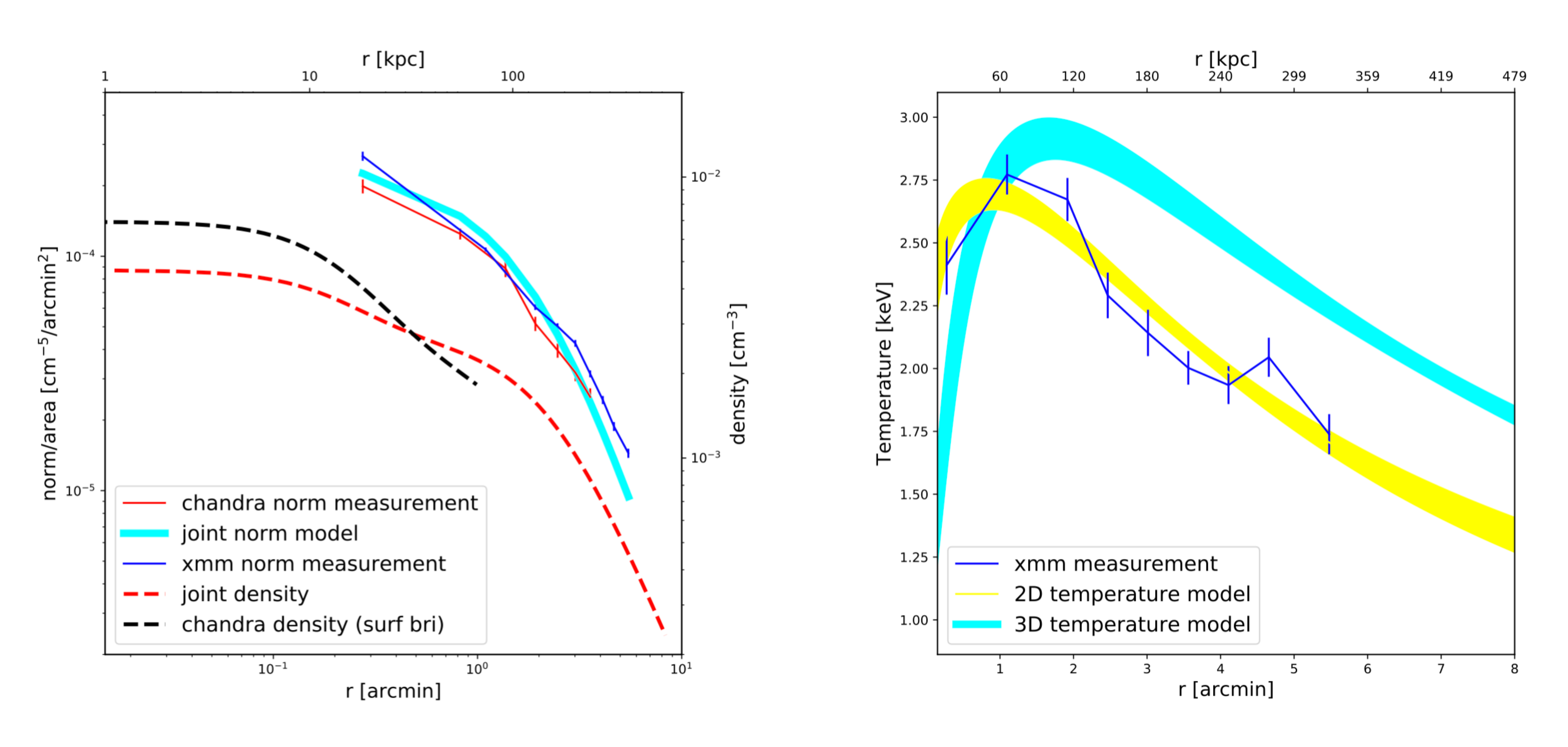}
\caption{{\it Left:}
Profiles of \texttt{apec} normalization per area measured with \xmm (blue solid) and \chandra (red solid). Red dashed lines indicate the best-fit 3D density profiles
for an empirical density profile of Equation~\ref{eq:d3d} through a joint-fit of \xmm and \chandra normalization profile. The corresponding normalization per area profiles derived with the best-fit 3D density profiles are shown in the cyan line, displaying good agreement with the measurements. Black dashed line is the deprojected density profile  derived from \chandra surface brightness profile and the \chandra spectral analysis of the innermost bin.
{\it Right} :
Projected temperature profile measured with \xmm (blue solid line). The best-fit deprojected temperature profile is shown in cyan. The corresponding projected temperature profile derived from Equation~\ref{eq:t3d} is shown in yellow, in good consistent with the measurement.}
\label{fig:deproj}
\end{figure*}
\subsubsection{Central AGN}
We notice a metallicity drop in the core of the Northern Clump. While it is commonly believed that this phenomena, which is often observed in galaxy groups and clusters with X-ray cavities, is the result of the AGN mechanical feedback and/or depletion of Fe into cold dust grains \citep{Panagoulia_2015, Lakhchaura_2019, Liu_2019}, it could also be an artifact that surfaces as the central AGN is not accounted properly in the spectral fitting \citep{Mernier_2017}. To investigate this, we therefore implemented the second and third fitting methods (Subsect. \ref{sec:xmm_spectralanalysis}). We present the best-fit parameter values of the three methods in Table \ref{tab:AGN} and the spectra with their best-fit model are displayed in Fig.~\ref{fig:xmm_AGN_spectra}.
\par
Since the on-axis encircled energy fraction (EEF) for MOS1 camera is 68\% at a $15''$ radius\footnote{\href{https://xmm-tools.cosmos.esa.int/external/sas/current/doc/eupper/node4.html}{\texttt{XMM-Newton Science Analysis System: Encircled energy function}}}, this indicates that there might be some AGN emission left in the source region while performing the fit using the first method. While 92\% of the total energy are encircled within a $60''$ radius, having a mask of this size in the core region is undesirable. The second and third methods are a better solution to this problem. However, as noticed from the metallicity profiles, the low metallicity persists even when the AGN component is added.
\par
The apparent low core metallicity is better explained by the above-mentioned mechanical processes, for instance by the AGN feedback, through which a fraction of the core metal-rich ICM gas is distributed outwards \citep{Sanders_2016, Mernier_2017} and/or depletion of core Fe into dust grains \citep{Panagoulia_2015, Lakhchaura_2019, Liu_2019}.
\par
\begin{figure*}[!ht]
\centering
\subfloat{\includegraphics[width=\textwidth]{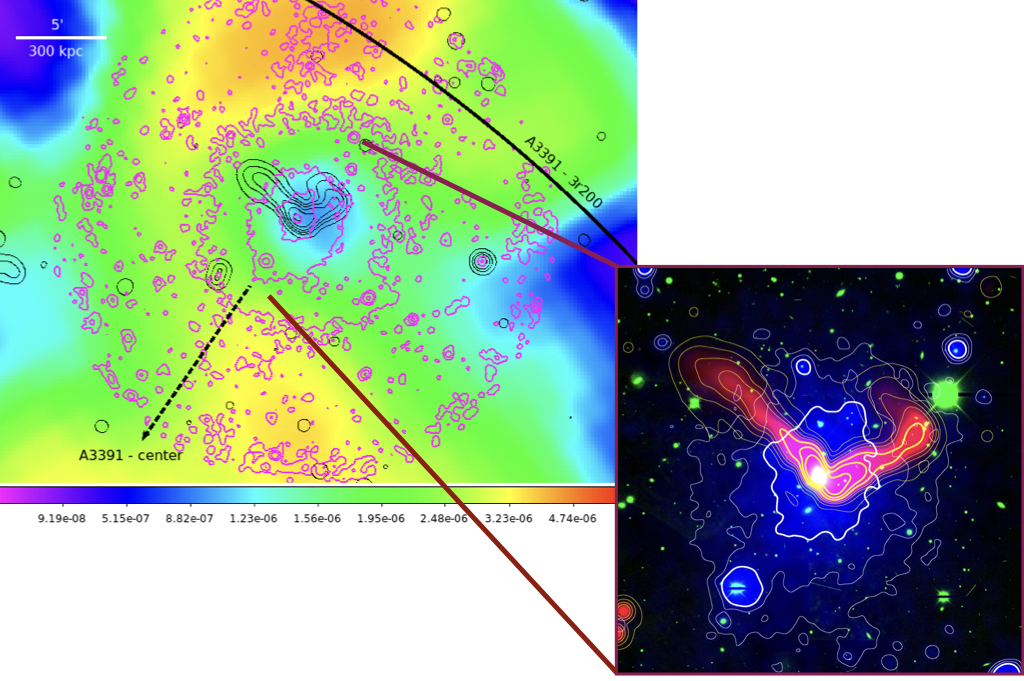}}
\caption{\textit{Left:} \textit{Planck y}-map overlaid with X-ray (from \xmm, in magenta) and radio (ASKAP/EMU, in black) contours. Black line depicts the $3\times R_{200}$ of A3391 and black dashed-line points toward the center of A3391. \textit{Right:} Radio+X-ray+optical composite image of the central region of the Northern Clump. The image is generated utilizing ASKAP/EMU image (red color and yellow contour), \xmm image (blue color and white contour), and DECam Sloan g-band image (green color).}
\label{fig:xmm_SZ_askap}
\end{figure*}

Additionally, we also observe lower Sunyaev-Zel'dovich (SZ) signal at the center of the cluster, as shown in Fig.~\ref{fig:xmm_SZ_askap}. The decrease of the SZ signal at this region is likely due to the core of the radio AGN, and similarly seen at the center of the Perseus Cluster (\citealp[Figure 5 in][]{Erler_2019}. \citealp[For more detailed review on AGN contamination of SZ clusters see][]{Lin_2009,Gupta_2017}). Unless special effort is done, such clusters may be missed in SZ surveys.

\subsubsection{Deprojection analysis and central entropy}
We obtain the best-fit \texttt{apec} normalization profiles for \chandra spectra extracted from concentric annuli ranging from $r=20$\:kpc to $r=200$\:kpc and for \xmm ranging from $r=20$\:kpc to $r=400$\:kpc as shown in Fig.~\ref{fig:deproj} (left).
We assume the deprojected density profile takes the form of 
\begin{equation}
n(r)=\sqrt{{n_{01}}^2\left(1+\frac{r^2}{{r_{c1}}^2}\right)^{-3\beta_1}+{n_{02}}^2\left(1+\frac{r^2}{{r_{c2}}^2}\right)^{-3\beta_2}}\,.
\label{eq:d3d}
\end{equation}
A normalization profile can be derived by substituting the 3D density profile (Equation~\ref{eq:d3d}) into Eq.~\ref{eq:norm}. We can obtain the best-fit parameters for the deprojected density profiles by fitting the normalization profiles measured with \chandra and \xmm to Equation~\ref{eq:norm}, as shown in Fig.~\ref{fig:deproj} (left). 
We use the $R^2$-score to evaluate the goodness of the fit, defined as
\begin{equation}
R^2(y,\hat{y})=1-\frac{\Sigma_{i=1}^n(y_i-\hat{y_i})^2}{\Sigma_{i=1}^n(y_i-\bar{y_i})^2}
\end{equation}
where $\hat{y_i}$ is the best-fit value of the $i$-th data point and ${y_i}$ is the measured value for total $n$ data points, and $\bar{y}=\frac{1}{n}\Sigma_{i=1}^{n}y_i$. We obtain a $R^2$-score of 0.95 for the deprojected density analysis. 

We also derived the deprojected density profiles using the \chandra surface brightness profile following \citet{Hudson_2010}. 
We fit the surface brightness profile of the west direction to a double-$\beta$ model, which takes the form of
\begin{equation}
    S(r)=S_{01}\left(1+\frac{r^2}{{r_{c1}}^2}\right)^{(-3\beta_1+0.5)}+S_{02}\left(1+\frac{r^2}{{r_{c2}}^2}\right)^{(-3\beta_2+0.5)},
\label{eq:surf}
\end{equation} 
corresponding to a density profile of Eq.~\ref{eq:d3d}.
The central density $n_0=\sqrt{{n_{01}}^2+{n_{01}}^2}$ can be estimated from 
\begin{equation}
n_0=\left[\frac{10^{14}4\pi(\Sigma_{12}\rm LI_2+LI_1){\it D_A D_L}\zeta N}{\Sigma_{12}\rm LI_2EI_1+LI_1EI_2}\right]^{\frac{1}{2}},
\end{equation}
where $\Sigma_{12}=S_{01}/S_{02}$ is the ratio of the central surface brightness as in Eq.~\ref{eq:surf} and N is the normalization of the \texttt{apec} model for the innermost bin of \chandra projected spectral analysis. Additionally, $D_A$ and $D_L$ are the angular diameter distance and the luminosity distance, respectively, and $\zeta=1.2$ is
the ratio of electrons to protons. ${\rm EI}_{i}$ is the emission
integral divided by the central density for component $i$:
\begin{equation}
    {\rm EI}_i\equiv\int\left(\frac{n}{n_0}\right)^2 dV = 2\pi \int_{-\infty}^{\infty}\int_{0}^{R}x\left(1+\frac{x^2+l^2}{r_{ci}^2}\right)^{-3\beta_i}dxdl,
\end{equation}
where $R$ is the outer radius of the spectrum extraction region. The line emission measure for component $i$, ${\rm LI}_i$, is defined as
\begin{equation}
    {\rm LI}_i\equiv\int_{-\infty}^{\infty}\left(1+\frac{l^2}{{r_{ci}}^2}\right)^{-3\beta_i}dl,
\end{equation}
and $n_{01}$ and $n_{02}$ can be further related by,
\begin{equation}
    \frac{{n_{01}}^2}{{n_{02}}^2}=\rm\frac{\Sigma_{12}LI_2}{LI_1}.
\end{equation}
We can solve these equations using the surface brightness parameters to obtain a density profile of Eq.~\ref{eq:d3d} as shown in Fig.~\ref{fig:deproj} (left). 

We adopted an empirical formula to model the deprojected temperature profile \citep{Andrade_Santos_2017,Su2019} 
\begin{equation}
T_{\rm 3D}(r)=\frac{T_0}{[1+(r/r_t)^2]}\times\frac{[r/(0.075r_t)]^{1.9}+T_m/T_0}{[r/(0.075r_t)]^{1.9}+1}.
\label{eq:t3d}
\end{equation}
The corresponding projected temperature can be obtained by projecting the gas density weighted $T_{\rm 3D}$ along the line of sight. 
We fit the azimuthally averaged projected temperature profile measured with \xmm to the 2D formula, 
 \begin{equation}
T_{\rm 2D}=\frac{\int {{\rho}_g}^2{T_{\rm 3D}^{1/4}dz}}{\int {{\rho}_g}^2{T_{\rm 3D}^{-3/4}dz}},
 \end{equation}
for which the density profile derived with \xmm is applied.
The best-fit 2D and 3D temperature profiles are plotted in Fig.~\ref{fig:deproj} (right). 
The 2D temperature model and the measured projected temperature profile give a $R^2$-score of 0.88. The discrepancies between \xmm and \chandra normalizations may be caused by, such as, different bin size, scattering, point source detection, as well as instrumental calibration uncertainties.
 
The Northern Clump has an extrapolated central electron density of $\lesssim\!7\times10^{-3}$\:cm$^{-3}$, falling short of $\sim\!1.5\times10^{-2}$\:cm$^{-3}$, a conventional threshold for cool-core clusters~ \citep{Andrade_Santos_2017,2020MNRAS.498.5620S}. 
Based on the deprojected density and temperature profiles, we extrapolate an entropy ($K \equiv k_B T_{\rm gas}/n_e^{2/3}$) of 30 keV\:cm$^{2}$ for the central $r<0.004R_{500}$. According to \cite{Hudson_2010}, the Northern Clump can be best described as a weak cool-core (WCC) cluster. 

\subsubsection{Northern Filament}
We consider the region between the virial radii of the Northern Clump and the A3391 cluster as an inter-cluster filament as shown in Fig.~\ref{fig:northclump_ero}. We used the measured \texttt{apec} normalization to derive the electron density, $n_e$, of the filament. We assume a simple geometry, such that the filament is a cylinder with its axis in
the plane of the sky and located at the mean redshift of both clusters, $z=0.0533$, such that Eq. \ref{eq:norm} can be rewritten as,

\begin{equation}
\begin{split}
    n_e = &\left[1.52\times10^{-10}~\mathrm{cm}^{-1}\times norm\times(1+z)^2 \right.\\
& \left. \times \left(\frac{D_A}{\mathrm{Mpc}}\right)^2 \times \left(\frac{r}{\mathrm{Mpc}}\right)^{-2} \times \left(\frac{h}{\mathrm{Mpc}}\right)^{-1} \right]^\frac{1}{2},
\end{split}
\label{eq:n_e}
\end{equation}
where $D_A(z=0.0533) = 214.09~\mathrm{Mpc}$, radius of the cylinder $r = 1.56~\mathrm{Mpc}$, and its height $ h = 1.81~\mathrm{Mpc}$. The Hydrogen density is taken to be $n_H \approx n_e/1.17$.
\par
We present the results of the \rosi spectral analysis for the Northern Filament in Table \ref{tab:eRO_fil}. Comparing our results to the WHIM properties predicted by the cosmological hydrodynamic simulations (\citealp[e.g.,][]{Dave_2001}. \citealp[For observational constraints of WHIM properties see, e.g.,][]{Nicastro_2018, Kovac_2019}), we find that the best-fit projected temperature obtained from the one-temperature (1T) component fit is beyond the expected temperature range of $T=10^5-10^7~\mathrm{K}$, while the electron density is within the expected range of $n_e\approx10^{-6}-10^{-4}~\mathrm{cm}^{-3}$.
\par
We further performed a two-temperature (2T) component fit, for which we fixed the abundances of the filament to 0.1 of \cite{Asplund_2009} due to the statistics. We identify a cooler component whose temperature and electron density are consistent with the properties of WHIM. The hotter component is likely to be cluster emission from either or both the Northern Clump and the A3391 cluster. Further study, for instance, constraining the properties of the ICM in the outskirts of A3391, as well as calculating the surface brightness profile along the filament, is needed to draw more conclusions (Veronica et al. in prep.).

\begin{table}[!h]
    \centering
    \caption{\rosi properties of Northern Filament.}
    \begin{tabular}{c c c c c}
    \hline
    \hline
\multirow{2}{*}{Fitting} & $norm$ & $k_BT$ & $Z$ & $n_e$\\
 & $[10^{-3}$ cm$^{-5}]$ & $[$keV$]$ & $[Z_{\odot}]$ & $[10^{-5}$ cm$^{-3}]$\\[5pt]
\hline
1T & $1.34_{-0.30}^{+0.64}$ & $1.52_{-0.54}^{+0.57}$ & $0.06_{-0.06}^{+0.19}$ & $4.85_{-0.58}^{+1.05}$ \\[5pt]
\hline
\multirow{2}{*}{2T} &  $1.06_{-0.29}^{+0.25}$ & $1.72_{-0.29}^{+1.02}$ & \multirow{2}{*}{$0.1$} & $4.32_{-0.65}^{+0.48}$\\[5pt]
 & $0.22_{-0.19}^{+0.24}$ & $0.68_{-0.64}^{+0.38}$ & & $1.99_{-1.24}^{+0.88}$\\[5pt]
    \hline
    \hline
    \end{tabular}
    \label{tab:eRO_fil}
\end{table}

\section{Insights from the Magneticum Simulation}
\begin{figure}
    \centering
    \includegraphics[width=.9\columnwidth,trim=90 0 70 10,clip]{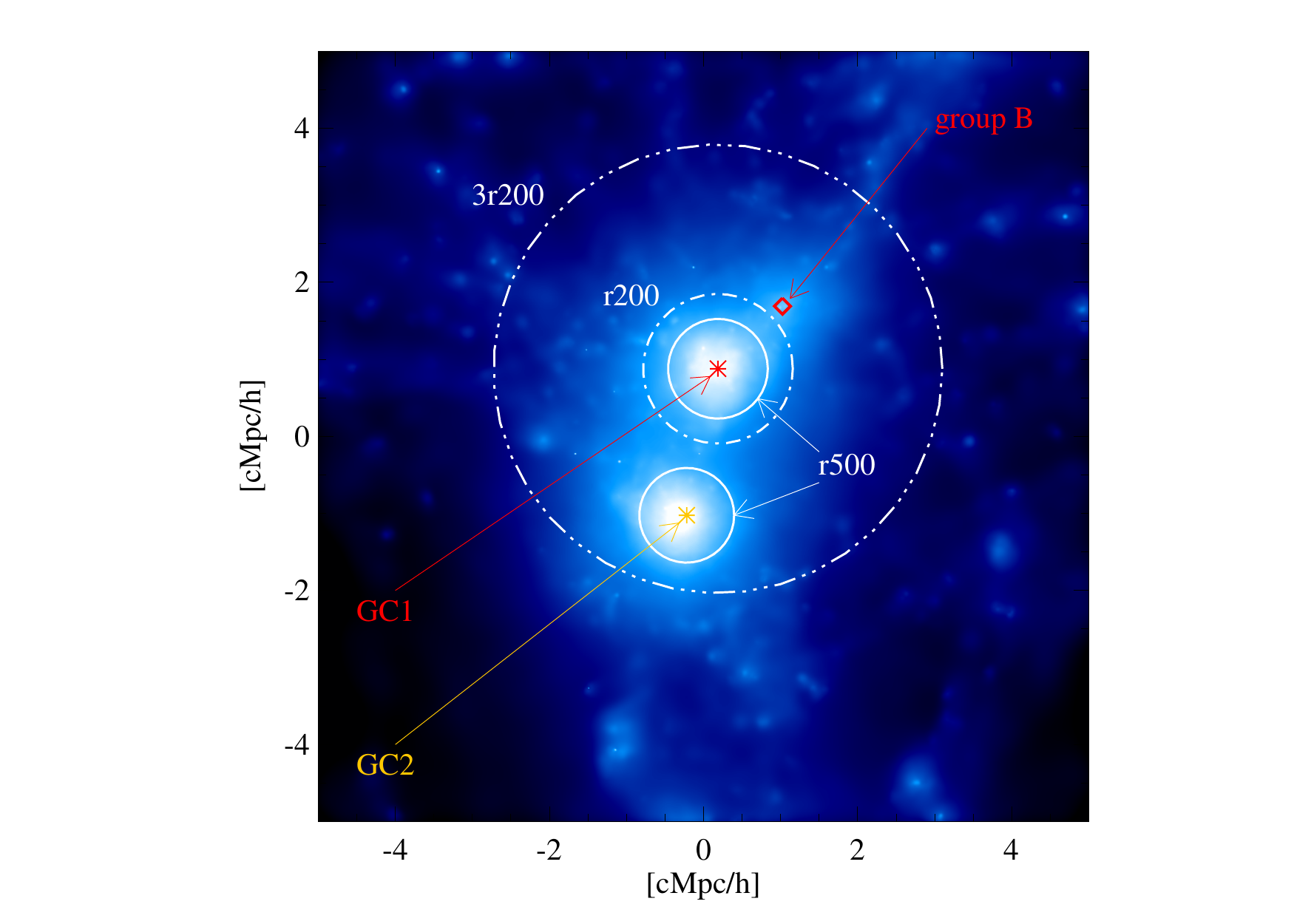}
    \caption{Gas density map of the A3391/95 analog pair from the Magneticum Simulation, at $z=0.07$. The region shown comprises a cubic volume of $10\:$cMpc$/h$ per side centered on the pair center of mass. The two main clusters GC1 and GC2 are marked, with their $R_{200}$ and $R_{500}$ (defined w.r.t.\ the critical density of the Universe), as well as the position of group B.
    }
    \label{fig:sim_map}
\end{figure}

In the $20\:$cMpc$/h$-scale region extracted from the Magneticum Simulation volume, that comprises an analog of the A3391/95 system~\cite[see][]{Reiprich_2021} at $z=0.07$, we also identify various structures with the size of galaxy groups \citep{Biffi_2021}.
Among these additional structures, we consider in particular a small halo (hereafter, ``group B"), in order to better interpret the history of the observed Northern Clump.
Group B is a group-size object that has entered the atmosphere of the upper pair member~\cite[marked as ``GC1";][]{Biffi_2021} at about $z=0.16$, and has reached its outskirts (at $d_{\rm 3D} \sim 1.5 \times R_{200}^{\rm GC1}$) by $z=0.07$. 
The configuration at redshift $z=0.07$ of the simulated pair and the group B is shown in Figure~\ref{fig:sim_map}. The Figure shows the gas density map projected along the $z$ axis of the simulation box, and comprises a volume of $10\:$cMpc$/h$ per side, centered on the pair center of mass. There we mark the positions of the two pair members, GC1 and GC2, and of group B. Furthermore, we mark cluster radii that are of interest for the present discussion.

Group B has a mass of $M_{500} \sim 3.6\times 10^{13}\:{\rm M}_{\odot}$ at redshift $z\sim 0.16$, just before the merging starts. At lower redshifts, ($0.07< z\lesssim 0.16$) the group is identified as a gravitationally-bound substructure of the GC1 cluster.
By tracking the history of its most-massive progenitor, we investigate its properties along the trajectory towards the node of the Cosmic Web where the cluster pair is finally settling by $z=0.07$.

\begin{figure}
    \centering
    \includegraphics[width=.9\columnwidth,trim=0 0 0 0,clip]{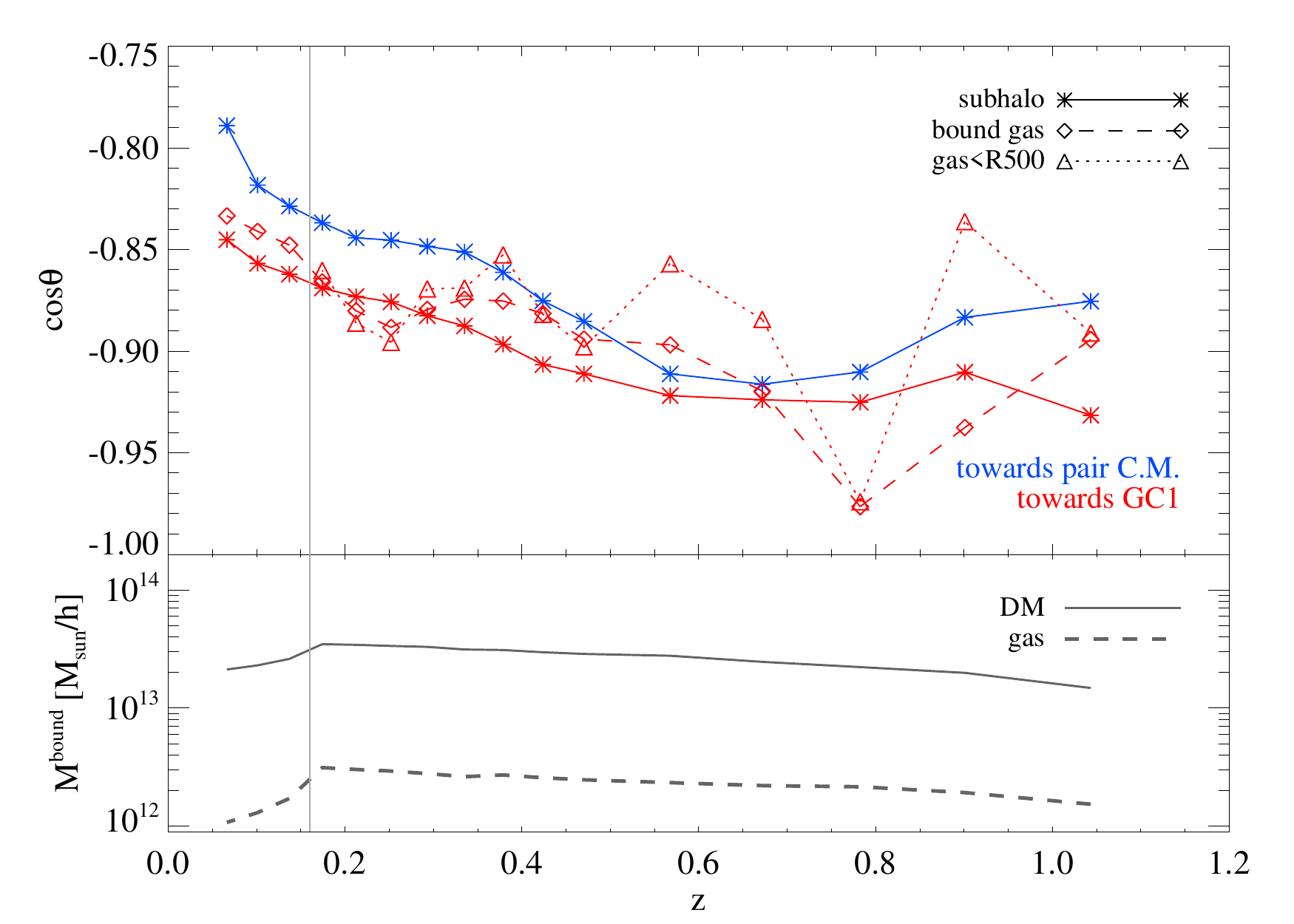}
    \caption{\label{fig:north-clump-vrad-1}
    Redshift evolution of the cosine of the angle between the radial direction towards the pair (center of mass, blue; GC1, red) and the velocity of the subhalo 
    (bulk, solid; bound gas, dashed; all gas within $R_{500}$, dotted). The gas velocity alignment is reported only for the infall direction towards GC1. Data points for all the gas within $R_{500}$ are only defined for the redshifts when the clump is an independent group from GC1 ($z\gtrsim 0.16$).
    We also report the redshift evolution of the subhalo DM (solid lines) and gas (dashed lines) bound mass in the lower inset of the figure.
    }
    \end{figure}

In the simulation analog, we find that the group movement towards the GC1 cluster is well defined below $z\lesssim 1$, according to several indicators.
In Fig.~\ref{fig:north-clump-vrad-1} (upper inset) we show the redshift evolution of the alignment between the subhalo velocity and the infall direction, quantified by the cosine of the angle defined by those two directions. Different colors refer to the direction towards the final positions at $z=0.07$ of the pair center of mass (c.m.; blue) and the center of GC1 (red), as in the legend.
We note that values of the cosine below $z\sim1$ are always comprised between $-0.8$ and $-0.95$, indicating a preferential alignement towards the GC1 final position: this confirms that group B is actually moving towards the GC1 cluster. This result holds for both the total bound dark matter (DM; solid line) and gas (dashed line) components and for all the gas (either bound or not) located within $R_{500}$ (dotted lines). While the total bound gas follows more closely the DM trend, the gas within $R_{500}$ is more subject to oscillations due to the fact that the selection is purely geometrical and can therefore comprise matter not yet gravitationally bound to the main halo potential.
At redshift $z\lesssim0.2$ the cosine absolute value decreases and marks a weaker alignment, and is also accompanied by a decrease of the total mass of the DM and gas that are gravitationally bound to group B (lower inset). This indicates a stripping process occurring while the group enters the cluster virial radius: the bound gas mass, in particular, decreases by a factor of $\sim3$ between $z\sim0.16$ and $z=0.07$.

\begin{figure}
    \centering
    \includegraphics[width=.9\columnwidth,trim=0 0 0 0,clip]{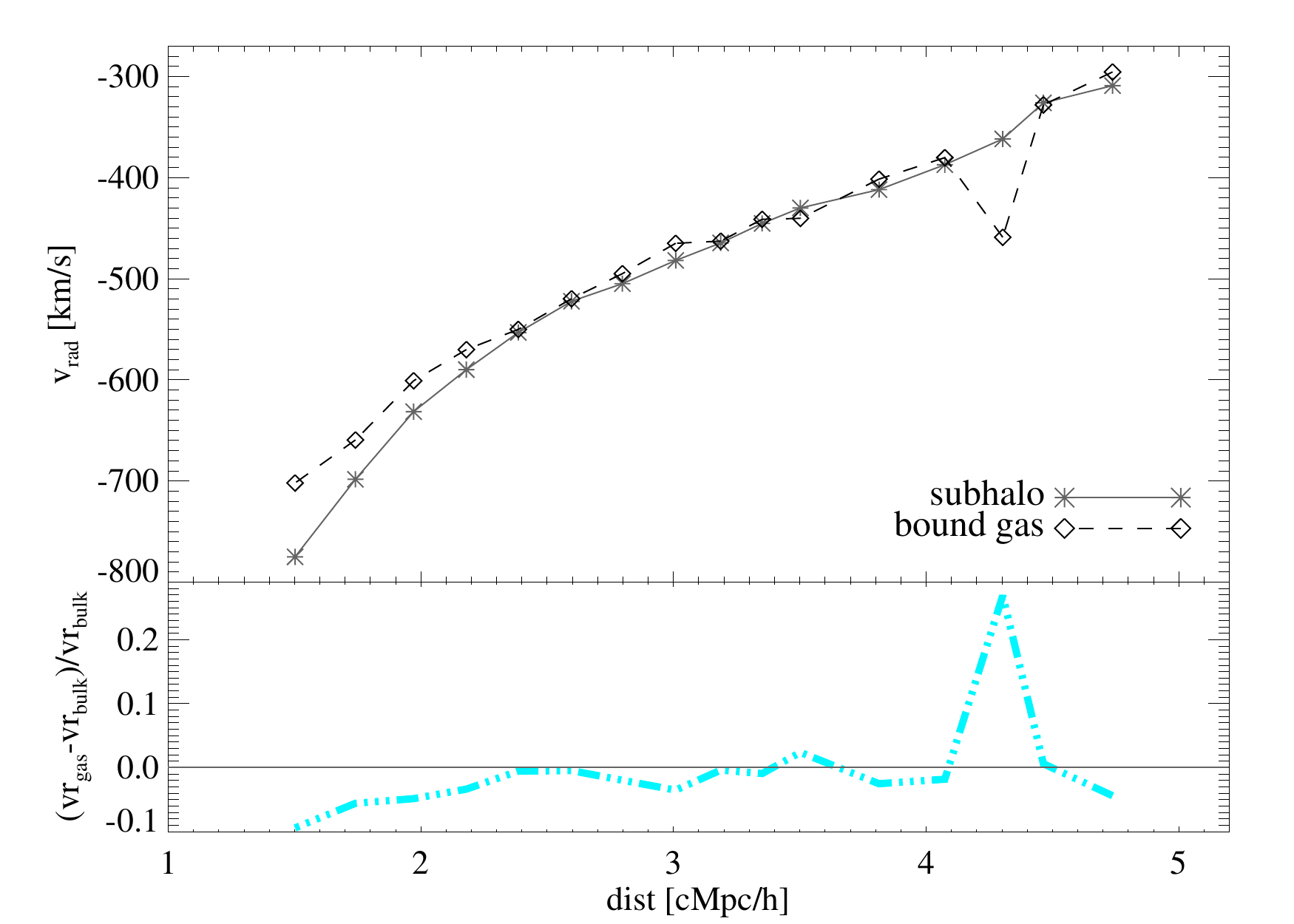}    \caption{\label{fig:north-clump-vrad-2}
    {\it Top:} radial component of the bulk (solid lines) and bound-gas velocity (dashed lines) as a function of the radial distance from the final position of GC1. 
    {\it Bottom:} relative difference between bound-gas and bulk 
    radial velocity, as a function of the radial distance.
    }
    \end{figure}

The infall of group B towards GC1 is furthermore characterized by an increasing difference between the radial components of gas and subhalo bulk velocities below redshift $z\sim0.25$. This is shown in Fig.~\ref{fig:north-clump-vrad-2}, where we report the radial components of bulk (solid) and bound gas (dashed) velocity, as a function of the distance (comoving units) between the position of group B at each redshift and the final position of GC1 (upper inset) --- here larger radial distances correspond to earlier times. 

From the Figure, one can note the slowing down of the gas component with respect to the DM (traced by the halo bulk velocity) at the smaller distances, corresponding to the redshifts $0.07 < z < 0.25$. 
This is better quantified in the lower inset of Fig.~\ref{fig:north-clump-vrad-2}, where we report the relative difference between gas and bulk radial velocities at each point. We note a monotonic increase of such difference, while group B is getting closer than $\sim 2.5\:$cMpc$/h$ from the final position of GC1, which corresponds to $z\lesssim 0.25$. 
At $z=0.25$, the actual distance between the center of group B and the GC1 progenitor corresponds to $d_{\rm 3D}\sim 3.5\:$cMpc$/h$, namely $d_{\rm 3D}\sim 1.7\times (R_{100}^{\rm B}+R_{100}^{\rm GC1}) \sim 4 \times R_{200}^{\rm GC1}$. 
In the reference projection~(\citealp[$xy$ plane; see][]{Reiprich_2021} and \citealp[][for more details]{Biffi_2021}), this corresponds to a distance of $\sim 3 \times R_{200}^{\rm GC1}$ from GC1.
The projected configuration and distance relative to the GC1 cluster found at $z=0.25$ are similar to the observed Northern Clump case, supporting the possibility for the bigger accreting cluster (A3391 in the observed system) and the infall movement to leave an observable signature on the Northern Clump properties (such as the emission tail --- see Subect.~\ref{sec:image_features}).
At such distances, we indeed expect to observe the effects of the closeness to the GC1 cluster. In fact, this is also the typical distance from clusters at which simulations predict to observe changes in the gas properties of main filaments connected to them~\cite[namely at three to four times the cluster virial radius; e.g.,][]{dolag2006}. 
A zoom-in onto the group is shown in Fig.~\ref{fig:groupB-maps} by the simulated map of the projected gas density within a $4\:$cMpc$/h$-side FoV centered on group B at redshift $z=0.25$ (left) and $z=0.07$ (right). 
In the maps, the movement of the group occurs roughly in the top-right to bottom-left direction and the dot-dashed circle indicates the radius corresponding to $\sim 3 \times R_{200}^{\rm GC1}$ from the GC1 progenitor at $z=0.25$. The black circles approximate the $R_{200}$ radius of group B at $z=0.25$ (solid line), and the $R_{200}$ radius of GC1 at $z=0.07$ (dot-dashed line), when group B is no more identified as a separate halo by our substructure-finding algorithm~\cite[SUBFIND,][]{springel2001,dolag2009}.

\begin{figure}
    \centering
    \includegraphics[width=.9\columnwidth,trim=15 0 0 21,clip]{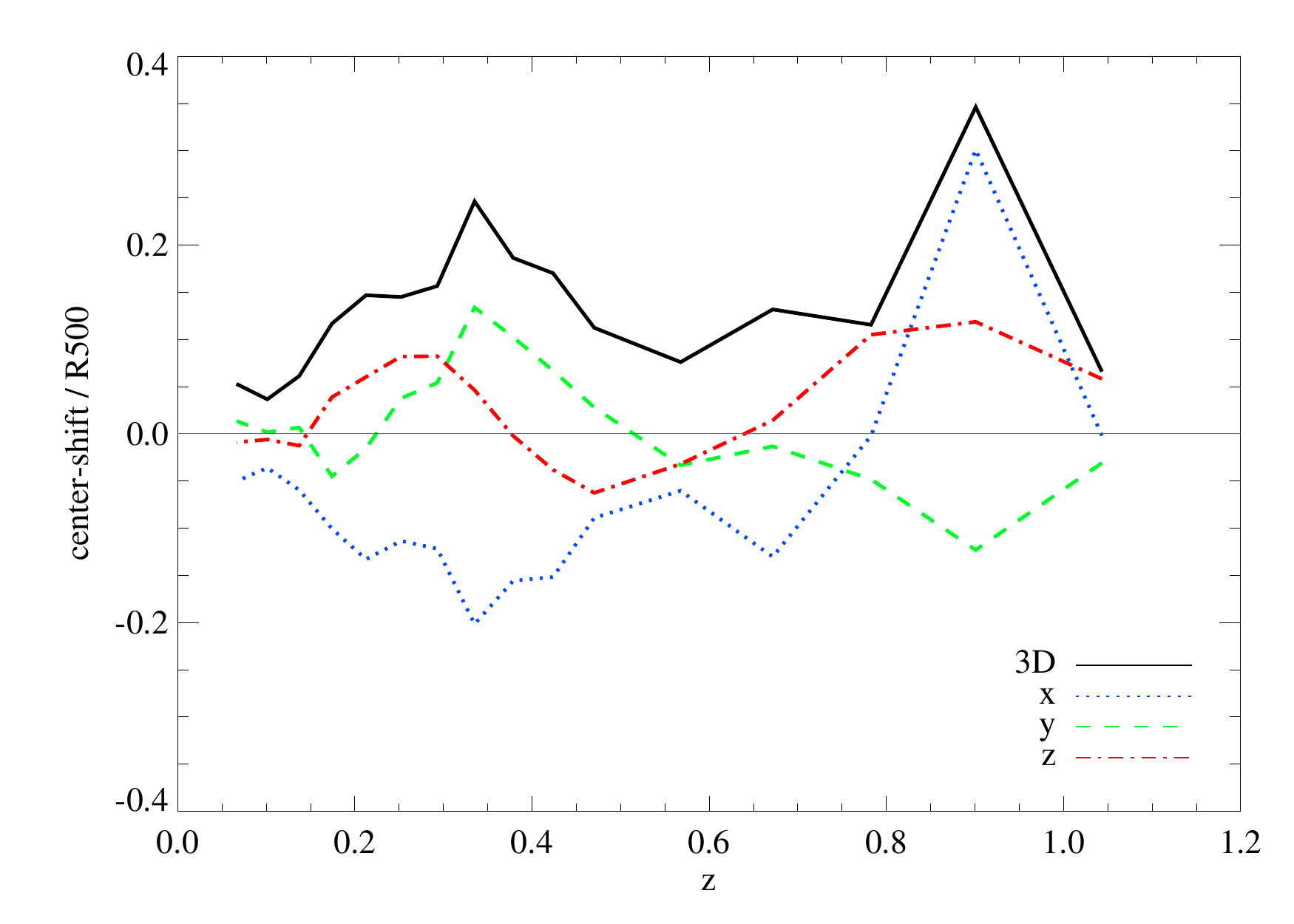}
    \caption{\label{fig:north-clump-vrad-3}
    Redshift evolution of the center shift between the position of the halo center (coinciding with the position of the minimum of the halo potential well) and the bound-gas center of mass, in units of $R_{500}$. The three-dimensional value of the shift and its three separate components along the major Cartesian axes of the simulations are reported, as in the legend.
    }
\end{figure}

\begin{figure*}
    \centering
    \includegraphics[width=.99\columnwidth,trim=60 0 40 22,clip]{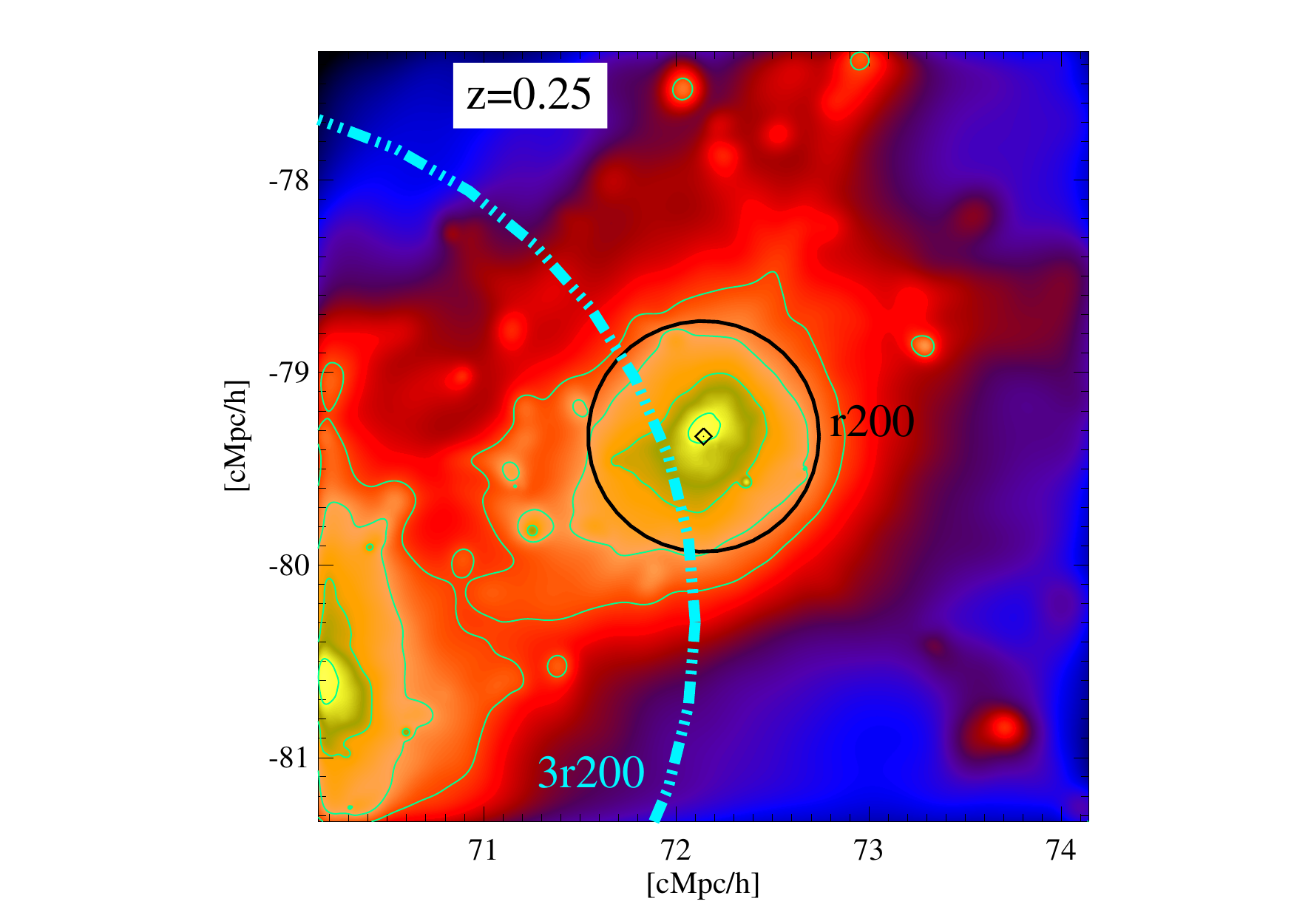}
    \includegraphics[width=.99\columnwidth,trim=60 0 40 22,clip]{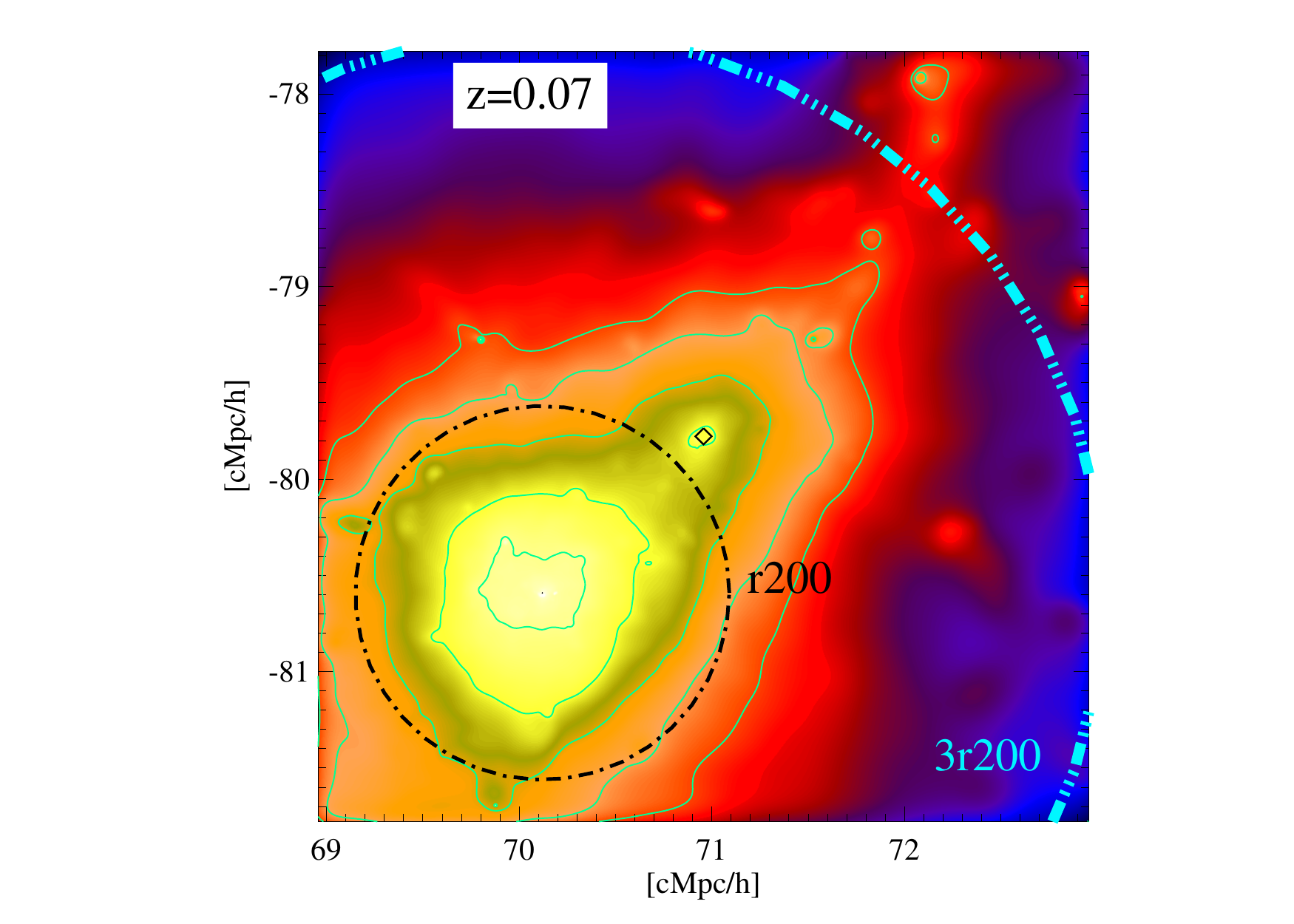}
    \caption{
    Simulated projected maps of gas surface density, centered on the infalling group B (black diamond marking its center), at $z=0.25$ (left) and $z=0.07$ (right). The gas surface density map encloses a comoving volume of $(4\,$cMpc$/h)^3$.
    In both panels, we mark the circles corresponding to three times the GC1 cluster virial radius ($3\times R_{200}^{\rm GC1}$) and $R_{200}$ of the group at $z=0.25$ (solid) and of the GC1 cluster at $z=0.07$ (dot-dashed; when group B is a substructure of GC1).
    }
    \label{fig:groupB-maps}
    \end{figure*}

We also find indications of halo sloshing throughout its evolution. This can be inferred from the oscillations described in the radial velocity cosine (upper inset in Fig.~\ref{fig:north-clump-vrad-1}), when all the gas within $R_{500}$ is considered (dotted curve) in comparison to the corresponding DM trend.
This effect is further confirmed by Fig.~\ref{fig:north-clump-vrad-3}, where we report the redshift evolution of the center shift for group B between $z\sim1$ and $z=0.07$. The center shift is computed as the three-dimensional difference between the halo center\footnote{The halo center is identified with the position corresponding to the minimum of the potential well.} and the bound gas center of mass, in units of the group $R_{500}$. In the Figure, we also report the separate components along the major axes of the simulation volume, from which we note an oscillating feature, especially between the $x$ and $y$ components --- i.e. in the plane perpendicular to the chosen l.o.s. 
This effect is likely connected to the merging and accretion of smaller galaxy-size substructures (such as between $z\sim1$ and $z\sim 0.8$), which perturb the gas component in the group and also produces elongated density contours.


\section{Discussion and conclusions}
We presented results of the X-ray imaging and spectral analyses of the MCXC J0621.7-5242 galaxy cluster, also known as the Northern Clump of the A3391/95 system, utilizing \rosi, \xmm, and \chandra observations. We used the ASKAP/EMU radio data, {\it Planck y}-map, and DECam data to study the influence of the wide-angle tail (WAT) radio source to the central intracluster medium (ICM) gas. We compared the observations with an analog of the A3391/95 system found in the Magneticum simulation. Our findings include:
\begin{itemize}
    \item The Northern Clump is not a relaxed system and its morphology deviates from the common spherical or elliptical shapes. In fact, it has an interesting pronounced boxy shape.
    \item We identified a southern density enhancement, which indicates motion of the filament gas relative to the Northern Clump resulting in compression, e.g., by filament gas being accreted towards the Northern Clump while the whole system, filament and the Northern Clump is being accreted towards A3391. The temperature and density enhancement at around the same distance support this.
    \item We detected a tail of emission north of the Northern Clump that supports the relative motion as well. The tail starts at $\sim\!7'$ ($419.16\mathrm{\:kpc}$) from the center with a projected length of $\sim\!5.3'$ ($317.4\mathrm{\:kpc}$). We speculate that the gas is either stripped from the Northern Clump and lagging behind, or infalling gas from a possible continuation of the filament north of the Northern Clump. While a more detailed analysis is required, the first picture is nevertheless consistent with findings from the analog system identified in the Magneticum simulations, where the gas in the clump start to lag behind with its infall velocity reduced compared to dark matter, while getting closer than $3\times R_{200}$ from the accreting cluster.
    \item A bright WAT radio galaxy occupies the center of the Northern Clump. A weak hint of X-ray surface brightness depressions coincident with the radio lobes is found, suggesting  interaction of the radio plasma with the Northern Clump ICM.
    \item We note that the morphology of the denser X-ray gas is asymmetric with a sharp edge that points towards the A3391 center, in the southeast direction. The lower density gas at around $\sim\!5'$ appears boxy, which may be the result of ram pressure stripping as it enters the atmosphere of A3391 and/or sloshing. However, the northeast bend of both lobes is not quite consistent with them being “swept back” by that global motion. We speculate that the lobes might have “escaped” some hot-gas confinement and the change of direction is due to motion through a lower-density ambient medium. The central radio source may be moving locally within the cluster toward the southwest as a result of sloshing, while the Northern Clump is moving toward the southeast.
    \item The electron density and entropy derived for the central region of the Northern Clump indicates that the Northern Clump is a weak cool-core (WCC) cluster. The temperature flattening towards the center further supports this indication. The WCC nature could indicate some disturbance of the core either through effects caused by entering the sphere of influence of A3391, through infalling filament gas, or through the central radio AGN.
    \item The low core metallicity derived for the core using three different methods signifies AGN feedback blowing the metal-rich ICM gas outwards the central region. This is consistent with the strong radio emission of the central galaxy and with the fact that the cooling flow in this cluster is suppressed, which as a result only forming a weak cool-core.
    \item We performed spectral analyses in the region between the $R_{200}$ of the Northern Clump and the A3391 cluster, the Northern Filament, where we assumed a simple cylindrical geometry on the plane of the sky. Through a 2T fit, we identified a cooler component whose temperature ($k_BT=0.68_{-0.64}^{+0.38}~\mathrm{keV}$) and electron density ($n_e=1.99_{-1.24}^{+0.88}\times10^{-5}~\mathrm{cm}^{-3}$) are consistent with the expected ranges of WHIM properties.
    \item Analysis of a similar system found in the Magneticum simulation reveals that an infalling clump (equivalent to the observed Northern Clump) starts to ``feel'' the influence of the main accreting system (equivalent to the observed A3391) at around $3\times R_{200}$, possibly related to accretion shocks expected to be present at such distance. For instance, the infall velocity of the gas component starts to decrease compared to the radial velocity of the dark matter. 
    Once the group approaches the cluster virial radius, it undergoes further stripping processes with a decrease by a factor of $\sim\!3$ in the gravitationally-bound gas mass between $z=0.16$ and $z=0.07$. Throughout its redshift evolution, we find evidences for halo sloshing marked by oscillations of the infall velocity and the center shift of the gas component compared to dark matter. In general, the findings in the Magneticum simulation are consistent with the properties of the observed system.
\end{itemize}

Thanks to the wide \rosi FoV and its superior soft response, significant emission was observed in the A3391/95 system beyond the virial radii, $R_{100}$, with good resolution. This has allowed us to discover the soft X-ray emission from filaments connected to the system, as well as some infalling clumps residing in these filaments, one of which is the Northern Clump. This infalling galaxy cluster appears to have just entered the atmosphere, or sphere of influence, of the A3391 cluster.
\par
In this work, we demonstrated how \rosi is complementary not only to other X-ray instruments, e.g., \xmm and \chandra, but also instruments from other wavelengths, e.g., the ASKAP radio telescope, the DECam optical survey, and the {\it Planck} microwave satellite. We also found that the \rosi image of the A3391/95 system and the results of our Northern Clump analysis are consistent with each other, as well as with the analog simulated system from the Magneticum simulations.

\begin{acknowledgements}
      We would like to thank the anonymous referee for their valuable feedback that helped improve the manuscript.
      We thank Kaustuv moni Basu for discussion about the {\it Planck y}-map.
      Funded by the Deutsche Forschungsgemeinschaft (DFG, German Research Foundation) – 450861021.
      This research was supported by the Excellence Cluster ORIGINS which is funded by the Deutsche Forschungsgemeinschaft (DFG, German Research Foundation) under Germany's Excellence Strategy – EXC-2094 – 390783311.
      VB acknowledges funding by the Deutsche Forschungsgemeinschaft (DFG, German Research Foundation) --- 415510302.
      YS acknowledges support from Chandra grants AR8-19020A and GO1-22126X.
      AV is a member of the  Max-Planck International School for Astronomy and Astrophysics (IMPRS) and of the Bonn-Cologne Graduate School for Physics and Astronomy (BCGS), and thanks for their support.
      This work is based on data from eROSITA, the soft X-ray instrument aboard SRG, a joint Russian-German science mission supported by the Russian Space Agency (Roskosmos), in the interests of the Russian Academy of Sciences represented by its Space Research Institute (IKI), and the Deutsches Zentrum für Luft- und Raumfahrt (DLR). The SRG spacecraft was built by Lavochkin Association (NPOL) and its subcontractors, and is operated by NPOL with support from the Max Planck Institute for Extraterrestrial Physics (MPE). The development and construction of the eROSITA X-ray instrument was led by MPE, with contributions from the Dr. Karl Remeis Observatory Bamberg and ECAP (FAU Erlangen-Nuernberg), the University of Hamburg Observatory, the Leibniz Institute for Astrophysics Potsdam (AIP), and the Institute for Astronomy and Astrophysics of the University of Tübingen, with the support of DLR and the Max Planck Society. The Argelander Institute for Astronomy of the University of Bonn and the Ludwig Maximilians Universität Munich also participated in the science preparation for eROSITA. The eROSITA data shown here were processed using the eSASS/NRTA software system developed by the German eROSITA consortium.
      Partly based on observations obtained with XMM-Newton, an ESA science mission with instruments and contributions directly funded by ESA Member States and NASA.
      We acknowledge the Wajarri Yamatji people as the traditional owners of the Murchison Radio-astronomy Observatory, where ASKAP is located.
\end{acknowledgements}

%
%
\bibliographystyle{aa}
\bibliography{list_bib}
\begin{appendix}
\onecolumn
\section{\xmm ~-- Lightcurves} \label{app:xmm}

\begin{figure}[h!]
\centering
\includegraphics[width=0.49\textwidth]{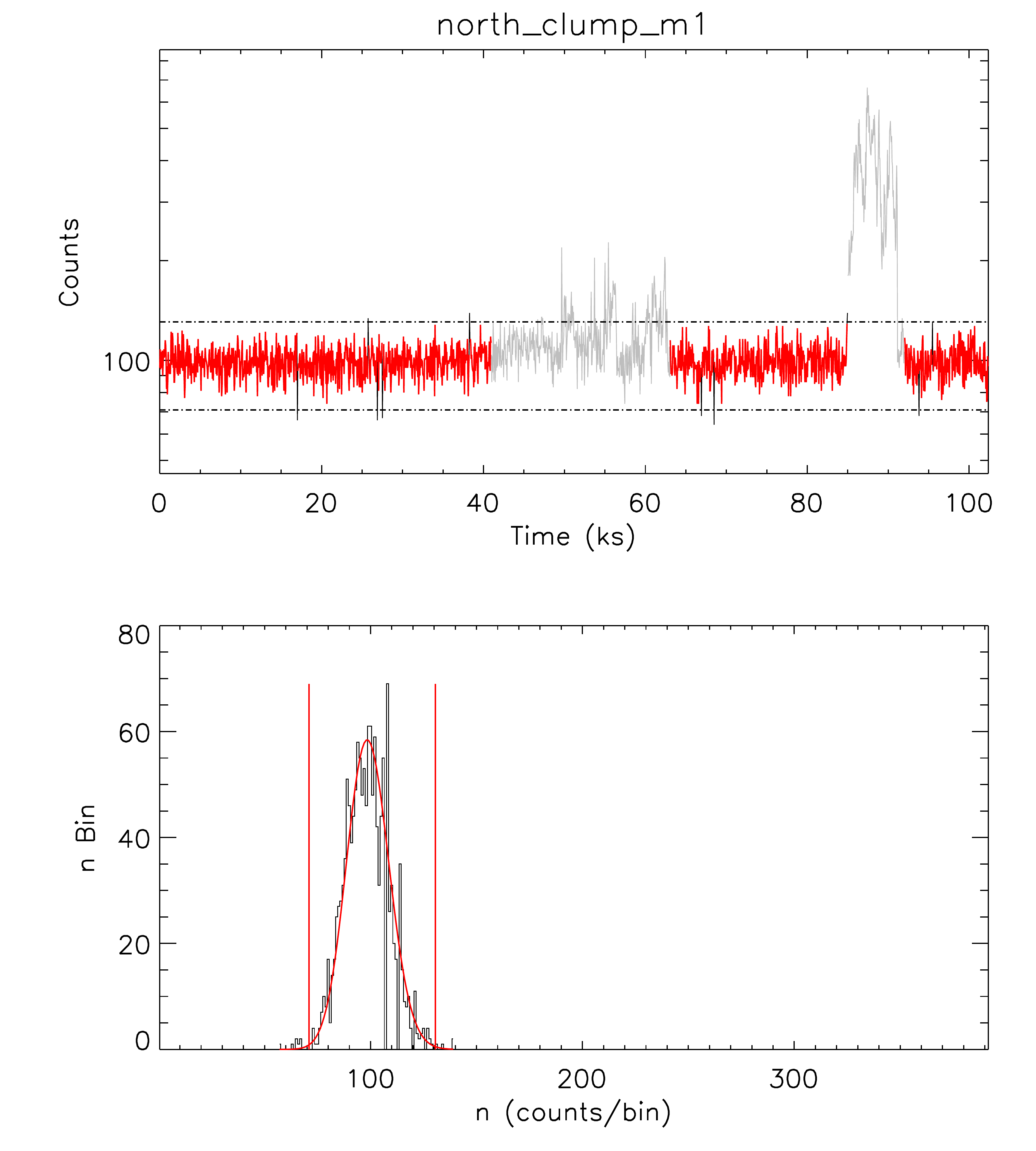}
\includegraphics[width=0.49\textwidth]{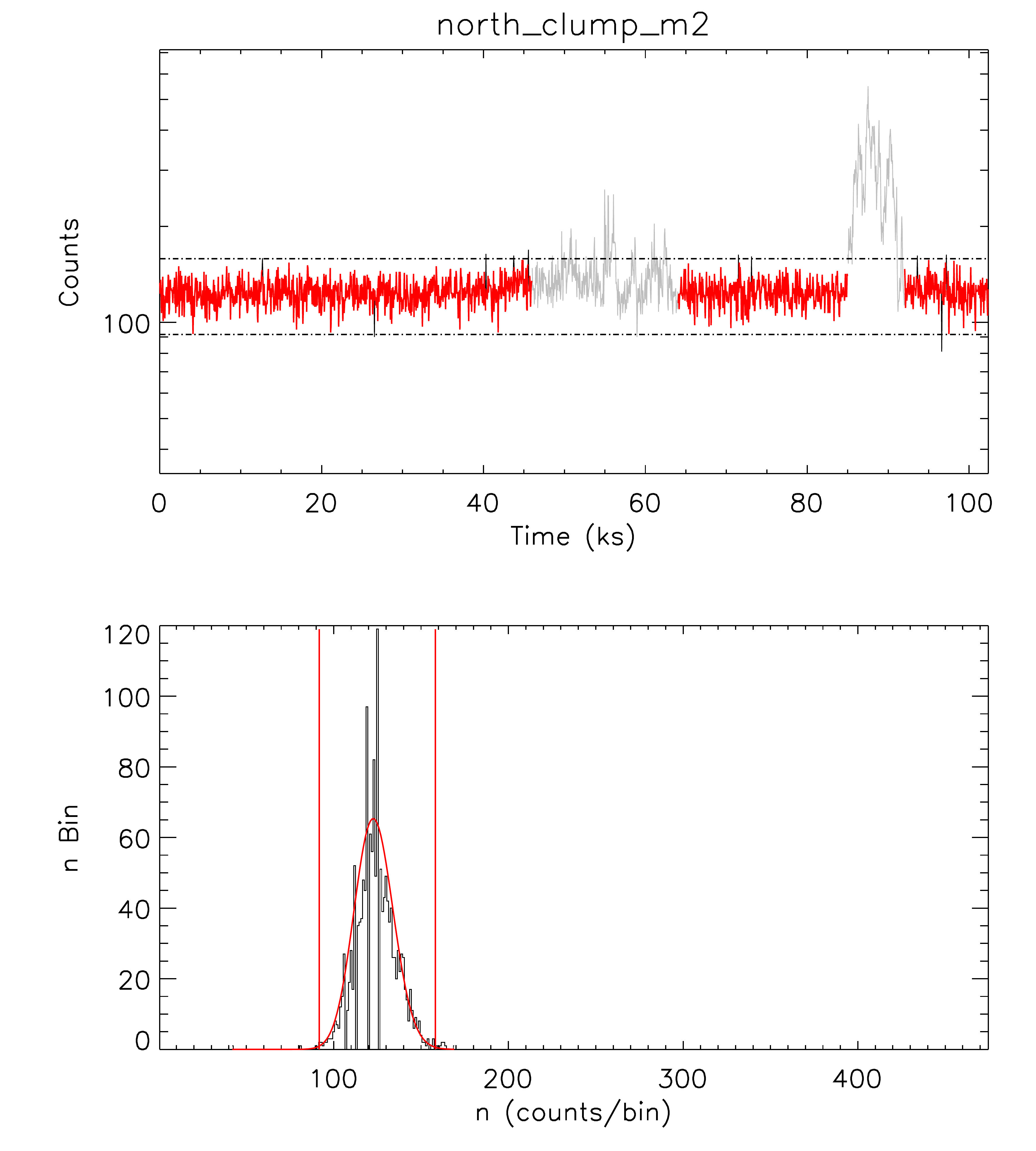}
\includegraphics[width=0.49\textwidth]{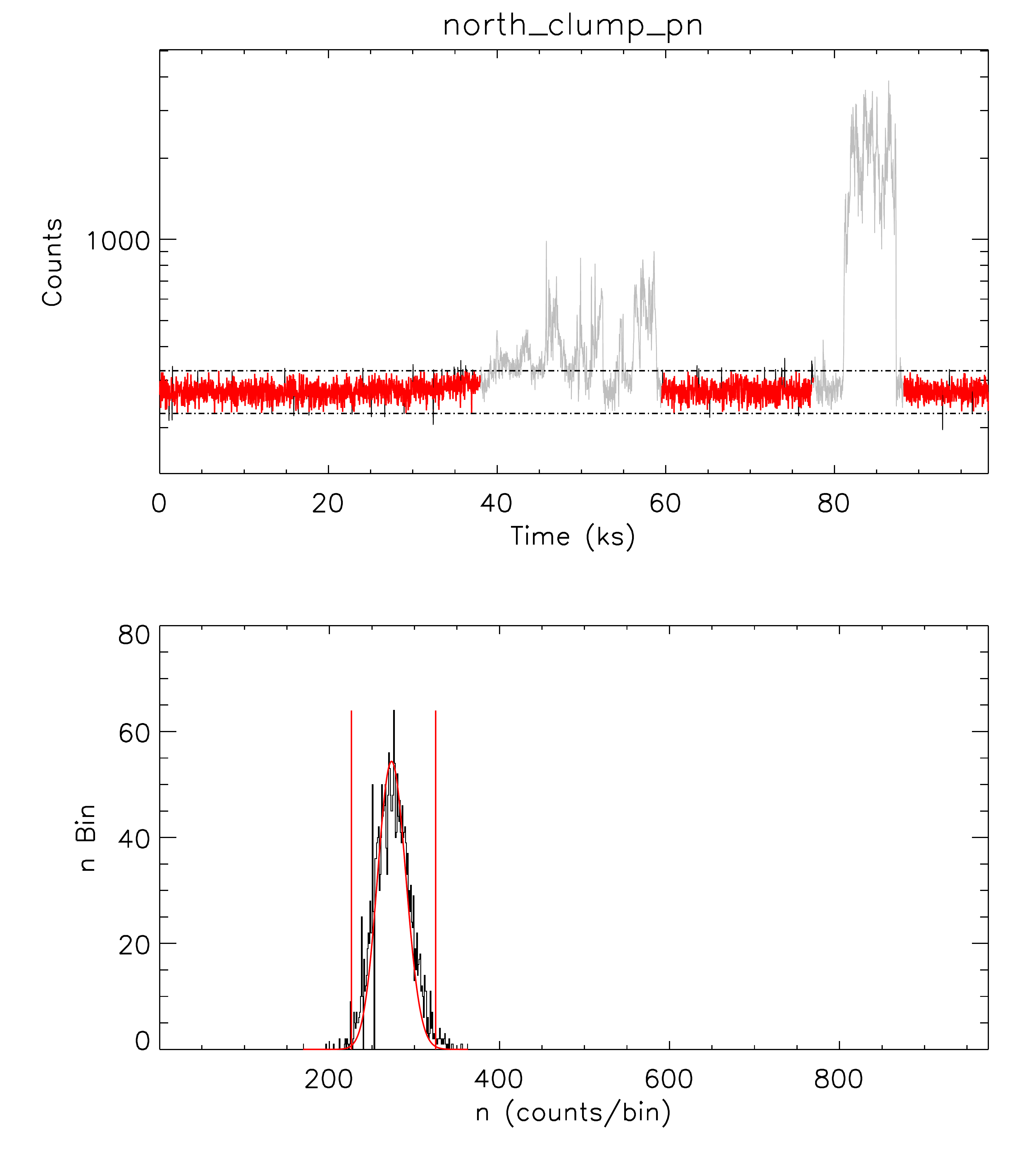}
\caption{Top panels show the lightcurves of MOS1 (\textit{top left}), MOS2 (\textit{top right}), and pn (\textit{bottom}) of the Northern Clump observation (ObsID: 0852980601). The red line depicts the accepted portion of the observation, while the grey is the rejected portion due to SPF. The dashed lines mark the $3\sigma$. The bottom panels show the histograms of the associated lightcurves, fitted with Gaussian (red curve). The red vertical lines mark the $3\sigma$.}
\label{fig:xmm_LC_north_clump}
\end{figure}

\clearpage
\twocolumn
\section{\xmm ~-- Images}\label{App:residuals}
\begin{figure}[!ht]
   \centering
   \includegraphics[width=\columnwidth]{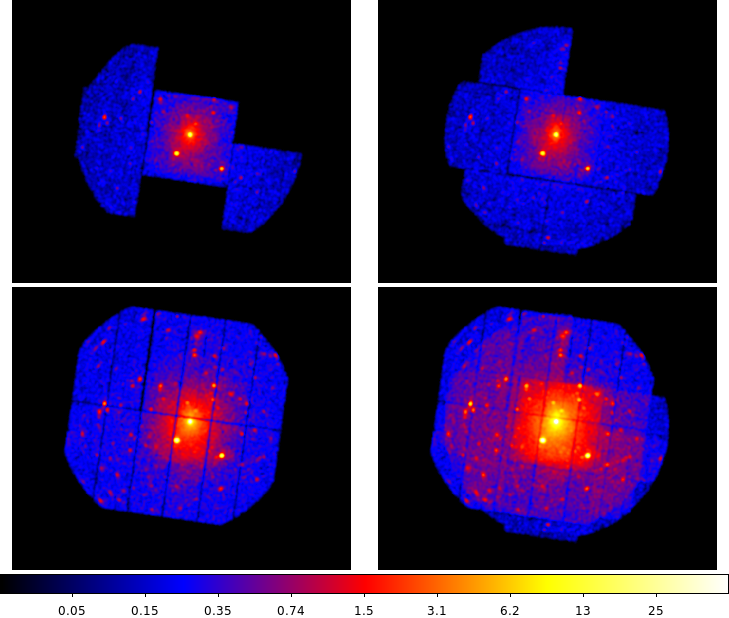}
   \caption{\xmm cleaned photon images, i.e., SPF filtered, QPB subtracted, anomalous CCDs removed, in the energy band $0.5-2$ keV. {\it Top left:} MOS1 image. {\it Top right:} MOS2 image. {\it Bottom left:} pn image. {\it Bottom right:} MOS1+2+pn image.}
\label{fig:XMM_photonimages}%
\end{figure}

\section{\xmm ~-- Surface brightness profile}\label{App:SB_sec_circ}
\begin{figure}
\centering
\includegraphics[width=0.48\textwidth]{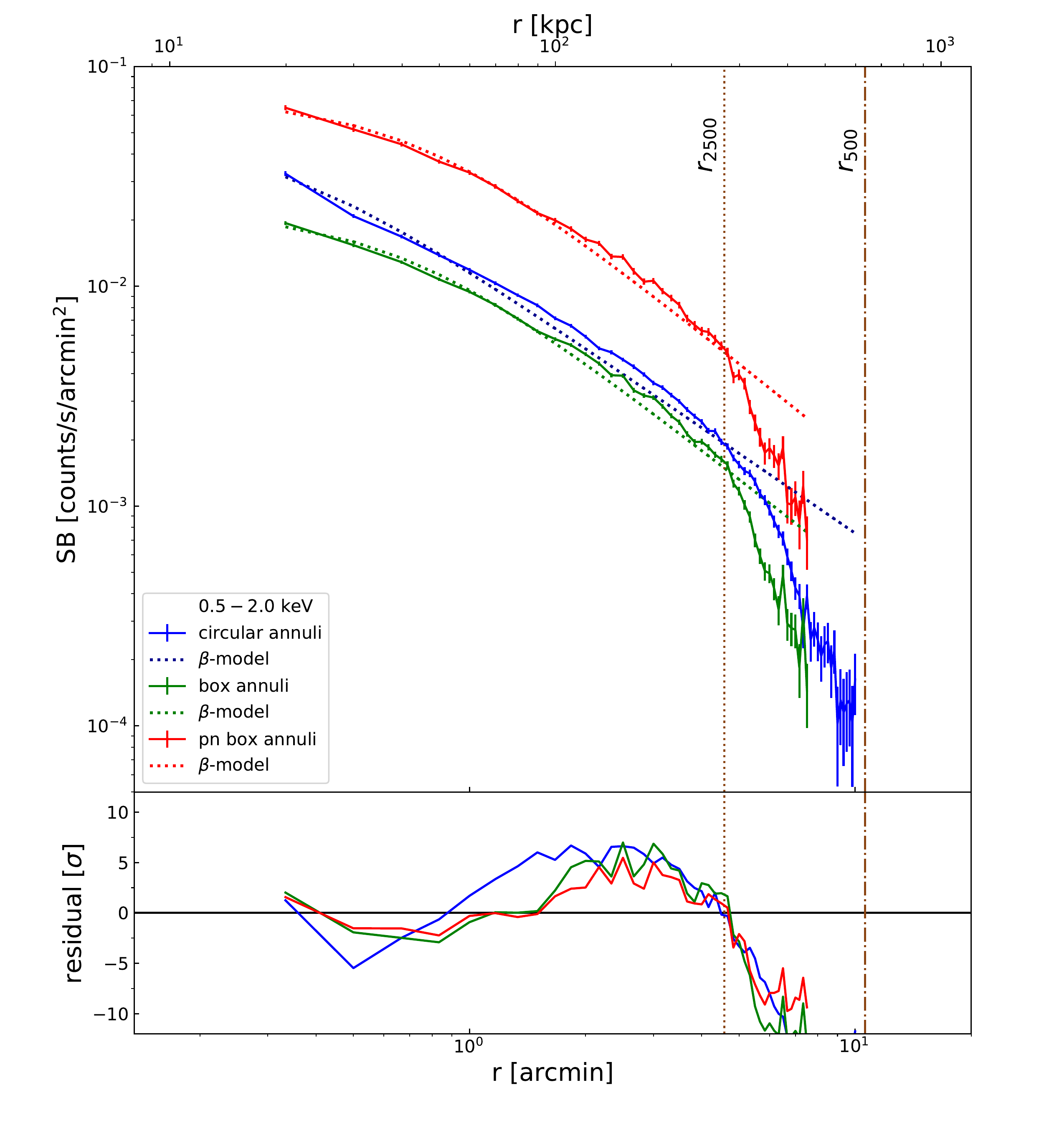}
\includegraphics[width=0.48\textwidth]{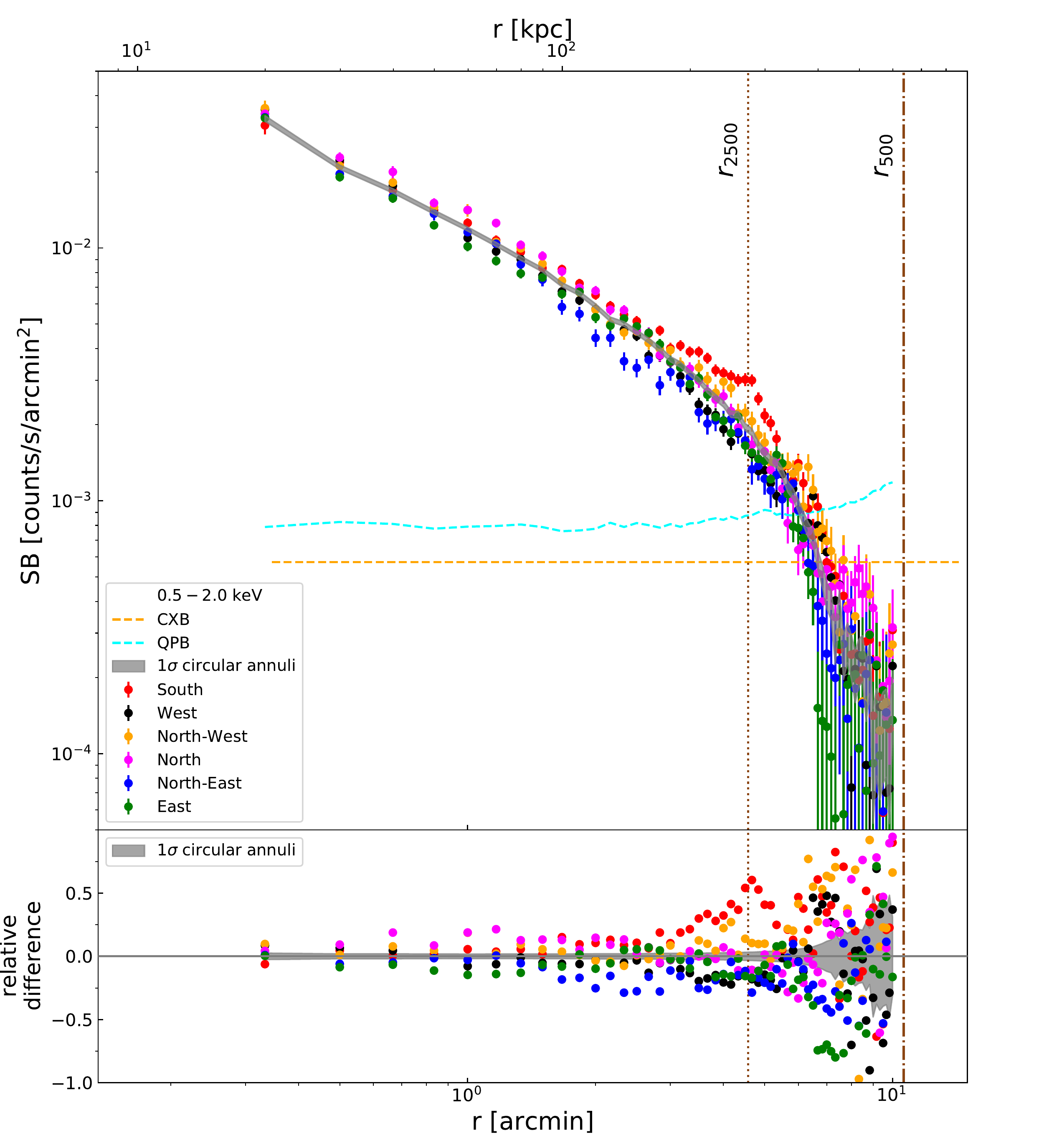}
\caption{\xmm $S_X$ profiles calculated from \xmm data in the energy band $0.5-2.0$ keV. {\it Top:} The full azimuthal $S_X$ profiles. The top panel shows the $S_X$ profiles (solid lines) and the corresponding $\beta$-model fits (dotted lines) of all \xmm detectors with circular annuli setup (blue) and box annuli setup (green), as well as the pn-only box annuli setup (red). The bottom panel shows the residuals in unit of $\sigma$. {\it Bottom:} \xmm $S_X$ profiles of different sectors \textit{top} and their relative differences \textit{bottom} from circular annuli setup. The grey shaded areas are the $1\sigma$ confidence regions of the full azimuthal $S_X$. The brown vertical lines indicate the various radii.}
\label{fig:xmm_SB_fullannuli}
\end{figure}

The $S_X$ profiles of the three setups are compared in Fig.~\ref{fig:xmm_SB_fullannuli} (left). The blue, green, and red solid lines are the $S_X$ profiles for circular annuli, box annuli, and pn-only box annuli setup. The dotted lines of the respective colors are their $\beta$-model fits. The $\beta$-model best-fit parameters can be found in Table \ref{tab:beta_model}. In the bottom panel, their residuals are plotted in the unit of $\sigma$.

\begin{table}[!h]
    \centering
    \caption{The $\beta$-model best-fit parameters.}
    \begin{tabular}{c c c c}
    \hline
    \hline
\multirow{2}{*}{Setups} & $S_X(0)$ & \multirow{2}{*}{$\beta $} & $r_c$\\
 & $[$cts/s/arcmin$^{2}]$ & & $[']$ \\
\hline
circular & $0.05\pm0.004$ & $0.368\pm0.007$ & $0.310\pm0.04$\\[5pt]
box & $0.0218\pm0.001$ & $0.396\pm0.010$ & $0.660\pm0.053$\\[5pt]
pn box & $0.0715\pm0.002$ & $0.404\pm0.011$ & $0.718\pm0.053$\\[5pt]
    \hline
    \hline
    \end{tabular}
    \label{tab:beta_model}
\end{table}

\clearpage
\section{Spectral analysis}\label{App:NC_profles}
\begin{figure}[!h]
   \centering
   \includegraphics[width=\columnwidth]{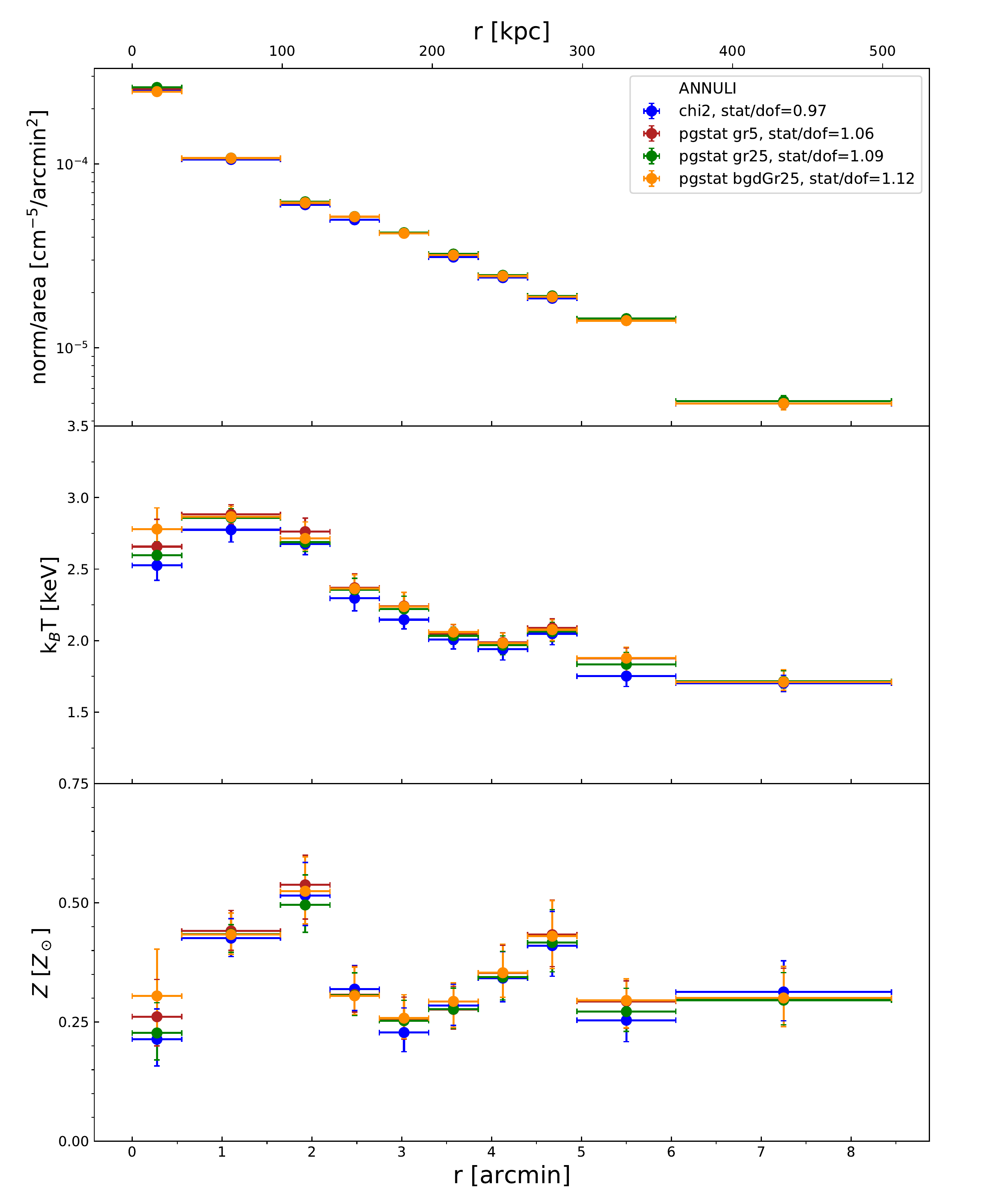}
   \caption{Normalization (top), temperature (middle), and metallicity (bottom) profiles of the Northern Clump derived in the energy band $0.7-7.0$ keV. Blue points are estimated using $\chi^2-$statistics, while the red, green, and orange points are estimated using pgstat statistic with source spectrum grouping 5, 25, and background spectrum grouping 25.}
\label{fig:pgstat}
\end{figure}

We performed several tests using pgstat statistic provided in \texttt{XSPEC}, where the \xmm source data were treated as Poisson data (C-statistic) and the background as Gaussian. Initially, we fit spectra that are grouped to have a minimum of 5 counts (grouping5) and 25 counts (grouping25) per bin. Then, we grouped the source spectra based on the background spectra, as such we would have 25 counts per bin for the background spectra (bgdGrouping25). This ensures the background spectra to be Gaussian. These tests were carried out for all the Northern Clump spectral analysis regions. We show the results from the full azimuthal region in Fig.~\ref{fig:pgstat} as a representative. In the plots we compare the fitting results using $\chi^2-$statistic (blue points, stat/dof=0.97), pgstat grouping5 (red, stat/dof=1.06), pgstat grouping25 (green, stat/dof=1.09), and pgstat bgdGrouping25 (orange, stat/dof=1.12). The resulting cluster parameters of these tests in various regions are always consistent within the $1\sigma$ error bars of the $\chi^2$-statistic fitting results.

\begin{figure}[!h]
   \centering
   \includegraphics[width=\columnwidth]{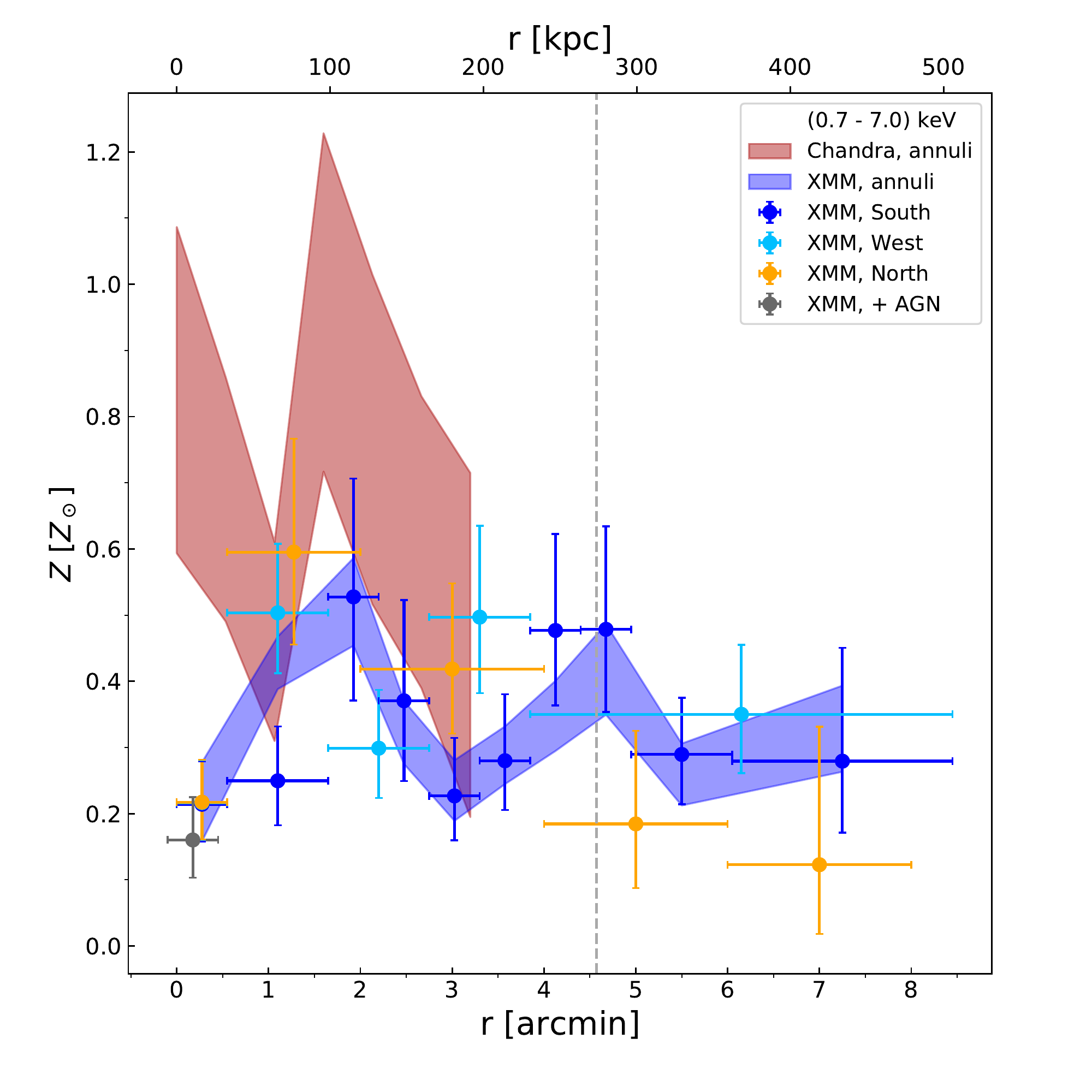}
   \caption{Metallicity profiles of various regions in the Northern Clump cluster in the energy band $0.7-7.0$ keV. The blue and red shaded areas represent \xmm and \chandra metallicity in full azimuthal direction. Due to modest photon counts, \chandra cannot constrain the metallicity. The dark blue, cyan, and magenta data points represent \xmm metallicity profiles in the south, west, and north directions, respectively. The grey dashed lines indicate the $R_{2500}$ of the Northern Clump. The x-axis error bars are not the $1\sigma$ error range, but the full width of the bins.}
\label{fig:profiles_allZ}%
\end{figure}

\begin{figure}[h!]
\centering
\includegraphics[width=0.5\textwidth]{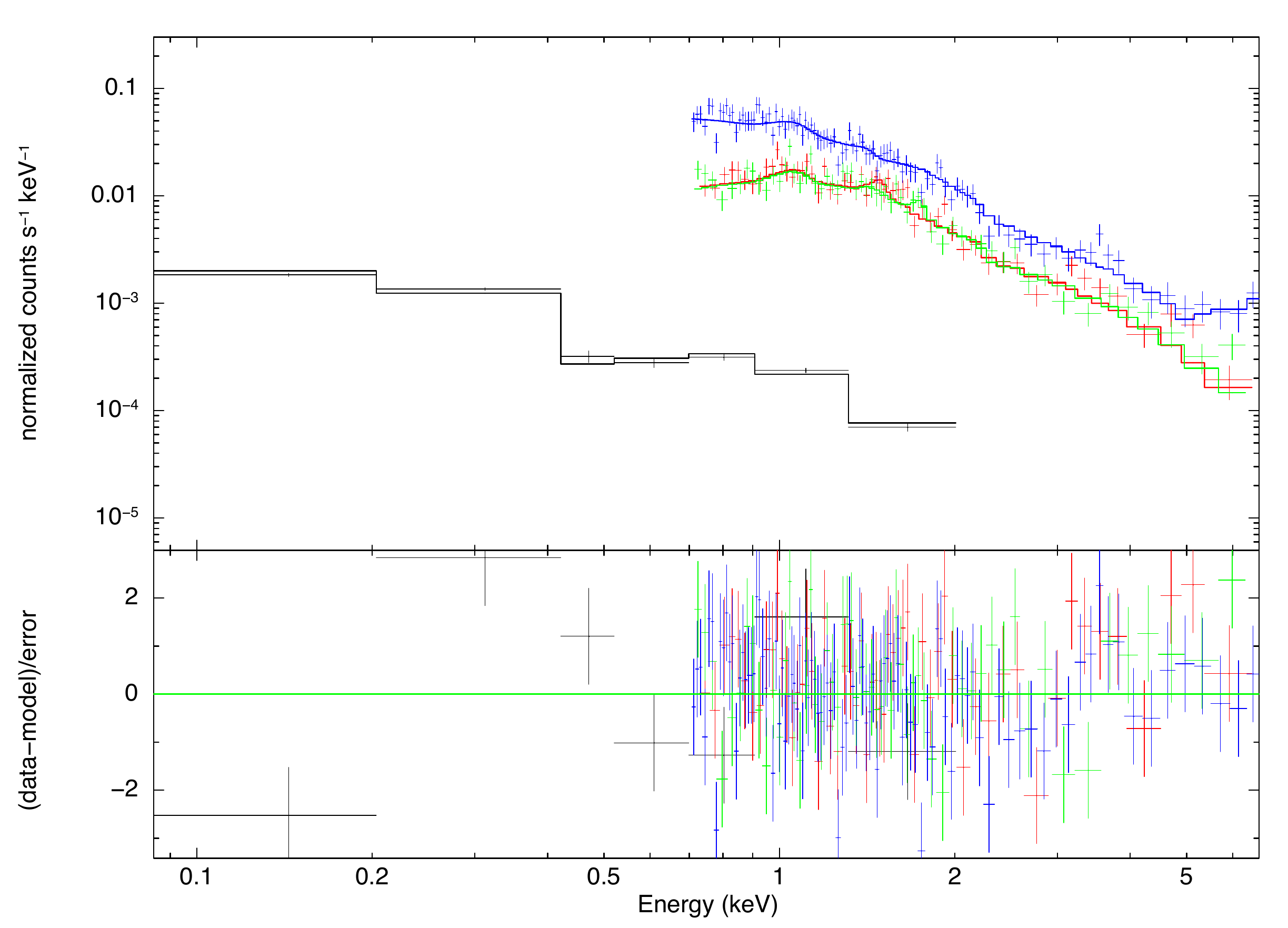}
\includegraphics[width=0.5\textwidth]{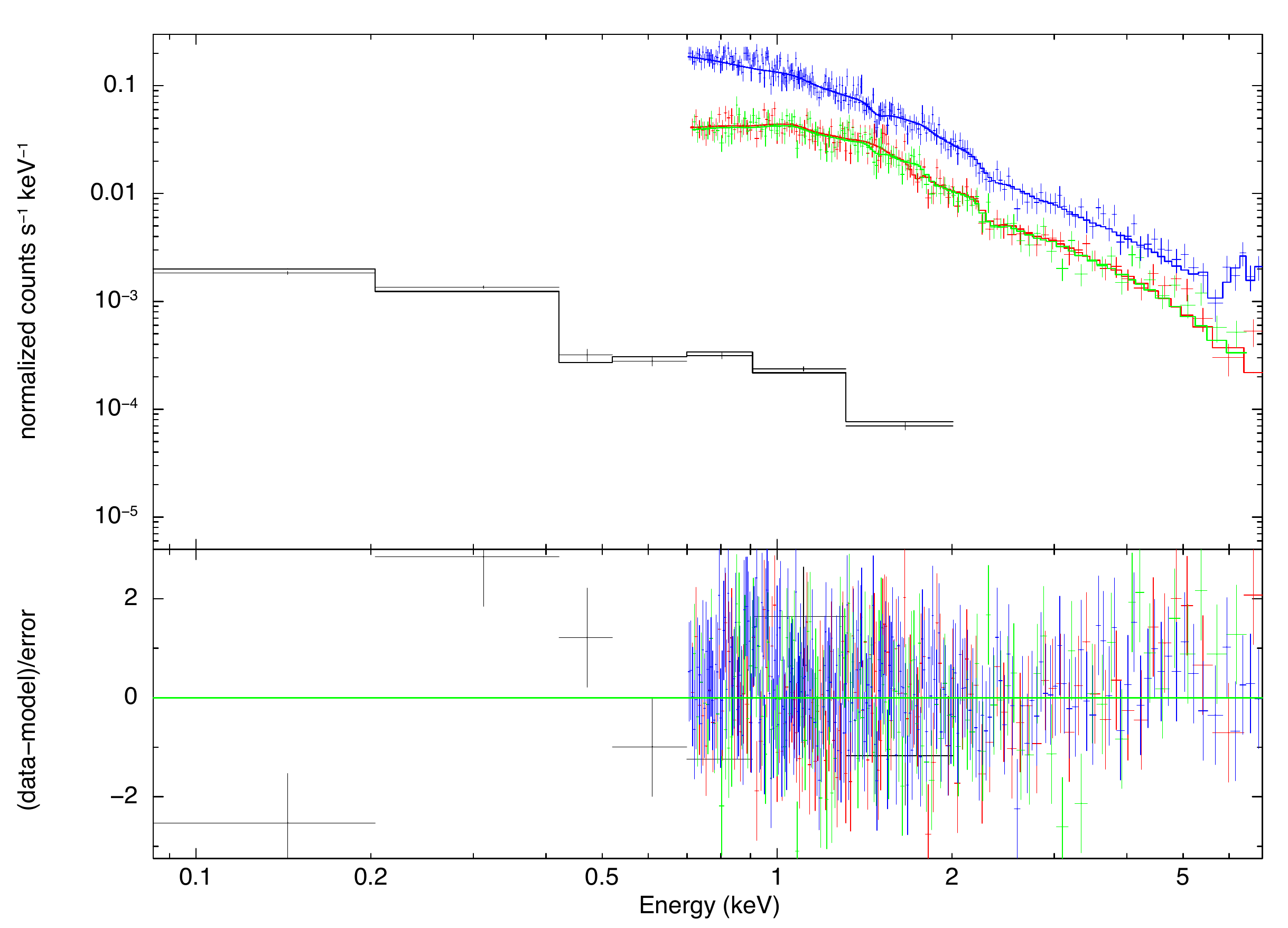}
\includegraphics[width=0.5\textwidth]{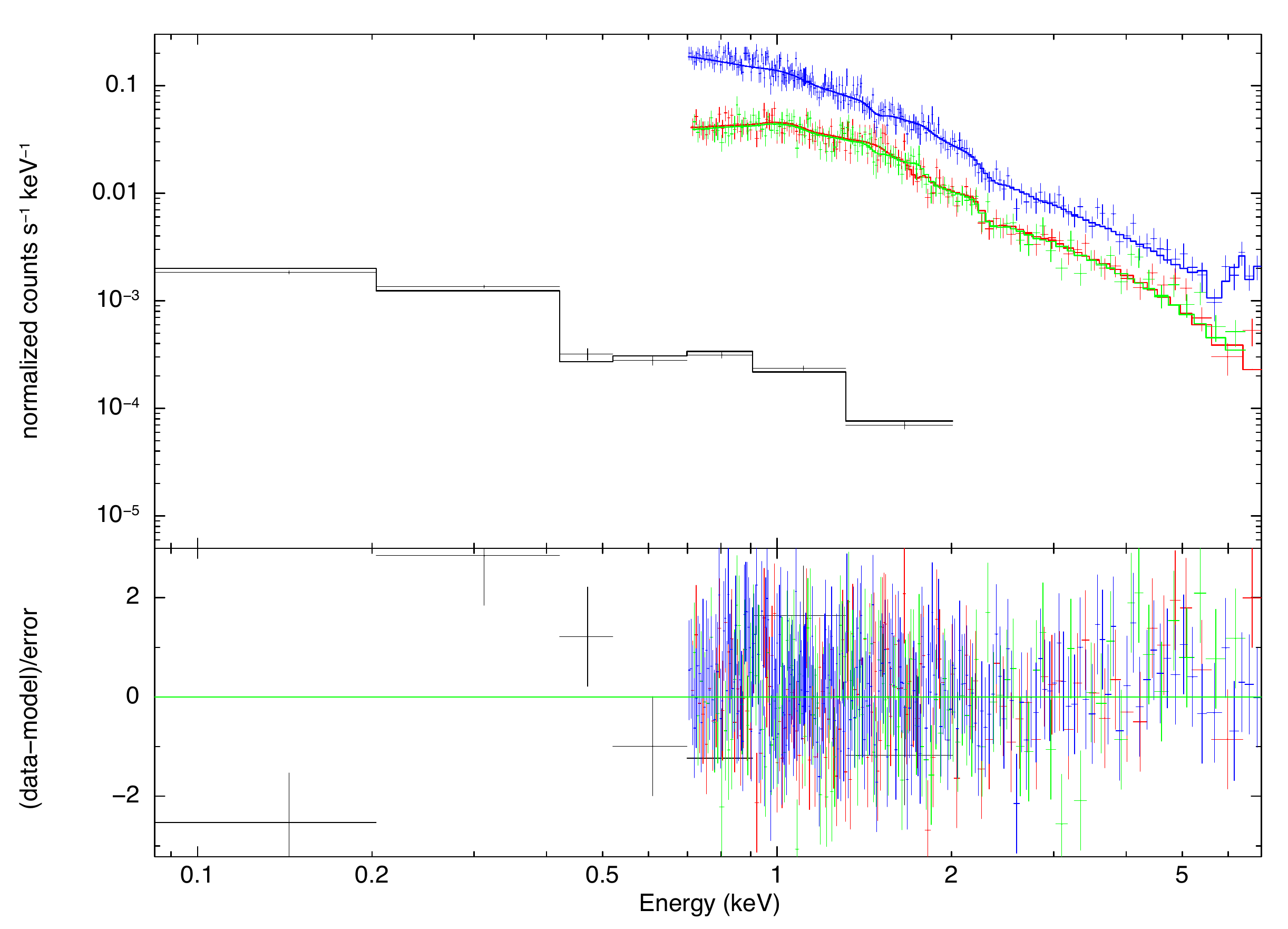}
\caption{\xmm and RASS background spectra (black) of the central bin fitted with the three different central AGN models in the energy range of $0.7-7.0$ keV. \textit{Top:} Method 1 -- with $15''$ mask to mask the central AGN. \textit{Middle:} Method 2 -- AGN component is modeled from \chandra best fit values. \textit{Bottom:} Method 3 -- AGN is modeled and additional thermal component is added.}
\label{fig:xmm_AGN_spectra}
\end{figure}

\begin{table}
    \centering
    \caption{\xmm best-fit values of central bin with different methods in energy $0.7-7.0$ keV.}
    \begin{tabular}{c c c c}
    \hline
    \hline
AGN & \bf $norm^\dagger$ & $k_BT$ & $Z$\\
method &  & $[$keV$]$ & $[Z_{\odot}]$\\[5pt]
\hline
1 & $2.535_{-0.095}^{+0.096}$ & $2.532\pm0.105$ & $0.217_{-0.056}^{+0.064}$\\[5pt]
2 & $2.651_{-0.118}^{+0.12 }$ & $2.411_{-0.116}^{+0.118 }$ & $0.16_{-0.057}^{+0.065}$\\[5pt]
3 & $1.682_{-0.277}^{+0.268}$ & $3.055_{-0.221}^{+0.208}$ & $0.155_{-0.05}^{+0.06}$\\[5pt]
& $0.969_{-0.308}^{+0.344}$ & $1.528$ & -- \\[5pt]
\hline
\multicolumn{4}{l}{\footnotesize $^\dagger$ the presented values have been normalized} \\
\multicolumn{4}{l}{\footnotesize ~~ by their respective source areas} \\
\multicolumn{4}{l}{\footnotesize ~~ $[10^{-4}$ cm$^{-5}/$arcmin$^{2}]$} \\
    \hline
    \hline
    \end{tabular}
    \label{tab:AGN}
\end{table}

\begin{table*}[!h]
    \centering
    \caption{\xmm and \chandra best-fit spectral fitting values in the south, west, and north in the energy band $0.7-7.0$ keV.}
    \begin{tabular}{c c c c | c c}
    \hline
    \hline
    \multicolumn{4}{c|}{\xmm} & \multicolumn{2}{c}{\chandra} \\
\hline
annulus &$norm^\dagger$ & $k_BT$ & $Z$ & $norm^\dagger$ & $k_BT$\\
$[']$ &  $[10^{-4}$ cm$^{-5}/$arcmin$^{2}]$ & $[$keV$]$ & $[Z_{\odot}]$ &  $[10^{-4}$ cm$^{-5}/$arcmin$^{2}]$ &  $[$keV$]$\\ [5pt]
\hline
\multicolumn{6}{c}{South}\\
\hline
$0.0-0.55$ & $2.539_{-0.094}^{+0.096}$ & $2.524_{-0.106}^{+0.107}$ & $0.214_{-0.056}^{+0.065}$ & $2.787_{-0.22}^{+0.22}$ & $2.74_{-0.36}^{+0.5}$ \\[5pt]
$ 0.55-1.65$ & $1.212_{-0.052}^{+0.051}$ & $2.593_{-0.123}^{+0.125}$ & $0.25_{-0.067}^{+0.082}$ & $1.137_{-0.046}^{+0.046}$ & $3.35_{-0.31}^{+0.43}$ \\[5pt]
$ 1.65-2.2$ & $0.66_{-0.043}^{+0.046}$ & $2.791_{-0.232}^{+0.25}$ & $0.528_{-0.157}^{+0.179}$ & $0.911_{-0.073}^{+0.073}$ & $1.87_{-0.19}^{+0.32}$ \\[5pt]
$ 2.2-2.75$ & $0.536_{-0.034}^{+0.034}$ & $2.732_{-0.178}^{+0.256}$ & $0.371_{-0.121}^{+0.152}$ & $0.636_{-0.041}^{+0.041}$ & $2.71_{-0.34}^{+0.52}$ \\[5pt]
$ 2.75-3.3$ & $0.482_{-0.033}^{+0.033}$ & $1.93_{-0.134}^{+0.131}$ & $0.227_{-0.067}^{+0.088}$ & $0.467_{-0.027}^{+0.033}$ & $1.86_{-0.22}^{+0.23}$ \\[5pt]
$ 3.3-3.85$ & $0.385_{-0.031}^{+0.029}$ & $1.756_{-0.098}^{+0.148}$ & $0.28_{-0.075}^{+0.1}$ & $0.395_{-0.028}^{+0.023}$ & $2.24_{-0.24}^{+0.4}$ \\[5pt]
$ 3.85-4.4$ & $0.295_{-0.026}^{+0.026}$ & $1.845_{-0.147}^{+0.131}$ & $0.477_{-0.113}^{+0.146}$ & $0.323_{-0.024}^{+0.024}$ & $1.89_{-0.2}^{+0.25}$ \\[5pt]
$ 4.4-4.95$ & $0.252_{-0.022}^{+0.024}$ & $1.9_{-0.155}^{+0.134}$ & $0.479_{-0.125}^{+0.156}$ & $0.324_{-0.022}^{+0.022}$ & $2.15_{-0.24}^{+0.44}$ \\[5pt]
$ 4.95-6.05$ & $0.157_{-0.013}^{+0.014}$ & $1.631_{-0.083}^{+0.067}$ & $0.29_{-0.075}^{+0.086}$ & $0.185_{-0.013}^{+0.013}$ & $2.17_{-0.24}^{+0.41}$ \\[5pt]
$ 6.05-8.45$ & $0.059_{-0.007}^{+0.008}$ & $1.732_{-0.069}^{+0.279}$ & $0.279_{-0.108}^{+0.171}$ & $0.047_{-0.004}^{+0.004}$ & $2.17_{-0.26}^{+0.58}$ \\[5pt]
\hline
$ 4.0-6.0$ & $0.211_{-0.011}^{+0.011}$ & $1.703_{-0.042}^{+0.103}$ & $0.344_{-0.054}^{+0.041}$ & * & * \\[5pt]
$ 6.0-8.0$ & $0.068_{-0.009}^{+0.009}$ & $1.756_{-0.098}^{+0.261}$ & $0.3_{-0.109}^{+0.181}$ & * & * \\[5pt]
$ 8.0-10.0$ & $0.024_{-0.004}^{+0.003}$ & $1.72_{-0.144}^{+0.326}$ & 0.3 & * & * \\[5pt]
$ 10.0-12.5$ & $0.011_{-0.002}^{+0.003}$ & $1.709_{-0.368}^{+0.663}$ & 0.3 & * & * \\[5pt]
\hline
\multicolumn{6}{c}{West} \\
\hline
$ 0.0-0.55$ & $2.536_{-0.095}^{+0.096}$ & $2.531_{-0.105}^{+0.105}$ & $0.217_{-0.056}^{+0.064}$ & $1.99_{-0.161}^{+0.108}$ & $3.55_{-0.55}^{+0.73}$ \\[5pt]
$ 0.55-1.65$ & $1.015_{-0.041}^{+0.041}$ & $2.811_{-0.144}^{+0.151}$ & $0.503_{-0.091}^{+0.104}$ & $1.129_{-0.034}^{+0.04}$ & $3.43_{-0.27}^{+0.3}$ \\[5pt]
$ 1.65-2.75$ & $0.532_{-0.023}^{+0.023}$ & $2.629_{-0.133}^{+0.153}$ & $0.299_{-0.075}^{+0.088}$ & $0.518_{-0.03}^{+0.034}$ & $3.5_{-0.69}^{+1.02}$ \\[5pt]
$ 2.75-3.85$ & $0.266_{-0.015}^{+0.016}$ & $2.548_{-0.168}^{+0.159}$ & $0.497_{-0.115}^{+0.138}$ & $0.291_{-0.013}^{+0.013}$ & $2.71_{-0.28}^{+0.33}$ \\[5pt]
$ 3.85-8.45$ & $0.081_{-0.006}^{+0.006}$ & $1.861_{-0.166}^{+0.142}$ & $0.35_{-0.089}^{+0.105}$ & $0.064_{-0.005}^{+0.005}$ & $2.79_{-0.47}^{+0.67}$ \\[5pt]
\hline
$ 4.0-6.0$ & $0.133_{-0.006}^{+0.006}$ & $1.958_{-0.098}^{+0.09}$ & $0.396_{-0.053}^{+0.082}$ & * & * \\[5pt]
$ 6.0-8.0$ & $0.058_{-0.005}^{+0.005}$ & $1.666_{-0.166}^{+0.196}$ & $0.241_{-0.077}^{+0.08}$ & * & * \\[5pt]
$ 8.0-10.0$ & $0.014_{-0.003}^{+0.002}$ & $1.363_{-0.128}^{+0.152}$ & 0.3 & * & * \\[5pt]
$ 10.0-12.5$ & $0.002_{-}^{+0.002}$ & $1.365_{-0.499}^{+0.607}$ & 0.3 & * & * \\[5pt]
\hline
\multicolumn{6}{c}{North} \\
\hline
$ 0.0-0.55$ & $2.537_{-0.095}^{+0.096}$ & $ 2.53_{-0.1}^{+0.11}$ & $ 0.22_{-0.06}^{+0.06}$ & * & * \\[5pt]
$ 0.55-2.0$ & $0.989_{-0.056}^{+0.056}$ & $ 2.95_{-0.21}^{+0.21}$ & $ 0.6_{-0.14}^{+0.17}$ & * & * \\[5pt]
$ 2.0-4.0$ & $0.363_{-0.024}^{+0.023}$ & $ 2.22_{-0.12}^{+0.19}$ & $ 0.42_{-0.1}^{+0.13}$ & * & * \\[5pt]
$ 4.0-6.0$ & $0.159_{-0.016}^{+0.018}$ & $ 2.01_{-0.28}^{+0.29}$ & $ 0.18_{-0.1}^{+0.14}$ & * & * \\[5pt]
$ 6.0-8.0$ & $0.069_{-0.016}^{+0.019}$ & $ 1.53_{-0.25}^{+0.38}$ & $ 0.12_{-0.1}^{+0.21}$ & * & * \\[5pt]
$ 8.0-10.0$ & $0.056_{-0.015}^{+0.015}$ & $ 1.87_{-0.36}^{+0.71}$ & $ 0.22_{-0.19}^{+0.53}$ & * & * \\[5pt]
$ 10.0-12.5$ & $0.063_{-0.022}^{+0.025}$ & $ 0.99_{-0.11}^{+0.08}$ & $ 0.11_{-0.07}^{+0.12}$ & * & * \\[5pt]

\hline
\multicolumn{6}{l}{\footnotesize $^\dagger$ the presented values have been normalized by their respective source areas} \\
\multicolumn{6}{l}{\footnotesize * \chandra observations do not cover these regions} \\
    \hline
    \hline
    \end{tabular}
    \label{tab:xmm_spectralparameters}
\end{table*}

\end{appendix}

\end{document}